\documentclass{article}
\usepackage[superscript]{cite}   
\usepackage{graphicx}
\usepackage{relsize,setspace}  
\usepackage{url}               
\usepackage{fancyhdr}
\usepackage{appendix}
\usepackage{titletoc}
\newcommand\DoToC{%
  \startcontents
  \printcontents{}{2}{\textbf{Contents}\vskip3pt\hrule\vskip5pt}
  \vskip3pt\hrule\vskip5pt
}

\usepackage{amsthm}
\newtheorem{result}{Result}

\theoremstyle{definition}
\newtheorem{assumption}{Assumption}
\newtheorem{proc}{Procedure}

\newtheorem{innercustomgeneric}{\customgenericname}
\providecommand{\customgenericname}{}
\newcommand{\newcustomtheorem}[2]{%
  \newenvironment{#1}[1]
  {%
   \renewcommand\customgenericname{#2}%
   \renewcommand\theinnercustomgeneric{##1}%
   \innercustomgeneric
  }
  {\endinnercustomgeneric}
}
\newcustomtheorem{customthm}{Theorem}

\renewenvironment{proof}{{\bfseries Proof.}}

\usepackage{hyperref}
\usepackage{dcolumn}
\usepackage[ruled,vlined]{algorithm2e}
\usepackage{algorithmic}
\usepackage{ulem} 
\usepackage[table,xcdraw]{xcolor}
\usepackage{graphicx,float}
\usepackage{booktabs}
\usepackage{amsmath,amsfonts,amssymb,mathtools}
\usepackage{makecell}
\newcommand\numberthis{\addtocounter{equation}{1}\tag{\theequation}}
\newcommand{\indep}{\perp \!\!\! \perp}
\allowdisplaybreaks
\usepackage{enumerate}
\usepackage{authblk}
\usepackage{bbm}

\newcommand{\reals}{\mathbbm{R}}

\usepackage{float}
\DeclareGraphicsExtensions{.pdf, .jpg, .png}
\setkeys{Gin}{width=0.85\textwidth}

\title{Regression-Based Proximal Causal Inference}
\date{\vspace{-10ex}}
\author[1]{Jiewen Liu\thanks{jiewen.liu@pennmedicine.upenn.edu}}
\author[2]{Chan Park\thanks{chanpk@wharton.upenn.edu}}
\author[3]{Kendrick Li\thanks{Kendrick.Li@STJUDE.ORG}}
\author[2,1]{Eric J. Tchetgen Tchetgen\thanks{ett@wharton.upenn.edu}}

\affil[1]{Department of Biostatistics, Perelman School of Medicine, University of Pennsylvania}
\affil[2]{Department of Statistics and Data Science, Wharton School, University of Pennsylvania}
\affil[3]{Department of Biostatistics, St. Jude Children’s Research Hospital}

\begin{document}

\maketitle
\begin{abstract} 
Negative controls are increasingly used to evaluate the presence of potential unmeasured confounding in observational studies. Beyond the use of negative controls to detect the presence of residual confounding, proximal causal inference (PCI) was recently proposed to de-bias confounded causal effect estimates, by leveraging a pair of treatment and outcome negative control or confounding proxy variables. While formal methods for statistical inference have been developed for PCI, these methods can be challenging to implement as they involve solving complex integral equations that are typically ill-posed.  We develop a regression-based PCI approach, employing two-stage generalized linear regression models (GLMs) to implement PCI, which obviates the need to solve difficult integral equations. The proposed approach has merit in that (i) it is applicable to continuous, count, and binary outcomes cases, making it relevant to a wide range of real-world applications, and (ii) it is easy to implement using off-the-shelf software for GLMs. We establish the statistical properties of regression-based PCI and illustrate their performance in both synthetic and real-world empirical applications.
\end{abstract}
\begingroup
\let\clearpage\relax
\section{Introduction}
Identification of causal effects can be challenging in observational studies because the standard exchangeability or treatment unconfoundedness conditions upon which causal inference commonly relies is threatened by the presence of confounding by hidden factors \cite{hernan2010causal}. To achieve unconfoundedness in practice typically requires a complete understanding of key causes of the treatment selection mechanism that also predict the outcome in view, and that such common causes of the treatment and outcome variables can be measured with sufficient accuracy that one can fully account for them empirically. However, even in carefully designed observational studies, where all sources of confounding are known, key drivers of confounding can rarely if ever be measured without error. As a result, the most one can hope for in practice is that measured confounders are (error-prone) proxies of underlying confounders, thus residual unmeasured confounding may be present even in the most well-designed observational studies. For instance, in studies of the causal impact of HIV treatment, pre-treatment CD4 count measurement is commonly used to control for confounding by indication as it is invariably used by physicians for treatment initiation decision-making \cite{hernan2002estimating,hiv2010effect}.  However, CD4 count measurements are notoriously error-prone and are at most imperfect snapshots of the evolving immune system status driving disease progression. As such, in scenarios where CD4 count directly affects treatment, CD4 count measurement is a proxy of the underlying source of confounding which remains unmeasured \cite{zivich2022bridged}. Pre-exposure outcome measurements sometimes provide good candidate proxies of confounding, e.g. Asthma rates at a county level measured the week before PM2.5 measurements are good confounding proxies for the causal effect of PM2.5 exposure on the county's Asthma rate in the subsequent week in a typical air pollution time series study \cite{hu2023using}.
\\ \\
A common type of proxy variable routinely used in epidemiology studies are so-called negative control variables, typically used to evaluate the potential presence of unmeasured confounding \cite{shi2020selective}. Specifically, an observed exposure is said to be a negative control exposure (NCE) (more broadly a treatment confounding proxy) if it is a priori known not to causally impact the primary outcome of interest, 
and is relevant, i.e. associated with an unmeasured confounder. Thus, such an NCE can be used to detect residual confounding by assessing the extent to which it is empirically associated with the outcome given measured covariates. Likewise, a negative control outcome variable (NCO) (more broadly an outcome confounding proxy) must be known a priori not to be causally impacted by the primary treatment, and must be relevant and therefore associated with a hidden confounder. An extensive list of common examples of proxy variables encountered in observational studies is given in Lipsitch et al. (2010), Shi et al. (2023) and Zafari et al. (2023) \cite{lipsitch2010negative,shi2023current,zafari2023state}. The last paper provides a state of use of proxy methods in pharmacoepidemiology. Although such NCE and NCO variables have been used quite widely in epidemiology to detect confounding bias \cite{vandenbroucke2019test, lipsitch2010negative,zhu2022estimating}, only recently has formal NC and proxy theory and methods been developed to debias causal effect estimates \cite{li2023double}. Such methods have been referred to as proximal causal inference (PCI) \cite{tchetgen2020introduction,zivich2023introducing}.
\\ \\
Specifically, Miao et al. (2016) study the problem of identification using proxy variables, and establish that given a pair of treatment and outcome confounding proxy variables ($Z$ and $W$ respectively) that are sufficiently relevant for hidden confounders, it is possible to nonparametrically identify a causal effect in the presence of unmeasured confounding \cite{miao2018identifying}. Tchetgen Tchetgen et al. (2023) generalize PCI to complex longitudinal studies, and derive the proximal g-formula to identify the joint causal effects of time-varying treatments subject to both measured and hidden time-varying confounding \cite{tchetgen2020introduction}, thus effectively extending the work of James Robins to the proximal setting \cite{robins1997causal}. They also propose proximal g-computation, a proxy-based generalization of the g-computation algorithm. 
\\ \\
In the continuous outcomes case, under linear models for NCO $W$ and primary outcome $Y$, Tchetgen Tchetgen et al. (2023) establish that the corresponding proximal g-computation algorithm can be implemented by following a two-stage least-squares procedure \cite{tchetgen2020introduction}. A key limitation of the proposed two-stage least-squares approach is that it does not easily extend to binary, polytomous or count outcomes for which linear models are generally inappropriate.
\\ \\ 
In this paper, we extend proximal g-computation for a point treatment via a two-stage regression approach which accommodates continuous, count and binary $W$ and $Y$ via generalized linear models (GLMs) with canonical link  function (identity, log and logit) that appropriately account for the outcome type. The paper is organized as follows. In the next section, we briefly review causal Directed Acyclic Graphs (DAGs) encoding key assumptions of the proximal causal learning framework and we review the two-stage least-squares estimation given in Tchetgen Tchetgen et al. (2023) for continuous outcome and outcome proxy variables. Next, we consider more general settings, where the primary outcome or the outcome proxy is either a count, binary or polytomous variable and we propose appropriate two-stage regression-based PCI approaches to identify and estimate the corresponding causal effect. For each combination of link functions, we provide a detailed formulation of the corresponding estimation procedure, and formally establish its validity. Detailed proofs and derivations are relegated to the Appendix. We also provide annotated \texttt{R} and \texttt{SAS} codes to execute the two-stage regression procedure proposed in this paper in the Appendix \ref{demo}. Additionally, we have developed \texttt{R} package \texttt{pci2s}, to conveniently estimate causal effects with consistent standard error estimates and corresponding confidence intervals. For illustration, we re-analyze data from the Study to Understand Prognoses and Preferences for Outcomes and Risks of Treatments (SUPPORT)\cite{connors1995controlled}, an observational study that evaluates the causal effect of right heart catheterization (RHC) on survival, length of hospitalization, intensity of care, and costs.
\section{Methodology}
\subsection{Notation and Causal Structure}
Throughout, we suppose that measured and unmeasured variables consist of independent and identically distributed (i.i.d.) samples $(Y,A,X,Z,W,U)$: $Y$ denotes the primary outcome; $A$ denotes the primary treatment; $X$ denotes measured confounders; $Z$ denotes treatment confounding proxies; $W$ denotes outcome confounding proxies; $U$ denotes unmeasured confounders. The causal structures for these variables can be represented by the three DAGs in Figure \ref{fig:1} of increasing complexity. Comparing DAG (1) to DAG (2), the latter allows for direct causal links from $Z$ to $A$ and from $W$ to $Y$. DAG (3) further includes measured confounders $X$. 
\begin{figure}[h]
\centering
\includegraphics[width=1\textwidth]{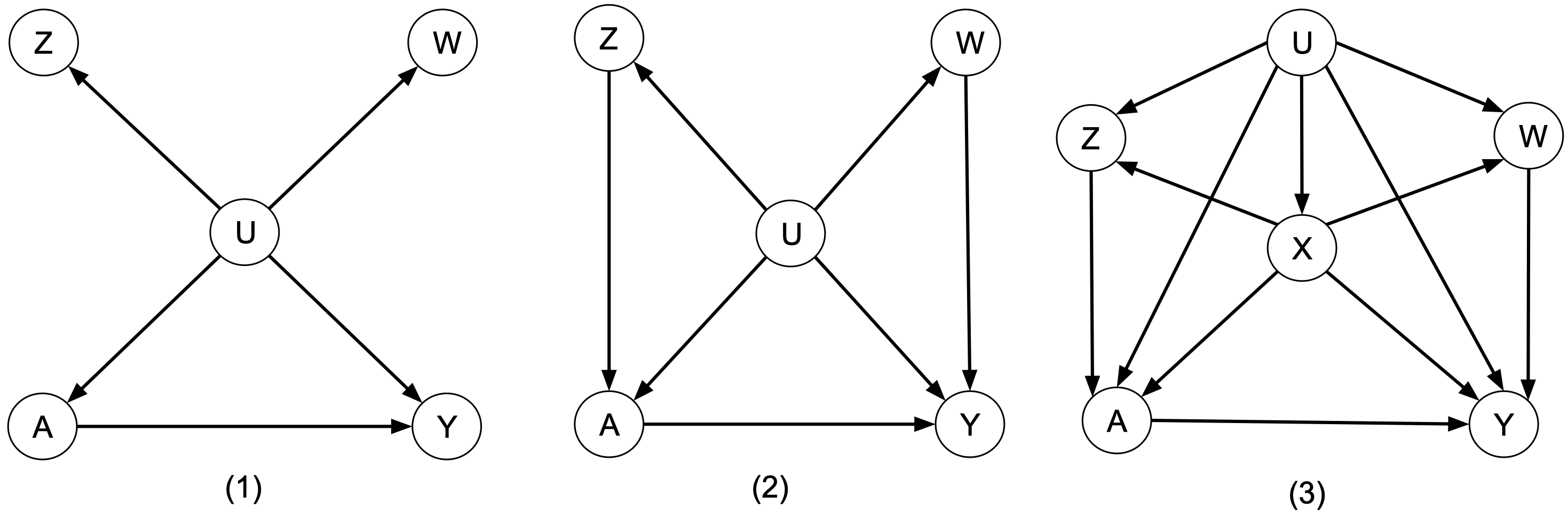}
\caption{Three Common DAGs in Classical Settings of Negative Control}
\label{fig:1}
\end{figure}
\\ \\Based on DAGs (3), we formalize key assumptions proxy variables must satisfy \cite{miao2018identifying,tchetgen2020introduction}. Let $Y_a$ denote a person's potential outcome had they, possibly contrary to fact, received treatment $A=a$. In the following, $A$ can be binary, polytomous, or continuous. Throughout, we assume that $U$ includes all unmeasured common causes of $A$ and $Y$ given $X$:
\begin{align*}
	Y_{a} \indep A &|X,U \numberthis \label{cf1}
\end{align*}
throughout, W and Z are considered valid outcome and treatment confounding proxies for the $A-Y$ association, respectively, if they satisfy the following key conditions:
\begin{align*}
	Z \indep Y &| A,X,U \numberthis \label{cod1} \\
	W \indep (A,Z) &| X,U \numberthis \label{cod2}
\end{align*}
Assumption \eqref{cod1} encodes the condition that as valid treatment confounding proxy, $Z$ must not be associated with $Y$ given $(A,X,U)$; while Assumption \eqref{cod2} encodes the condition that as valid outcome confounding proxy $W$ must not be associated with $(A,Z)$ given $(X,U)$. Of note, $Z$ and $A$ may be causally associated and so can $W$ and $Y$ provided \eqref{cod1} and \eqref{cod2} hold. Notably, proxies that are conditionally independent of both $A$ and $Y$ given $X,U$ may be specified as either treatment or outcome confounding proxy. 
\\ \\
In subsequent sections, consistent with DAG (2) we suppress all measured covariates $X$ in order to simplify the exposition of the proposed two-stage regression approach. We focus on model specification, fitting and estimation of causal effects via two-stage regression with GLMs on outcomes $Y$ and $W$ with identical link functions, deferring details on the use of different link functions, large sample behavior, statistical inference and simulation studies to Section \ref{uni} of the Appendix. The approach readily extends to the setting depicted in DAG (3) by incorporating measured covariates $X$ in both regression models of the two-stage approach. All results hold in DAG (1) a submodel of DAG (3), where $W$ and $Y$, and $A$ and $ Z$ are conditionally independent given $U$, respectively. For inference, one may either use the nonparametric bootstrap, or an empirical estimate of the asymptotic variance we derive in  Section \ref{theori} of the Appendix.

\subsection{Continuous $(Y,W)$}
\label{conti}
We briefly review the two-stage least-squares approach for continuous $(Y,W)$ in which case the identity link function is appropriately specified \cite{tchetgen2020introduction}. It may be reasonable to assume that the following linear structural models hold: 
\begin{align*}
	&E[Y|A,Z,U] = \beta_0+\beta_aA+\beta_uU  \numberthis \label{c1} \\ 
	&E[W|A,Z,U] = \alpha_0 + \alpha_uU \numberthis \label{c2}
\end{align*}
Note that, compatible with DAG (2), $(Z,W)$ are valid treatment and outcome confounding proxies, and therefore $Z$ does not appear on the right-hand side of model \eqref{c1}, and $(A,Z)$ do not appear on the right-hand side of model \eqref{c2}. In addition, model \eqref{c1} excludes any $A-U$ interactions. Our main objective is to estimate $\beta_a$ which encodes the causal effect of $A$ on $Y$ on the additive scale: 
$$ E[Y^{a=1}-Y^{a=0}|U]=E[Y^{a=1}|U]-E[Y^{a=0}|U]= E[Y|A=1,Z,U] - E[Y|A=0,Z,U]= \beta_a$$
An analogous counterfactual interpretation holds for continuous and polytomous treatments, details are omitted. \\ \\
However, model \eqref{c1} involves conditioning for the unobserved confounder $U$ and therefore,  $\beta_a$ cannot be recovered by directly estimating the regression model by standard least-squares fit. To address this, we leverage our proxies and instead evaluate the conditional expectation on both sides of models \eqref{c1} and \eqref{c2}, conditional on $A$ and $Z$:
\begin{align*} 
	&E[Y|A,Z] = \beta_0+\beta_aA+\beta_u E[U|A,Z]   \numberthis \label{c3}\\
	&E[W|A,Z] = \alpha_0 + \alpha_u E[U|A,Z]   \numberthis \label{c4} 
\end{align*}
These equations thus establish that $E[W|A,Z]$ and $E[Y|A,Z]$ are both linearly related with $E[U|A,Z]$, implying equation \eqref{c5} below, which identifies the causal effect $\beta_a$ upon rearranging a few terms as previously reported in Tchetgen Tchetgen et al.:\\
\begin{result} \label{result-1}
Under assumptions \eqref{c1}and \eqref{c2}
one obtains    
\begin{align*}
	&E[Y|A,Z] = \beta_0^* + \beta_a^*A + \beta_u^* E[W|A,Z],  \numberthis \label{c5} \\
	&\text{where } 
	\beta_0^*= \beta_0 - \beta_u \frac{\alpha_0}{\alpha_u}, \ \beta_a^* =\beta_a, \ \beta_u^* = \frac{\beta_u}{\alpha_u}, \text{ provided that } \alpha_u \neq 0. 
\end{align*}
\end{result} 
The condition $\alpha_u \neq 0$ encodes the requirement that $W$ is relevant for, and therefore associated with $U$. Result \ref{result-1} states that if $E[W|A,Z]$ were known, one could in principle estimate $\beta_a$ by regressing $Y$ on $(A,E[W|A,Z])$ via standard least-squares fitting. In practice, as $E[W|A,Z]$ will not generally be known, Tchetgen Tchetgen et al. proposed to estimate it in the first stage by fitting a regression of $W$ on $(A,Z)$, to obtain fitted values for each observation, $\hat{E}[W|A,Z]$ from, say regression equation \eqref{c6} below, although more flexible models can in principle be used (see Section \ref{uni} for further details): 
\begin{align*}
	& E[W|A,Z] = \alpha_0^* + \alpha_a^*A +\alpha_z^*Z  \numberthis \label{c6} 
\end{align*}
With these assumptions, Tchetgen Tchetgen et al. formulated the following two-stage least-squares regression procedure to estimate $\beta_a$:
\begin{enumerate}
  \item Specify a linear regression function for $E[W|A,Z]$. Say $W\sim A+Z$.
  \item Perform the first-stage linear regression $W\sim A+Z$.
  \item Compute $\hat{E}[W|A,Z]= \hat{\alpha}_0^* +\hat{\alpha}_a^*A + \hat{\alpha}_z^*Z$ using the estimated coefficients from the first-stage regression fit.
  \item Perform the second-stage linear regression $Y \sim A + \hat{E}[W|A,Z]$ to obtain $\hat{\beta}_a$.
\end{enumerate}
As they establish, the resulting estimator $\hat{\beta}_a$ is consistent and asymptotically normal under the stated assumptions, provided that $W$ and $Z$ are $U-$relevant such that the coefficient $\alpha_z^* \neq 0$. Furthermore, as they point out, this two-stage least-squares procedure is reminiscent of an analogous two-stage least-squares approach routinely used in the instrumental variable literature. This analogy is useful as it implies that one can implement the above proximal two-stage linear regression approach and obtain corresponding inference using off-the-shelf software such as the \texttt{R} function ``ivreg''\cite{tchetgen2020introduction}. Moreover, it is important to note that the regression equations \eqref{c1} and \eqref{c2} are not contingent upon the presence of direct causal links from $Z$ to $A$ and from $W$ to $Y$. This indicates that the same regression procedure to estimate $\beta_a$ applies to DAG (1). 
\\ \\
Regarding the violation of key assumptions for valid proxies, identification of the treatment effect would not be possible without invoking additional assumptions. For example, if exclusion condition \ref{cod2} $W \indep (A, Z) \mid X, U$ is violated and $A$ has a direct causal effect on $W$, the first-stage linear regression equation may be 
\begin{align*}
	& E[W|A, Z] = \alpha_0 + \alpha_a A + \alpha_u E[U|A, Z] \numberthis \label{vio1} 
\end{align*}
instead of equation \ref{c4}. And if the main effect of $A$ in $E[U|A,Z]$ is linear, for example $E[U|A,Z] = \gamma_0 + \gamma_a A+\gamma_uU $, it may take the form
\begin{align*}
	& E[W|A, Z] = (\alpha_0+\alpha_u\gamma_0) + (\alpha_a+ \alpha_u\gamma_a) A + \alpha_u\gamma_z Z \numberthis \label{vio2}.
\end{align*}
In this scenario, the result of second-stage linear regression equation \ref{c3} of $Y$ on $A$ and $E[W|A,Z]$ from \ref{vio1} will not be able to distinguish $\beta_a^*$ from $\beta_a$ and $\beta_a - \beta_u \frac{\alpha_a}{\alpha_u}$. 
\\ \\
Having reviewed two-stage least-squares, we are now ready to consider settings where $Y$ and $W$ may no longer be continuous, in which case linear models may no longer be appropriate. While we focus the exposition on the canonical case where both variables $W$ and $Y$ are of the same type (e.g. both are counts or binary), we later discuss how the methods can be adapted to accommodate settings where they are of different types.
\subsection{Count $(Y,W)$}
In this Section, suppose that $(Y,W)$ are both counts in which case a log link function is needed to model the mean of variables appropriately. In this vein, consider the corresponding structural models: 
\begin{align*}
	& \mathrm{log}(E[Y|A,Z,U]) =  \beta_0+\beta_aA+\beta_uU \numberthis \label{lo1} \\
	& \mathrm{log}(E[W|A,Z,U]) =  \alpha_0 + \alpha_uU \numberthis \label{lo2} \\
	& U|A,Z \sim E[U|A,Z] + \epsilon; \ E[\epsilon] = 0; \ \epsilon \indep A,Z; \numberthis \label{as2} \\
        & \textrm{the marginal distribution of $\epsilon$ is unrestricted.} 
\end{align*}
Under \eqref{lo1}, $\beta_a$ can be interpreted as a multiplicative causal effect of the treatment conditional on $U$, that is, in the case of binary $A$:
$$ \mathrm{log} \left( \frac{E[Y^{a=1}|U]}{E[Y^{a=0}|U]} \right) = \mathrm{log} \left( \frac{E[Y|A=1,Z,U]}{E[Y|A=0,Z,U]} \right)=\beta_a$$
Based on this model specification,  marginalizing over $U$ yields the following model relating the conditional expectation of $Y$ and $W$ given $(A,Z)$ (see Section \ref{lo-lo} of the Appendix for a detailed derivation of the result).\\
\begin{result} \label{result-2}
Under assumptions \eqref{lo1}, \eqref{lo2} and \eqref{as2}, one obtains
\begin{align*}
	& \mathrm{log}(E[Y|A,Z]) = \beta_0^*+\beta_a^*A+\beta_u^* \mathrm{log}(E[W|A,Z]), \numberthis \label{lo5} \\
	&\text{where } \beta_0^*= \tilde{\beta}_0 - \beta_u \frac{\tilde{\alpha}_0}{\alpha_u}, \ \beta_a^*=\beta_a, \ \beta_u^* = \frac{\beta_u}{\alpha_u}, \text{ provided that } \alpha_u \neq 0; \\
		&\text{$\tilde{\beta}_0$ and $\tilde{\alpha}_0$ are defined in Section \ref{lo-lo} of the Appendix.}
\end{align*}
\end{result} 
Therefore, analogous to the previous development in the case of linear models, we conclude that, here again, were one given access to $E[W|A,Z]$, one would recover a consistent estimator of $\beta_a$ by fitting a Poisson regression model for $Y$ given $A$, adjusting for $\mathrm{log}(E[W|A,Z])$ as a regressor.  In practice, the latter may not be feasible as one may not generally know $E[W|A,Z]$ which must therefore be estimated. Thus, the following two-stage regression can be implemented. 
\begin{enumerate}
  \item Specify a linear regression function for $\mathrm{log}(E[W|A,Z])$. Say $W\sim A+Z$.
  \item Perform the first-stage Poisson regression $W\sim A+Z$.
  \item Compute $\mathrm{log}(\hat{E}[W|A,Z])= \hat{\alpha}_0^* + \hat{\alpha}_a^*A + \hat{\alpha}_z^*Z$ using the estimated coefficients from the first-stage regression.
  \item Perform the second-stage Poisson regression $Y \sim A + \mathrm{log}(\hat{E}[W|A,Z])$ to obtain $\hat{\beta}_a$.
\end{enumerate}
This result formally establishes proximal two-stage Poisson regression as a practical approach for implementing PCI in the context of count primary outcome and outcome confounding proxy variables. For Poisson regression with overdispersion, one may either use a standard sandwich estimator for robust variance estimation or opt for a Negative Binomial regression. Both methods have been implemented in the \texttt{R} package ``pci2s''. Next, we consider the case of binary outcomes.
\subsection{Binary $(Y,W)$}
\label{bi}
Suppose $W$ and $y$ are dichotomous. In such a case, one may wish to posit models that appropriately account for the natural scale of the outcomes. As the $\mathrm{logit}$ link is routinely used for binary outcomes, thus suppose that a logistic structural equation model is appropriate to model such data. In this setting, we show that some care is required in specifying an appropriate two-stage regression approach which accounts for the nonlinear logit link function.
\\\\
To proceed, consider the following logistic structural equation models, compatible with DAG (2):
\begin{align*}
	&\mathrm{logit}(\mathrm{Pr}(Y=1|A,Z,W,U)) = \beta_0+\beta_aA+\beta_uU+\beta_w W  \numberthis \label{g1} \\
	&\mathrm{logit}(\mathrm{Pr}(W=1|A,Z,Y,U)) = \alpha_0+\alpha_uU+\alpha_y Y  \numberthis \label{g2} \\
	&U|A,Z,Y=0,W=0 \sim E[U|A,Z,Y=0,W=0] + \epsilon; \ E[\epsilon]=0; \ \epsilon \indep (A,Z) | Y=0,W=0; \numberthis \label{as3} \\
	&\textrm{the distribution of $\epsilon|Y=0,W=0$ is unrestricted.}  \\
	&\text{\eqref{g1} and \eqref{g2} $\implies$ }\beta_w=\alpha_y
\end{align*}
Unlike the identity and log links, we note that the logistic structural regression model for $Y$ conditions on $W$ and that for $W$ conditions on $Y$, which appears to be necessary to appropriately account for the fact that by symmetry of odds ratios, $\beta_w=\alpha_y$ because the odds ratio association between $Y$ and $W$ given $U$ and $A$  by definition must match that between $W$ and $Y$ given $U$ and $A$. Such restriction does not necessarily hold in the case of identity or log link functions and therefore was not required, leading to a simplification in the model formulation. Under this logistic structural equation model, the causal effect of interest is given by 
\begin{align*}
	& \mathrm{log} \left( \frac{\mathrm{Pr}(Y^{a=1}=1|W,U)/\mathrm{Pr}(Y^{a=1}=0|W,U)}{\mathrm{Pr}(Y^{a=0}=1|W,U)/\mathrm{Pr}(Y^{a=0}=0|W,U)} \right) \\
	&= \mathrm{logit}(\mathrm{Pr}(Y=1|A=1,Z,W,U)) - \mathrm{logit}(\mathrm{Pr}(Y=1|A=0,Z,W,U))=\beta_a 
\end{align*}
which encodes the conditional causal log-odds ratio relating $A$ to $Y$ given $(U,W)$. 
\\ \\
In Section \ref{linot-linot} of the Appendix, we establish that the specified structural model implies the following observed data model for $\mathrm{Pr}(Y=1|A,Z,W)$ and $\mathrm{Pr}(W=1|A,Z,Y)$. 
\\ 
\begin{result} \label{result-3}
    Under assumptions \eqref{g1}, \eqref{g2} and \eqref{as3}, it follows that    
\begin{align*}
&\mathrm{logit}(\mathrm{Pr}(Y=1|A,Z,W)) = \beta_0^* +\beta_a^*A + \beta_u^*\mathrm{logit}(\mathrm{Pr}(W=1|A,Z,Y=1)) + \tilde{\beta}_wW, \numberthis \label{g5} \\
&\text{where } \beta_0^*=\tilde{\beta}_0 - \beta_u \frac{(\tilde{\alpha}_0+\tilde{\alpha}_y)}{\alpha_u}, \ \beta_a^*=\beta_a, \ \beta_u^*=\frac{\beta_u}{\alpha_u}, \text{ provided that } \alpha_u \neq 0; \\
&\text{$\tilde{\beta}_0$, $\tilde{\alpha}_0$, $\tilde{\alpha}_y$ and $\tilde{\beta}_w$ are defined in Section \ref{linot-linot} of the Appendix.} 
\end{align*}
\end{result} 
Result \ref{result-3} states that $\beta_a$ can be estimated by fitting a logistic regression of $A$ and $W$ adjusting for $\mathrm{logit}({\mathrm{Pr}}(W=1|A,Z,Y=1))$ as a covariate. As the latter is not known,  one may proceed in two stages as before , first by estimating $\mathrm{logit}(\widehat{\mathrm{Pr}}(W=1|A,Z,Y=1))$ from a logistic regression for the conditional probability of $W=1$ conditional on $(A,X,Y)$.
The two-stage regression procedure to estimate $\beta_a$ is thus as follows: 
\begin{enumerate}
  \item Specify a linear regression function for $\mathrm{logit}(\mathrm{Pr}(W=1|A,Z,Y))$. Say $W \sim A+Z+Y$.
  \item Perform the first-stage logistic regression $W\sim A+Z+Y$.
  \item Compute $\mathrm{logit}(\widehat{\mathrm{Pr}}(W=1|A,Z,Y=1))= \hat{\alpha}_0^* + \hat{\alpha}_a^*A + \hat{\alpha}_z^*Z + \hat{\alpha}_y^*$ using the estimated coefficients from the first-stage regression.
  \item Perform the second-stage logistic regression $Y \sim A + \mathrm{logit}(\widehat{\mathrm{Pr}}(W=1|A,Z,Y=1))+W$ to obtain $\hat{\beta}_a$.
\end{enumerate}
The above model specification excludes interactions between $(Y,W)-U$, which restricts specifying more flexible regression functions. However, such a restriction can be relaxed by allowing for certain interactions as detailed in Section \ref{lit-lit} of the Appendix. We should also note that the approach requires no modification even if $W$ is a priori known not to cause $Y$ as in DAG (1).
\subsection{Polytomous $(Y,W)$}
\label{poly}
The proposed approach readily extends with a slight modification to accommodate polytomous outcome and outcome proxy by fitting polytomous logistic regression models rather than simple logistic regression models. Let us consider a scenario where 
$W$ has $K+1$ categories $(0,\ldots,K)$, $Y$ has $T+1$ categories $(0,\ldots,T)$ and scalar $U$. We set $0$ as the reference level for $W$ and $Y$ and proceed with the following parameterization:
\begin{align*}
	&\mathrm{logit} (\mathrm{Pr}(W=k))
	 = \mathrm{log} \left( 
	 \frac{\mathrm{Pr}(W=k)}{\mathrm{Pr}(W=0)}
	 \right)
	 \text{ and }
	 \mathrm{logit} (\mathrm{Pr}(Y=t))
	 = \mathrm{log} \left( 
	 \frac{\mathrm{Pr}(Y=t)}{\mathrm{Pr}(Y=0)}
	 \right), \\
    &\textrm{for $k=1,\ldots,K$ and $t=1,\ldots,T$.}
\end{align*}
Suppose then that the following structural regression models hold for $k=1,\ldots,K$ and $t=1,\ldots,T$: 
\begin{align*}
& \mathrm{logit} (\mathrm{Pr}(Y=t|A,Z,W,U)) = \beta_{0t} + \beta_{at}A  + \beta_{ut}U+ \sum_{k=1}^K  \beta_{wtk} I(W=k) \numberthis \label{p1} \\
& \mathrm{logit} (\mathrm{Pr}(W=k|A,Z,Y,U)) = \alpha_{0k} + \alpha_{uk} U + \sum_{t=1}^T \alpha_{ykt} I(Y=t)  \numberthis \label{p2} \\
&U|A,Z,Y=0,W=0 \sim E[U|A,Z,Y=0,W=0] + \epsilon;  \ E[\epsilon]=0;  \numberthis \label{as4} \\
	&\epsilon \indep (A,Z)|Y=0,W=0; \textrm{ the distribution of $\epsilon|Y=0,W=0$ is unrestricted.}  \\
	& \text{\eqref{p1} and \eqref{p2} $\implies$ } \beta_{wtk} = \alpha_{ykt}
\end{align*}
\begin{result} \label{result-4}
    Under assumptions \eqref{p1}, \eqref{p2} and \eqref{as4}, it follows that
\begin{align*}
	&\mathrm{logit} (\mathrm{Pr}(Y=t|A,Z,W))  \\
	&= \beta_{0t}^* + \beta_{at}^*A+ \beta_{ut}^*\sum_{k=1}^K\mathrm{logit} (\mathrm{Pr}(W=k|A,Z,Y=0))  +  \sum_{k=1}^K \tilde{\beta}_{wtk}I(W=k) \ \textrm{, where $\beta_{at}^*=\beta_{at}$.}  \numberthis \label{p5} 
\end{align*}
\end{result}
Result \ref{result-4} suggests the following two-stage regression approach as a straightforward generalization to polytomous outcomes:
\begin{enumerate}
  \item Specify a linear regression function for $\mathrm{logit}(\mathrm{Pr}(W=k|A,Z,Y))$. Say $W \sim A+Z+Y$.
  \item Perform the first-stage polytomous logistic regression $W\sim A+Z+Y$.
  \item Compute $\sum_{k=1}^K \mathrm{logit}(\widehat{\mathrm{Pr}}(W=k|A,Z,Y=0))= \sum_{k=1}^K \left(  \hat{\alpha}_{0k}^* + \hat{\alpha}_{ak}^*A + \hat{\alpha}_{zk}^*Z \right)$ using the estimated coefficients from the first-stage regression.
  \item Perform the second-stage polytomous logistic regression $Y \sim A + \sum_{k=1}^K \mathrm{logit}(\widehat{\mathrm{Pr}}(W=k|A,Z,Y=0))+W$ to obtain $\hat{\beta}_{at}$.
\end{enumerate} 
Multidimensional \(U\) can be accommodated with certain rank conditions satisfied for high-dimensional \((Z, W)\). In Section \ref{lip-lip} of the Appendix, we establish a more general version of Result 4 and outline the corresponding regression procedure for multidimensional \(U\).

\section{Application: Right Heart Catheterization Treatment Effect}
To illustrate the proposed proximal two-stage regression approach in practice, we revisit the SUPPORT study. The objective is to estimate the treatment effect of right heart catheterization (RHC) performed within the first 24 hours of ICU stay on 30-day survival outcome. The SUPPORT dataset comprises information on 5735 patients from a large prospective cohort study conducted at 5 medical centers. The treatment $A$ indicates receiving RHC treatment, and the outcome $Y$ is the survival time up to 30 days. The study also collected rich information encoded in 73 covariates including basic demographics (e.g. age, sex, race, education, income and insurance status), estimated probability of survival, comorbidity, vital signs, physiological status and functional status. Ten covariates measuring a patient's overall physiological status were measured from a blood test drawn within 24 hours in the ICU: serum sodium, serum potassium, serum creatinine, bilirubin, albumin, PaO2/(.01*FiO2) ratio, PaCO2, serum PH (arterial), white blood cell count, and hematocrit. These variables may be subject to substantial measurement error and as single snapshots of the underlying physiological state over time can be considered potential confounding proxies.
\\ \\
Following Tchetgen Tchetgen et al. (2023), we specify biomarker measurements $Z= \textrm{(pafi1, paco21)}$ and $W=\textrm{(ph1, hema1)}$ \cite{tchetgen2020introduction} as proxies, and include all remaining physiological status measurements, medical history, and baseline variables such as age and gender in $X$. Subsequently, the authors applied the proximal two-stage least-squares approach to estimate an effect of RHC on survival days equal to $\hat{\beta}_a(\text{Proximal}) = -1.80$ with 95\% C.I. $[-2.64,-0.96]$, such that on average a patient receiving RHC would live 1.8 days less than had he or she not received RHC. This result was significantly larger than that obtained by standard linear regression, adjusting for all covariates $\hat{\beta}(OLS) = -1.25$ days with 95\% C.I. $[-1.80,-0.70]$.
\\
\begin{table}[!thp]
\centering
\begin{tabular}{ll}
\hline  
\multicolumn{1}{c}{\textbf{Variable }} & \multicolumn{1}{c}{\textbf{Description}}                                       \\
\hline  
Y (Binary)         &  1 if the patient is alive at 30th day and 0 otherwise. \\ \hline
A (Binary)         &  1 if the RHC treatment is performed and 0 otherwise. \\ \hline
X            & Remaining 68 measured covariates including other physiological measurements.  \\ \hline
Z            & pafi1, paco21.        \\ \hline
W   (Polytomous)        & ph1, hema1 encoded by 1 if greater than the median and 0 otherwise.      \\ 
				        & W=0 if (ph1=0, hema1=0); W=1 if (ph1=1, hema1=0); W=2 if (ph1=0, hema1=1);        \\ 
				        & W=3 if (ph1=1, hema1=1);      \\ \hline
U            & Disease Severity Status.                \\ \hline
\end{tabular}
\caption{\label{table:2} Definition of variables included in models. }
\end{table}

In our illustration of the method in the binary outcome case, we consider the dichotomized primary outcome $Y$ indicating whether a patient survived at least 30 days. Furthermore, outcome confounding proxies $W=\textrm{(ph1, hema1)}$ are likewise dichotomized according to their median value (0, lower than the median; 1, higher than the median) and then further combined into a polytomous variable $W$. For data analysis, following the proposed two-stage approach for the binary primary outcome, in the first stage, we fit a polytomous logistic regression for $W|A,Z,X,Y$ under linear specification on the logit scale, i.e. $W \sim A+Z+X+Y$ from which we obtain fitted values 
$\sum_{k=1}^3\mathrm{logit}(\widehat{\mathrm{Pr}}(W=k|A,Z,Y=0))$ 
which is included as a covariate in the second stage logistic regression for $Y|A,Z,X,W$ as described in Section \ref{poly}. 
\begin{align*}
&\mathrm{logit}(\widehat{\mathrm{Pr}}(W=k|A,Z,X,Y)) = \hat{\alpha}_{0k}^* + \hat{\alpha}_{ak}^*A +\hat{\alpha}_{zk}^*Z + \hat{\alpha}_{xk}^*X + \hat{\tilde{\alpha}}_{yk}Y, \ \text{where } k\in\{1,2,3\},  \\ 
&\mathrm{logit}(\widehat{\mathrm{Pr}}(Y=1|A,Z,X,W)) = \hat{\beta}_0^* + \hat{\beta}_a^* A + \hat{\beta}_x^*X  + \hat{\beta}_u^*\sum_{k=1}^3\mathrm{logit}(\widehat{\mathrm{Pr}}(W=k|A,Z,Y=0)) + \sum_{k=1}^3 \hat{\tilde{\beta}}_{wk} I(W=k), \\
&\text{where } \hat{\beta}_a^* = \hat{\beta}_a.
\end{align*}
The approach yields an estimated treatment effect $\hat{\beta}_a(Proximal) = -0.40$ with 95\% C.I. $[-0.56,-0.26]$, such that on average a 0.40 decrease in the log odds of 30-day survival is anticipated for patients receiving RHC.
\\ \\
In comparison, a standard logistic regression model, which includes treatment and adjusts for all measured covariates gives a slight attenuation of the estimated causal effect $\hat{\beta}_a(MLE) = -0.36$ with 95\% C.I. $[-0.51,-0.21]$. 
\begin{align*}
&\mathrm{logit}(\widehat{\mathrm{Pr}}(Y=1|A,Z,W,X)) = \hat{\beta}_0 + \hat{\beta}_a A + \hat{\beta}_z Z + \hat{\beta}_xX + \sum_{k=1}^3 \hat{\beta}_{wk} I(W=k)
\end{align*}
Similar to Tchetgen Tchetgen et al.\cite{tchetgen2020introduction}, who found that the effect estimate using proximal two-stage least-squares was significantly larger than the standard OLS estimate, our proximal two-stage logistic regression results likewise suggest that RHC may exert a relatively more adverse effect on 30-day survival than estimated via standard adjusted logistic regression analysis. 
\section{Discussion}
In this paper, we have reviewed proximal G-computation via proximal two-stage least-squares regression approach due to Tchetgen Tchetgen et al. (2023), which posits linear structural models for both outcome confounding proxies $W$ and the primary outcome $Y$\cite{tchetgen2020introduction}. The approach delivers valid causal inferences in the presence of hidden confounders for which proxies are available. Extending the approach, we have established analogous two-stage regression approaches for structural causal models with either log, and logit links apply to either $W$ or $Y$. For all possible combinations of canonical link functions for the primary outcome and outcome confounding proxies, we have formulated an appropriate two-stage regression estimator detailed in Section \ref{uni} of the Appendix, thus making it possible for epidemiologists to use the proximal causal approach by fitting familiar generalized linear models for the most common settings encountered in practice. \\ \\
While the current paper focuses on the implementation of proximal causal inference in settings where the use of GLMs provides an adequate fit to the data at hand, PCI has been shown to apply in more general cases, where nonparametric models may provide a better fit to the observed data. However, as mentioned in the introduction, PCI in such more general cases typically requires solving certain linear integral equations that can be ill-posed and, as a result, numerically unstable. We refer the interested reader to the works of Miao et al. (2016) and Tchetgen Tchetgen et al. (2023), Kallus et al. (2021), Cui et al. (2023), Ying et al. (2023) and Zhang et al. (2023) \cite{miao2018identifying,tchetgen2020introduction,kallus2021causal,cui2023semiparametric,ying2023proximal,zhang2023proximal}. These studies provide in-depth discussion and an extensive treatment of the proximal nonparametric identification problem and elaborate on nonparametric and semiparametric estimation techniques. 
\\ \\
In the application considered in this paper, we followed the ad-hoc proxy allocation of the prior work by Tchetgen Tchetgen et al \cite{tchetgen2020introduction}. Their paper further considered the robustness of such allocation by evaluating the extent to which the estimator remains stable upon permuting the roles of outcome and treatment confounding proxies. The approach is intuitive, however, it lacks formal performance guaranties. Future work might consider more formal proxy selection algorithms with theoretical guaranties; see, for instance, the DANCE algorithm for structural linear models which represents initial progress\cite{kummerfeld2022data}. It is worth noting that throughout the paper we have explicitly assumed no treatment by unmeasured confounder interaction on the GLM scale. While the proposed proximal two-stage regression approach does not easily handle such interaction, it is worth noting that more flexible proximal methods that accommodate the presence of interactions with unmeasured confounders are available, however at the cost of computational complexity, see Cui et al. (2023) and Ghassami et al. (2020) for additional details \cite{cui2023semiparametric, ghassami2022minimax}.  Finally, right-censored survival data are of common occurrence in epidemiological practice, and evaluating causal effects for such outcomes using proximal causal inference techniques presents special challenges worth considering in future developments.

%
%

\endgroup
\bibliography{Maintext/feh}
\setcounter{page}{1}
\appendix
\begin{appendices}
\section{Appendix}
\DoToC

\newpage

\subsection*{Appendix Outline}
In Section \ref{uni}, we introduce a unified formulation of the proximal two-stage regression approach under different combinations of GLMs with different link functions for $Y$ and $W$, which allows for addressing a variety of scenarios, such as the case where $Y$ is continuous while $W$ is binary.  In the review of negative control methods by Shi et al. (2020) \cite{shi2020selective}, several applications using negative controls in epidemiological studies were discussed, including examples with $Y$ for offspring weight, BMI, and blood pressure, and $W$ for paternal smoking \cite{howe2012maternal,brion2007similar}. In these scenarios, $Y$ may be encoded as a continuous variable, while $W$ is encoded as a binary variable.
\\ \\
In Sections \ref{id-id}-\ref{lo-li}, we provide details for proximal two-stage regression for all cases outlined in the table of Section \ref{uni}, including each combination of three link functions. We also present more flexible model setups that allow for interaction terms in model specification in cases where the logit link is involved in the first stage, the second stage, or both. The formulation for polytomous regression is considered in \ref{lip-lip}. We suppress measured covariates $X$ for simplicity in derivation, which is consistent with DAG (2) in Figure \ref{fig:1}. Note that all formulations and corresponding estimation procedures accommodate DAG (1) in Figure \ref{fig:1} where there is no direct causal link from $W$ to $Y$ and from $Z$ to $A$.
\\ \\
The proofs center on the mathematical property of link functions used to define GLMs. Consistent with Section \ref{conti}, we implicitly assume proxies are valid and thus $W$ and $Z$ are $U-$relevant such that the coefficients encoding the relationship between them are non-zero throughout the proofs. We use $\mathbf{M}(*)$ to denote the moment-generating function (MGF) with input $*$. 
In addition, $\propto_L$ is used to denote the linear transformation, say $Y \propto_L X$ means $Y=aX+b$ with constants $a$ and $b$ where $a \neq 0$, and $\propto$ is used to denote the proportionality, say $Y \propto X$ means $Y = aX$ with a non-zero constant $a$. 
\\ \\ 
In Section \ref{ss}, we provide the details of the simulation study for the logit link case. In Section \ref{theori}, we establish the asymptotic normality of the two-stage regression estimators and statistical inference based on standard M-estimator theory. Additionally, we provide closed-form representations for the asymptotic variance of the estimators in Section \ref{theori}.
\\ \\
The superscript/hat notations are defined as follows:
\begin{table}[h]
\renewcommand{\arraystretch}{1.4} \centering
\begin{tabular}{lll}
\hline  
\multicolumn{1}{c}{\textbf{Superscript }} & \multicolumn{1}{c}{\textbf{Example }} & \multicolumn{1}{c}{\textbf{Meaning}}                                       \\
\hline  
None   & $\beta_0$     &  The true parameter in the data generation process assumptions \\ \hline
$\sim $ & $\tilde{\beta}_0 = \beta_0 + b$     &  The parameter with a constant shift related to MGFs \\ \hline
$*$     & ${\beta}_0^*  = a \beta_0 + b $ ($a \neq 0$)     &  The linearly transformed parameter  \\ \hline
$\verb!^!$  & $\hat{\beta}_0$      &  The estimated parameter \\ \hline
\end{tabular}
\caption{\label{table:3} The description of superscripts used.}
\end{table}	
\\
Proximal control variable $S$ is defined to be the conditional expectation $W$ (up to a link function) on $(A,Z)$ or $(A,Z,Y=1)$, depending on the model setup, i.e. $S = g(E[W|A,Z])$ or $S=g(E[W|A,Z,Y=1])$ where $g$ is a link function. The expression $(Y \sim X_1 + X_2 + \ldots X_p)$ means a (generalized) linear regression model in which $Y$ is a response variable and $(1,X_1,X_2,\ldots,X_p)$ are regressors. Note that this notation is motivated by the syntax of the \texttt{lm} function in the \texttt{R} programming language.
 
\newpage
\subsection{Unification of Proximal inference for GLMs}
\label{uni}
\begin{figure}[H]
\centering
\includegraphics[width=1\textwidth]{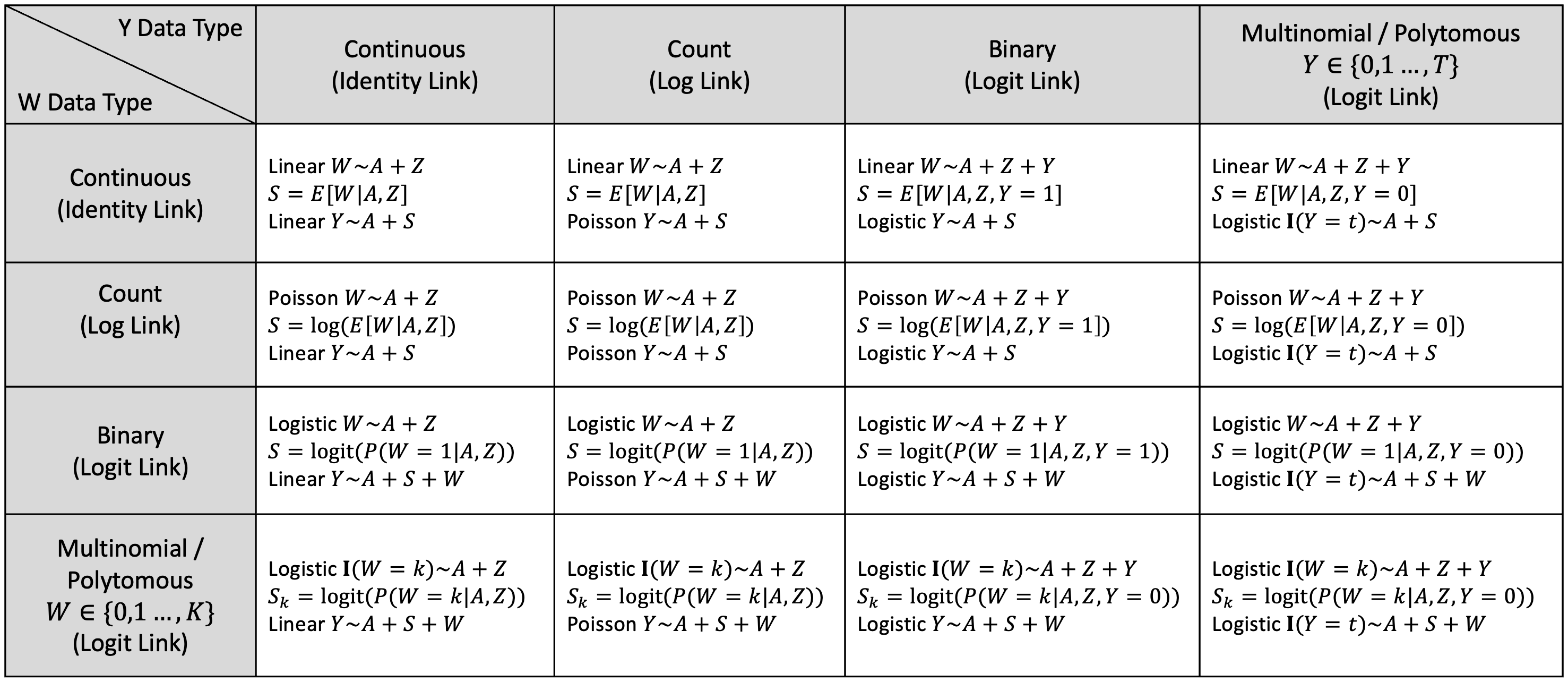}
\caption{Estimation procedures for each combination of three link functions. $S$ denotes proximal control variable for $U$.}
\label{fig:uni_1}
\end{figure}

Assuming linear regression functions for the first- and the second-stage regression equations, the table of Figure \ref{fig:uni_1} summarizes estimation procedures for each combination of three link functions introduced in Sections \ref{id-id}-\ref{lo-li}. Importantly, each instance delivers a procedure that produces a consistent estimator of the causal effect of interest, whether or not there is a direct causal link from either $Z$ to $A$ or from $W$ to $Y$ because either the corresponding model specification is not contingent upon the presence of these direct causal links or the same estimation procedure applies to the model specification accounting for the presence of these direct causal.
\\ \\
For simplicity, thus far the exposition has ignored measured confounders $X$ and interaction terms, we now describe the ease with which they can be incorporated into the proposed methodology, by simply including them as covariates in regressions at both stages. Specifically, suppose that $Y$ and $W$ are both continuous and therefore modeled using linear models compatible with DAG (3) which now also involves measured confounders $X$, thus the first-stage linear model may take the form:
\begin{align*}
	E[W|A,Z,X] = \alpha_0^* + \alpha_a^*A + \alpha_z^*Z + \alpha_{az}^*AZ + \alpha_{x}^*X;
\end{align*} 
while the second-stage may take the form:
\begin{align*}
	&E[Y|A,Z,X] = \beta_0^* + \beta_a^*A + \beta_u^* E[W|A,Z,X] + \beta_x^* X, \ \text{where } \beta_a^*=\beta_a 
\end{align*}
A similar approach applies to all other cases under different combinations of link functions, including those involving logit links where $(Y,W)-(A,Z,X)$ interactions might be considered by implementing a slightly more different estimation procedure (see Sections \ref{lit-lit}, \ref{id-li-int}, \ref{id-li-sym-int}, \ref{lo-li-int} and \ref{lo-li-int-sym}).
\\ \\
For inference, the bootstrap may be convenient in practice; however, we also provide analytic expressions for the asymptotic variance of parameter estimates obtained via proximal two-stage regression which we have implemented in \texttt{R} package "pci2s". The proof of asymptotic normality and standard error computation are also provided in Section \ref{theori}.
\newpage
\subsection{Identity Links Combination}
\label{id-id}
In this section, we present details on the continuous $Y$ and continuous $W$ case, which corresponds to the gray cell in the table below.
\begin{figure}[H]
\centering
\includegraphics[width=1\textwidth]{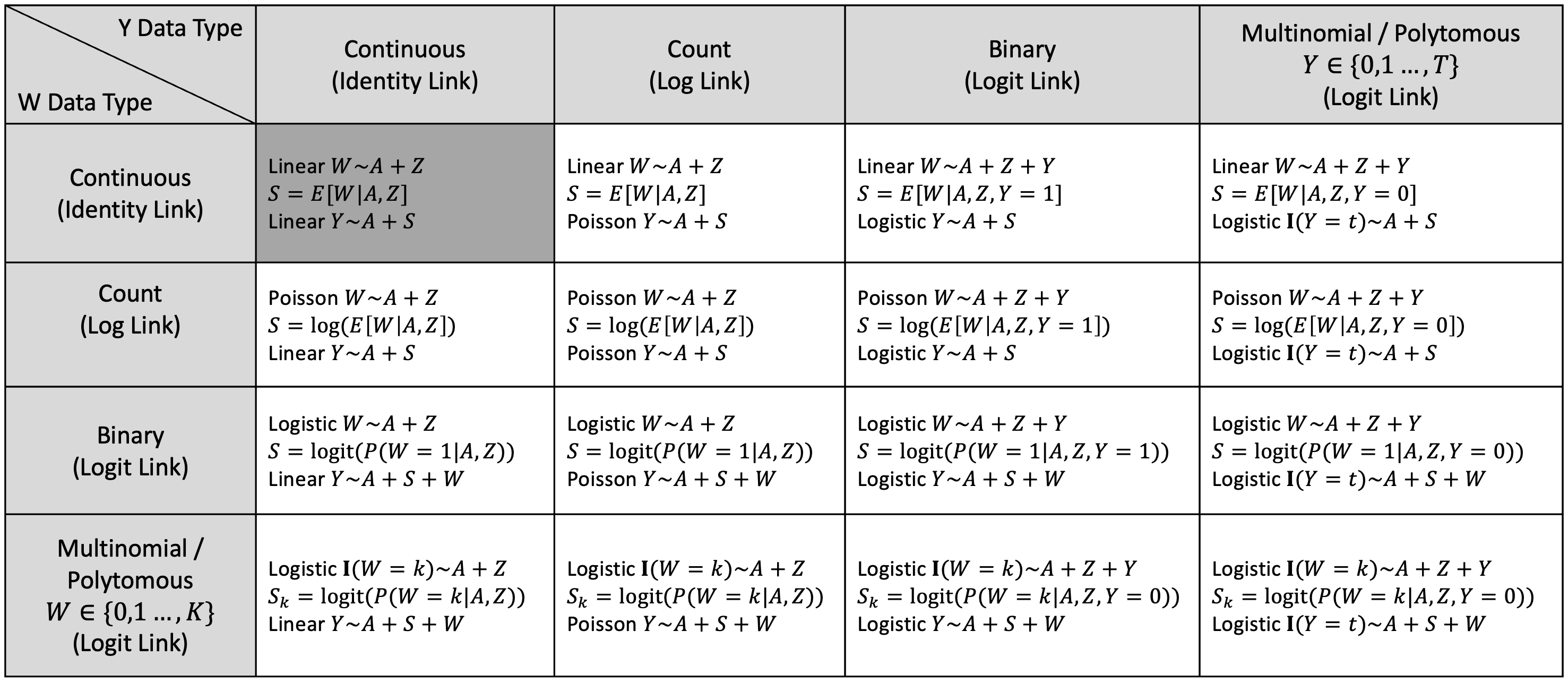}
\label{fig:uni_2}
\end{figure}
We investigate the ($Y$-Identity, $W$-Identity) case and assume that the data are generated from the following model:

\begin{assumption} \label{assumption:linY-linW}
\begin{align*}
	&E[Y|A,Z,U] = \beta_0+\beta_aA+\beta_uU  \numberthis \label{idid1} \\
	&E[W|A,Z,U] = \alpha_0 + \alpha_uU \numberthis \label{idid2} \\
	&|E[U|A,Z]| < \infty \numberthis \label{idid-ass}
\end{align*}    
\end{assumption}

We now prove Result \ref{result-1} under Assumption \ref{assumption:linY-linW}.

\begin{proof}
\\ \\
By taking the expectation of regression equations \eqref{idid1} and \eqref{idid2} for $U$ conditional on $(A,Z)$, we obtain the following two regression equations:
\begin{align*}
	&E[W|A,Z] 
	= \alpha_0 + \alpha_u E[U|A,Z] \numberthis \label{idid3} \\
	&E[Y|A,Z] = \beta_0+\beta_aA+\beta_u E[U|A,Z] \numberthis \label{idid4}
\end{align*}
Equations \eqref{idid3} and \eqref{idid4} imply Result \ref{result-1} in the main text as follows:
\begin{align*}
	&E[Y|A,Z] = \beta_0^* + \beta_a^*A + \beta_u^* E[W|A,Z], \\
	&\text{where } 
	\beta_0^*= \beta_0 - \beta_u \frac{\alpha_0}{\alpha_u}, \ \beta_a^* =\beta_a, \ \beta_u^* = \frac{\beta_u}{\alpha_u}.
\end{align*}
\end{proof}
\noindent Proximal control variable $S$ is a linear transformation of $E[U|A,Z]$, i.e.
$$S=E[W|A,Z] =\alpha_0 + \alpha_u E[U|A,Z] \propto_L E[U|A,Z]$$
The procedure below shows how $S$ and $\beta_a$ can be estimated.
\\ 
\begin{proc} \label{proc-1} Identity $Y$ Link, Identity $W$ Link Case
\begin{enumerate}
  \item Specify a linear regression function for $E[W|A,Z]$. Say $W\sim A+Z$.
  \item Perform the first-stage linear regression $W\sim A+Z$.
  \item Compute $\hat{S} = \hat{E}[W|A,Z]= \hat{\alpha}_0^* +\hat{\alpha}_a^*A + \hat{\alpha}_z^*Z$ using the estimated coefficients from the first-stage regression fit.
  \item Perform the second-stage linear regression $Y \sim A + \hat{S}$ to obtain $\hat{\beta}_a$.
\end{enumerate}
\end{proc}
\newpage

\subsection{Log Links Combination}
\label{lo-lo}
In this section, we present details on the count $Y$ and count $W$ case, which corresponds to the gray cell in the table below.
\begin{figure}[H]
\centering
\includegraphics[width=1\textwidth]{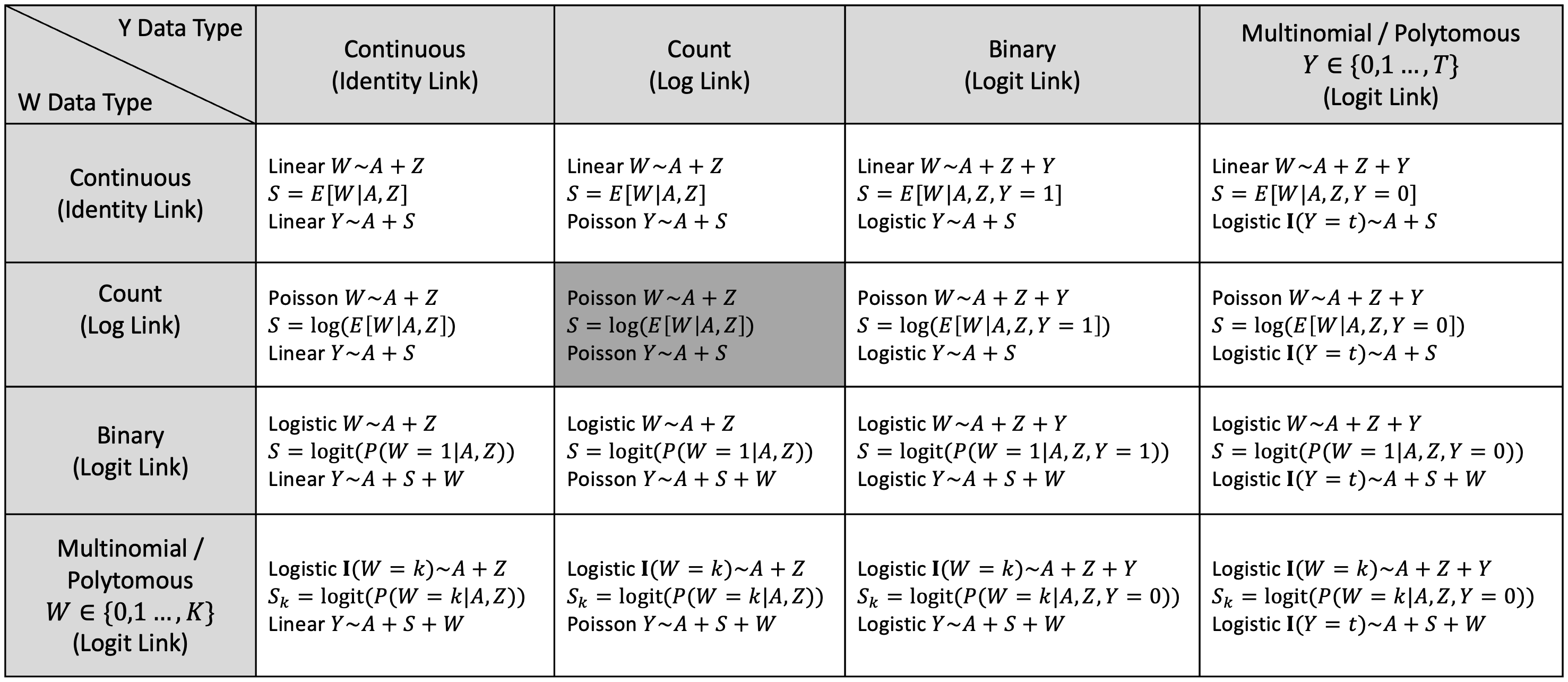}
\label{fig:uni_3}
\end{figure}

We investigate the ($Y$-Log, $W$-Log) case and assume that the data are generated from the following model:

\begin{assumption} \label{assumption:poisY-poisW}
\begin{align*}
	&\mathrm{log}(E[Y|A,Z,U]) = \beta_0+\beta_aA+\beta_uU  \numberthis \label{lolo1} \\
	&\mathrm{log}(E[W|A,Z,U]) = \alpha_0 + \alpha_uU \numberthis \label{lolo2} \\
	&U|A,Z \sim E[U|A,Z] + \epsilon; \ E[\epsilon] = 0; \ \epsilon \indep (A,Z); \numberthis \label{lolo-ass} \\
	&\textrm{$\mathbf{M}_\epsilon(*)$, the MGF of $\epsilon$, exists for any real value *.} 
\end{align*}
\end{assumption}
We now prove Result \ref{result-2} under Assumption \ref{assumption:poisY-poisW}.
\\ \\
\begin{proof}
\\ \\
We take the expectation of regression equations \eqref{lolo1} and \eqref{lolo2} for $U$ conditional on $(A,Z)$
\begin{align*}
	E[W|A,Z] &= E[exp\{ \alpha_0 + \alpha_u U\} |A,Z ] \\
	&=  exp\{ \alpha_0\} E[exp\{ \alpha_u U\}|A,Z]\\
	& \stackrel{(a)}{=}  exp\{ \alpha_0\} E[exp\{ \alpha_u (E[U|A,Z]+\epsilon)\}|A,Z]\\
	&=  exp\{ \alpha_0 + \alpha_u E[U|A,Z]\} E[exp\{ \alpha_u \epsilon\} | A,Z]\\
	&\stackrel{(b)}{=}  exp\{ \alpha_0 + \alpha_u E[U|A,Z]\} E[exp\{ \alpha_u  \epsilon\}]\\
	&\stackrel{(c)}{=}exp\{ \alpha_0 + \alpha_u E[U|A,Z] + \mathrm{log}\mathbf{M}_\epsilon(\alpha_u) \} \\
	&=exp\{ \tilde{\alpha}_0 + \alpha_u E[U|A,Z]\}  
 \end{align*} 
where $\tilde{\alpha}_0$ in the last line is defined as $\tilde{\alpha}_0= \alpha_0 + \mathrm{log}\mathbf{M}_\epsilon(\alpha_u)$. The equalities with $(a)$ and $(b)$ are from \eqref{lolo-ass}. The equality with $(c)$ is from the definition of the MGF $\mathbf{M}_\epsilon$, which is assumed to exist from \eqref{lolo-ass}. The other equalities are straightforward from \eqref{lolo2}.  
\\ \\
Likewise, we can obtain a similar result for $E[Y|A,Z]$:
 \begin{align*} 
	&E[Y|A,Z] = exp\{\tilde{\beta}_0+\beta_aA+\beta_u E[U|A,Z]\}, \\
        &\text{where } \tilde{\beta}_0=\beta_0+\mathrm{log}\mathbf{M}_\epsilon(\beta_u).
\end{align*}
Combining the two established results, we obtain the following two regression equations:
\begin{align*}
	&\mathrm{log}(E[W|A,Z]) = \tilde{\alpha}_0 + \alpha_u E[U|A,Z] \numberthis \label{lolo3} \\ 
	&\mathrm{log}(E[Y|A,Z]) = \tilde{\beta}_0+\beta_aA+\beta_uE[U|A,Z] \numberthis \label{lolo4}
\end{align*}
Equations \eqref{lolo3} and \eqref{lolo4} imply Result \ref{result-2} in the main text. 
\begin{align*}
	& \mathrm{log}(E[Y|A,Z]) = \beta_0^*+\beta_a^*A+\beta_u^*\mathrm{log}(E[W|A,Z]),\\
	&\text{where } \beta_0^*= \tilde{\beta}_0 - \beta_u \frac{\tilde{\alpha}_0}{\alpha_u}, \ \beta_a^*=\beta_a, \ \beta_u^* = \frac{\beta_u}{\alpha_u}.
\end{align*}
\end{proof}
Proximal control variable $S$ is a linear transformation of $E[U|A,Z]$, i.e.
$$S=\mathrm{log}(E[W|A,Z])=\tilde{\alpha}_0 + \alpha_u E[U|A,Z] \propto_L E[U|A,Z]$$
The procedure below shows how $S$ and $\beta_a$ can be estimated.
\\
\begin{proc} \label{proc-2} Log $Y$ Link, Log $W$ Link Case  
\begin{enumerate}
  \item Specify a linear regression function for $\mathrm{log}(E[W|A,Z])$. Say $W\sim A+Z$.
  \item Perform the first-stage Poisson regression $W\sim A+Z$.
  \item Compute $\hat{S} = \mathrm{log}(\hat{E}[W|A,Z])= \hat{\alpha}_0^* + \hat{\alpha}_a^*A + \hat{\alpha}_z^*Z$ using the estimated coefficients from the first-stage regression.
  \item Perform the second-stage Poisson regression $Y \sim A + \hat{S}$ to obtain $\hat{\beta}_a$.
\end{enumerate}
\end{proc}
\newpage
\subsection{Logit Links Combination}
\label{li-li}
In this section, we present details on the binary $Y$ and binary $W$ case, which corresponds to the gray cell in the table below.
\begin{figure}[H]
\centering
\includegraphics[width=1\textwidth]{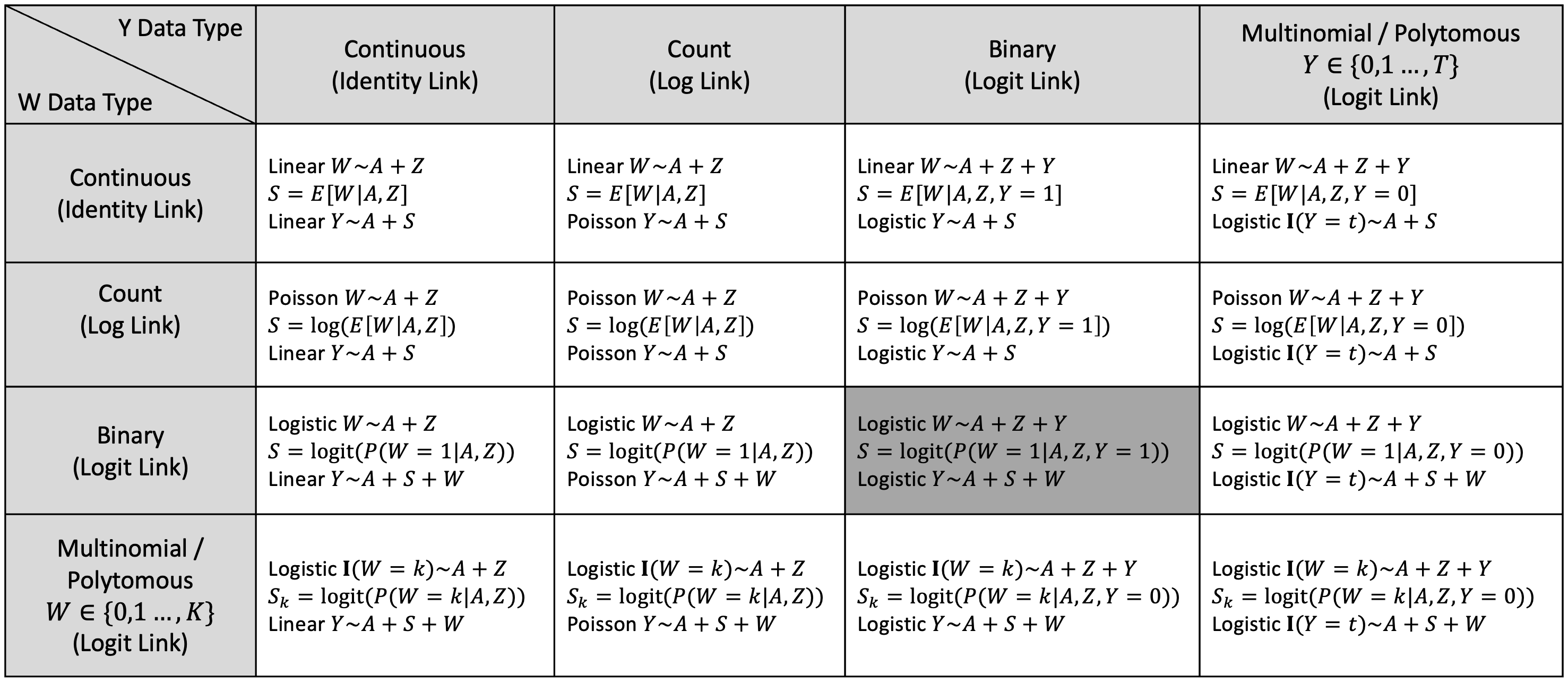}
\label{fig:uni_4}
\end{figure}
A notable difference between the binary $(Y,W)$ case from the previous two cases (i.e., continuous $(Y,W)$ and count $(Y,W)$ cases) is that the former is more complicated. Due to the nature of collapsibility shared by identity and log links, taking the expectation in previous regression equations does not entangle $U$ with other variables, say $(A,Z)$. However, we cannot employ similar steps as in the identity and log link cases to formulate the two-stage regression procedure because the mathematical properties of the logit link do not yield a manageable result like identity and log links after taking the expectation of $U$ conditional on $(A,Z)$, i.e.
\begin{align*}
	&\mathrm{logit}(\mathrm{Pr}(Y=1|A,Z)) \neq E[\mathrm{logit}(\mathrm{Pr}(Y=1|A,Z,U))|A,Z]=  \beta _0^* + \beta_u E[Y|A,Z], \\
	&\mathrm{logit}(\mathrm{Pr}(W=1|A,Z)) \neq E[\mathrm{logit}(\mathrm{Pr}(W=1|A,Z,U))|A,Z]=  \alpha_0^* + \alpha_u E[U|A,Z].
\end{align*}
Using logit links in modeling odds ratios necessitates exploring new compatible structural linear models and assumptions for the unmeasured confounder $U$ \cite{greenland1996absence,greenland1999confounding}. Therefore, instead of considering $f(Y|A,Z,U)$ and $f(W|A,Z,U)$ directly, we take a new path by considering the joint probability density function $f(Y,W|A,Z,U)$ and their marginalized counterparts $f(Y|A,Z,U)$ and $f(W|A,Z,U)$.
\\ \\
We characterize the joint probability density functions of random variables using the so-called odds ratio parameterization \cite{smi}. Specifically, the joint probability density function of random variables $A$ and $B$ given $C$ can be represented as follows:
\begin{align} \label{eq-odds ratio parametrization}
    f(A,B|C) = \frac{f(A|B=0,C) \cdot f(B|A=0,C) \cdot OR(A,B|C)}{\underbrace{\int_{\mathcal{(A,B)}} f(A=a|B=0,C) \cdot f(B=b|A=0,C) \cdot OR(A=a,B=b,C) \cdot d(a,b)}_{\textrm{\ \ Normalizing Constant}}} 
\end{align}
where $OR(A,B|C)$ is the odds ratio function relating $A$ and $B$ given $C$, and is defined as
\begin{align} \label{eq-odds ratio}
    OR(A,B|C)
    & = \frac{f(A,B|C)\cdot f(A=0,B=0|C)}{f(A=0,B|C)\cdot f(A,B=0|C)} 
    \nonumber
    \\
    & = \frac{f(A|B,C)\cdot f(A=0|B=0,C)}{f(A=0|B,C)\cdot f(A|B=0,C)} 
    \nonumber
    \\
    & = \frac{f(B|A,C) \cdot f(B=0|A=0,C)}{f(B|A=0,C)\cdot f(B=0|A,C)} \ .
\end{align}

\subsubsection{Logit Links Combination with No Interaction Term}
\label{linot-linot}
For the ($Y$-Logit, $W$-Logit) case with no $(Y,W)-U$ interactions, respectively, we make the following assumption:

\begin{assumption} \label{assumption:binY-binW}
    \begin{align*}
	&\mathrm{logit}(\mathrm{Pr}(Y=1|A,Z,W,U)) = \beta_0+\beta_aA+\beta_uU+\beta_w W  \numberthis \label{lili1} \\
	&\mathrm{logit}(\mathrm{Pr}(W=1|A,Z,Y,U)) = \alpha_0+\alpha_uU+\alpha_y Y  \numberthis \label{lili2} \\
	&U|A,Z,Y=0,W=0 \sim \underbrace{E[U|A,Z,Y=0,W=0]}_{=m(A,Z)} + \epsilon; \ E[\epsilon]=0; \ \epsilon \indep (A,Z)|Y=0,W=0; \numberthis \label{lili-ass} \\
	&\textrm{$\mathbf{M}_\epsilon(*)$, the MGF of $\epsilon|Y=0,W=0$ exists for any real value *.}\\
	&\text{\eqref{lili1} and \eqref{lili2} $\implies$ }\beta_w=\alpha_y
 \numberthis \label{lili-ass-1}
\end{align*}
\end{assumption}
We prove Result \ref{result-3} under Assumption \ref{assumption:binY-binW}. 
\\ \\
\begin{proof}
\\ \\
From \eqref{eq-odds ratio}, we obtain
    \begin{align*}
	OR(U,Y|A,Z,W) &= \frac{f(Y|U,A,Z,W)f(Y=0|U=0,A,Z,W)}{f(Y=0|U,A,Z,W)f(Y|U=0,A,Z,W)} \\
	&=\frac{\frac{exp\{Y(\beta_0+\beta_aA+\beta_uU +\beta_wW)\}}{1+exp\{\beta_0+\beta_aA+\beta_uU +\beta_wW\}}\frac{1}{1+exp\{\beta_0+\beta_aA + \beta_wW\}}}
	{\frac{1}{1+exp\{\beta_0+\beta_aA+\beta_uU+\beta_wW\}}\frac{exp\{Y(\beta_0+\beta_aA+\beta_wW)\}}{1+exp\{\beta_0+\beta_aA+\beta_wW\}}} = exp\{ \beta_u U Y\} 
\end{align*}

Based on similar algebra,
\begin{align*}
    &OR(U,W|A,Z,Y) = exp\{ \alpha_uUW\}  \\
	&OR(Y,W|A,Z,U) = exp\{ \beta_wYW\} = exp\{ \alpha_y YW \}
\end{align*} 
Equation \eqref{lili-ass-1} in Assumption \ref{assumption:binY-binW} guarantees that $OR(Y,W|A,Z,U)$ is well-defined. 
We use the odds ratio parametrization in \eqref{eq-odds ratio parametrization}. For notational brevity, we suppress normalizing constants whenever appropriate. The joint density $f(W,U|A,Z,Y=0)$ can be written as 
\begin{align*}
	f(W,U|A,Z,Y=0)
	&\propto  f(W|A,Z,Y=0,U=0) f(U|A,Z,Y=0,W=0) OR(U,W|Y=0,A,Z) \\
	& = f(W|A,Z,Y=0,U=0) f(U|A,Z,Y=0,W=0) exp\{\alpha_u UW\} 
\end{align*}
Therefore, $f(W|A,Z,Y=0)$ can be obtained by marginalizing $f(W,U|A,Z,Y=0)$ with respect to $U$:
\begin{align*}
    f(W|A,Z,Y=0)
	&\propto \int_\mathcal{U} f(W|A,Z,Y=0,U=0)  f(U=u|A,Z,Y=0,W=0) exp\{\alpha_u uW\} du \\
	&= f(W|A,Z,Y=0,U=0)  E[exp\{\alpha_u UW\}|A,Z,Y=0,W=0]  \\
	&= f(W|A,Z,Y=0,U=0)  E[exp\{\alpha_u W(m(A,Z)+\epsilon)\}|A,Z,Y=0,W=0]  \\
	&=  f(W|A,Z,Y=0,U=0)   exp\{\alpha_u Wm(A,Z)\} E[exp\{\alpha_uW \epsilon\}|Y=0,W=0] \\
	&= \frac{exp\{W(\alpha_0)\}}{1+exp\{W(\alpha_0)\}} exp\{\alpha_u Wm(A,Z)+ \mathrm{log}\mathbf{M}_\epsilon(\alpha_uW) \} \\
	&\propto exp\{W(\alpha_0 + \alpha_u m(A,Z)+ \mathrm{log}\mathbf{M}_\epsilon(\alpha_u)) \} \\
	&= exp\{W(\tilde{\alpha}_0 + \alpha_u m(A,Z))\}  
 \numberthis  \label{lili-tmp3}
\end{align*}
The first identity follows from \eqref{eq-odds ratio parametrization} and $OR(U,W|A,Z,Y) = exp\{ \alpha_uUW\}$. 
The second identity holds from the definition of conditional expectation.
The third and fourth identities hold from \eqref{lili-ass}. 
The fifth identity holds from the definition of, $\mathbf{M}_\epsilon$, the MGF of $\epsilon$. 
The sixth identity is straightforward.
The last identity holds where $\tilde{\alpha}_0$ is defined as $\tilde{\alpha}_0=\alpha_0+\mathrm{log}\mathbf{M}_\epsilon(\alpha_u)$. 
Analogously, we obtain the result for $f(Y,U|A,Z,W=0)$:
\begin{align}
	&f(Y|A,Z,W=0) \propto exp\{Y(\tilde{\beta}_0 + \beta_aA+\beta_um(A,Z))\}   \numberthis  \label{lili-tmp4}
\end{align}
Since $W$ is binary, \eqref{lili-tmp3}  and \eqref{lili-tmp4} imply the following results, respectively:
\begin{align}
	&f(W|A,Z,Y=0) = \frac{exp\{W(\tilde{\alpha}_0 + \alpha_u m(A,Z))\}}{1+exp\{\tilde{\alpha}_0 + \alpha_u m(A,Z)\}} \numberthis \label{lili3}  \\
	&f(Y|A,Z,W=0) = \frac{exp\{Y(\tilde{\beta}_0 + \beta_aA + \beta_u m(A,Z))\}}{1+exp\{\tilde{\beta}_0 + \beta_aA + \beta_u m(A,Z)\}} \numberthis \label{lili4} 
\end{align}

Next, we find an alternative representation of $f(W,U|A,Z,Y=1)$. From Bayes' rule, we have the following
\begin{align*}
	f(U|A,Z,Y=1,W=0)&=\frac{f(Y=1|A,Z,W=0,U)f(U|A,Z,W=0)}{f(Y=1|A,Z,W=0)}\\
	f(U|A,Z,Y=0,W=0)&=\frac{f(Y=0|A,Z,W=0,U)f(U|A,Z,W=0)}{f(Y=0|A,Z,W=0)}\\
	\frac{f(U|A,Z,Y=1,W=0)}{f(U|A,Z,Y=0,W=0)}&= \frac{f(Y=1|A,Z,W=0,U)f(Y=0|A,Z,W=0)}{f(Y=0|A,Z,W=0,U) f(Y=1|A,Z,W=0)} \\
	&=\frac{exp\{ \beta_0+\beta_aA+\beta_uU\}  }
	{exp\{\tilde{\beta}_0+\beta_aA+\beta_um(A,Z)\}} = exp\{ \beta_0 -\tilde{\beta}_0 +\beta_u(U-m(A,Z))\} 
\end{align*}
These results imply
\begin{align} \label{eq-bayes}
    f(U|A,Z,Y=1,W=0)&= exp\{ \beta_0 -\tilde{\beta}_0 +\beta_u(U-m(A,Z))\}f(U|A,Z,Y=0,W=0)
\end{align}
Combining the above result with \eqref{eq-odds ratio parametrization} and $OR(U,W|A,Z,Y) = exp\{ \alpha_uUW\}$, an alternative representation of $f(W,U|A,Z,Y=1)$ is given below
\begin{align*}
	&f(W,U|A,Z,Y=1) \\
	&\propto  f(W|A,Z,Y=1,U=0) f(U|A,Z,Y=1,W=0) OR(U,W|Y=1,A,Z) \\
	&= f(W|A,Z,Y=1,U=0) f(U|A,Z,Y=1,W=0) exp\{\alpha_u UW\}
 \end{align*} 
By marginalizing the above representation with respect to $U$, we have 
\begin{align*}
	f(W|A,Z,Y=1)
	&\propto \int_\mathcal{U} f(W|A,Z,Y=1,U=0)  f(U=u|A,Z,Y=1,W=0) exp\{\alpha_u uW\} du \\
	&=  f(W|A,Z,Y=1,U=0) 
         E[exp\{\alpha_u UW\}|A,Z,Y=1,W=0]
    \end{align*}
    When $W=0$, we have $exp\{\alpha_uUW\}=1$, which implies 
    \begin{align*}
        f(W=0|A,Z,Y=1) \propto 1 \cdot \int_\mathcal{U} f(U=u|A,Z,Y=1,W=0) du=1
    \end{align*}
    When $W=1$, we have $exp\{\alpha_uUW\}=\exp\{\alpha_uU\}$, which implies
    \begin{align*} 
    f(W=1|A,Z,Y=1) 
	&\propto \int_\mathcal{U} f(W=1|A,Z,Y=1,U=0) exp\{\alpha_u u\} f(U=u|A,Z,Y=1,W=0) du\\
	&\propto exp\{\alpha_0+\alpha_y \} \int_\mathcal{U} exp\{\alpha_u u\} exp\{ \beta_0 -\tilde{\beta}_0 +\beta_u(u-m(A,Z))\}f(U=u|A,Z,Y=0,W=0) du\\
	&\propto exp\{\alpha_0+\alpha_y \}
	E[exp\{\alpha_u U\} exp\{ \beta_0 -\tilde{\beta}_0 +\beta_u(U-m(A,Z))\}|A,Z,Y=0,W=0]
	\\
	&= exp\{\alpha_0+\alpha_y \} 
	E[exp\{\alpha_u (m(A,Z)+\epsilon) +  (\beta_0 -\tilde{\beta}_0 +\beta_u\epsilon )\}|Y=0,W=0] \\
	&= exp\{\alpha_0+\alpha_y \} exp\{\alpha_u m(A,Z)\} exp\{\beta_0 -\tilde{\beta}_0\}E[exp\{ (\alpha_u +\beta_u )\epsilon \}|Y=0,W=0] \\
	&= exp\{\alpha_0+\alpha_y +\alpha_u m(A,Z) \} 
	exp\{ -\mathrm{log}(\mathbf{M}_\epsilon (\beta_u))\} exp\{ \mathrm{log}(\mathbf{M}_\epsilon (\alpha_u+\beta_u))\}\\
	&= exp\{ 
	\alpha_0 + \mathrm{log}(\mathbf{M}_\epsilon (\alpha_u)) +
	\alpha_y +
	\mathrm{log}
        \left(
        \frac
	{\mathbf{M}_\epsilon (\alpha_u+\beta_u)}
	{\mathbf{M}_\epsilon (\alpha_u)\mathbf{M}_\epsilon (\beta_u)}
        \right) + 
	\alpha_um(A,Z)\}
        \\
	&= exp\{\tilde{\alpha}_0 + \tilde{\alpha}_y + \alpha_um(A,z) \}
\end{align*}
The first line holds from \eqref{eq-odds ratio parametrization} and $OR(U,W|A,Z,Y) = exp\{ \alpha_uUW\} = \exp\{\alpha_uU\}$ when $W=1$. The second line holds from \eqref{lili2} and \eqref{eq-bayes}. The third line holds from the definition of the conditional expectation. The fourth line is satisfied by the conditional independence from \eqref{lili-ass}. The fifth, sixth, and seventh lines are trivial from the definition of the MGF $\mathbf{M}_\epsilon$. The last line is satisfied where $\tilde{\alpha}_0$ and $\tilde{\alpha}_y$ are defined as 
\begin{align*}
    &
    \tilde{\alpha}_0=\alpha_0 + \log(\mathbf{M}_\epsilon(\alpha_u)) \ , 
    \quad 
    \tilde{\alpha}_y = \alpha_y + \mathrm{log} \left( 
    \frac
	{\mathbf{M}_\epsilon (\alpha_u+\beta_u)}
	{\mathbf{M}_\epsilon (\alpha_u)\mathbf{M}_\epsilon (\beta_u)}
    \right)
\end{align*}
Consequently, we have
\begin{align*}
	f(W|A,Z,Y=1) \propto exp\{ W(\tilde{\alpha}_0 + \tilde{\alpha}_y+  \alpha_u m(A,Z)) \} \numberthis \label{lili-tmp3w}
\end{align*}
From analogous algebra, we find that $f(W|A,Z,Y=0)$ and $f(W|A,Z,Y=1)$ have the same form except for intercept terms. Therefore, we can combine \eqref{lili-tmp3} and \eqref{lili-tmp3w}, and write
\begin{align*}
	f(W|A,Z,Y)  \propto exp\{W(\tilde{\alpha}_0 + \alpha_um(A,Z) + \tilde{\alpha}_yY)\}  \numberthis \label{lili-final1}
\end{align*} 

Likewise, we obtain a parallel result for $f(Y|A,Z,W=1)$ as follows:
\begin{align*}
	&f(Y|A,Z,W=1) \propto exp\{Y(\tilde{\beta}_0 + \beta_aA + \beta_um(A,Z) + \tilde{\beta}_wW)\},  \numberthis \label{lili-final2}\\
        &\text{where }
    \tilde{\beta}_w=\beta_w+\mathrm{log}\left( 
    \frac
	{\mathbf{M}_\epsilon (\alpha_u+\beta_u)}
	{\mathbf{M}_\epsilon (\alpha_u)\mathbf{M}_\epsilon (\beta_u)}
    \right).
\end{align*}
Note that $ \tilde{\beta}_w=\tilde{\alpha}_y$ from \eqref{lili-ass-1}.
Combining \eqref{lili-final1} and \eqref{lili-final2}, we get the following probability mass functions:
\begin{align}
	&f(W|A,Z,Y) = \frac{exp\{W ( \tilde{\alpha}_0 + \alpha_u m(A,Z) + \tilde{\alpha}_y Y)\}}{1+exp\{\tilde{\alpha}_0 + \alpha_u m(A,Z) + \tilde{\alpha}_y Y\}} \numberthis \label{lili-dup3}  \\
	&f(Y|A,Z,W) = \frac{exp\{Y(\tilde{\beta}_0 + \beta_aA + \beta_u m(A,Z) + \tilde{\beta}_wW)\}}{1+exp\{\tilde{\beta}_0 + \beta_aA + \beta_u m(A,Z) + \tilde{\beta}_wW\}} \numberthis \label{lili-dup4}  
\end{align}
From \eqref{lili-dup3} and \eqref{lili-dup4}, we obtain the following two regression equations:
\begin{align*}
	&\mathrm{logit}(\mathrm{Pr}(W=1|A,Z,Y)) = \tilde{\alpha}_0 + \alpha_u m(A,Z) + \tilde{\alpha}_y Y  \numberthis \label{lili5} \\
	&\mathrm{logit}(\mathrm{Pr}(Y=1|A,Z,W)) = \tilde{\beta}_0 + \beta_aA + \beta_u m(A,Z) + \tilde{\beta}_wW  \numberthis \label{lili6}
\end{align*}
Equations \ref{lili5} and \ref{lili6} imply Result \ref{result-3} of the main paper
\begin{align*}
&\mathrm{logit}(\mathrm{Pr}(Y=1|A,Z,W)) ={\beta}_0^* +\beta_a^*A + \beta_u^*\mathrm{logit}(\mathrm{Pr}(W=1|A,Z,Y=1)) + \tilde{\beta}_wW, \\
&\text{where }\beta_0^*=\tilde{\beta}_0 - \beta_u \frac{(\tilde{\alpha}_0+\tilde{\alpha}_y)}{\alpha_u},\  \beta_a^*=\beta_a, \ \beta_u^*=\frac{\beta_u}{\alpha_u}.
\end{align*}
\end{proof}

Proximal control variable $S$ is a linear transformation of $E[U|A,Z,W=0,Y=0]=m(A,Z)$, i.e.,  
$$S=\mathrm{logit}(\mathrm{Pr}(W=1|A,Z,Y=1)) \propto_L m(A,Z)$$
The procedure below shows how $S$ and $\beta_a$ can be estimated.
\\
\begin{proc} \label{proc-3} Logit $Y$ Link, Logit $W$ Link Case with No Interaction Term
\begin{enumerate}
  \item Specify a linear regression function for $\mathrm{logit}(\mathrm{Pr}(W=1|A,Z,Y))$. Say $W \sim A+Z+Y$.
  \item Perform the first-stage logistic regression $W\sim A+Z+Y$.
  \item Compute $\hat{S} = \mathrm{logit}(\widehat{\mathrm{Pr}}(W=1|A,Z,Y=1))= \hat{\alpha}_0^* + \hat{\alpha}_a^*A + \hat{\alpha}_z^*Z + \hat{\alpha}_y^*$ using the estimated coefficients from the first-stage regression.
  \item Perform the second-stage logistic regression $Y \sim A + \hat{S} + W$ to obtain $\hat{\beta}_a$.
\end{enumerate}
\end{proc}
We remark that the relationship $\tilde{\alpha}_y = \tilde{\beta}_w = \alpha_y + c  = \beta_w + c $ with a constant 
$c=\mathrm{log}\left( 
    \frac
	{\mathbf{M}_\epsilon (\alpha_u+\beta_u)}
	{\mathbf{M}_\epsilon (\alpha_u)\mathbf{M}_\epsilon (\beta_u)}
    \right)$
involves quantities related to $U$ mainly, the conditional MGF of $\epsilon$ and $\alpha_u$ and $\beta_u$. In particular, the coefficients $\alpha_u$ and $\beta_u$ encode the causal links from $U$ to $Y$ and $U$ to $W$, respectively, which are present by $U$-relevance in all DAGs in Figure \ref{fig:1}. Therefore, even though there is no link from $Z$ to $A$ and $W$ to $Y$ (i.e. $\alpha_y=\beta_w=0$) as in DAG (1), the procedure for the logit link case necessitates including $Y$ in the first-stage regression and $W$ in the second-stage regression, respectively. Consequently, the same procedure for estimating $\beta_a$ is applicable to DAG (1).
\\ \\
The above model specification excludes $(Y,W)-(A,Z)$ interactions in the estimation procedure. The more general model specification is detailed in the next section.
\newpage
\subsubsection{Logit Links Combination with Interaction Terms}
\label{lit-lit}
As previously mentioned, the model specification in the ($Y$-Logit, $W$-Logit) case given above excludes the interactions between $(Y,W)-U$, respectively. Consider an alternative model specification:
\begin{assumption} \label{assumption-4}
    \begin{align*}
	&\mathrm{logit}(\mathrm{Pr}(Y=1|A,Z,W,U)) = \beta_0+\beta_aA+\beta_uU+\beta_w W + \beta_{uw}UW   \numberthis \label{lili-it1} \\
	&\mathrm{logit}(\mathrm{Pr}(W=1|A,Z,Y,U)) = \alpha_0+\alpha_uU+\alpha_yY +\alpha_{uy}UY  \numberthis \label{lili-it2} \\
	&U|A,Z,Y=0,W=0 \sim \underbrace{E[U|A,Z,Y=0,W=0]}_{=m(A,Z)} + \epsilon; \ E[\epsilon]=0; \ \epsilon \indep (A,Z)|Y=0,W=0; \numberthis \label{lili-it-ass} \\
	&\textrm{$\mathbf{M}_\epsilon(*)$, the MGF of $\epsilon|Y=0,W=0$ exists for any real value *.}\\
	&\text{\eqref{lili-it1} and \eqref{lili-it2} $\implies$ } \beta_w=\alpha_y \ \text{and } \beta_{uw}=\alpha_{uy}
\end{align*}
\end{assumption}
The model encodes an interaction between $U$ and $W$ in the structural logistic model for $Y$ and an interaction between $U$ and $Y$ in the structural logistic model for $W$. The following result modifies the two-stage regression procedure to account for such interactions.
\begin{result}[Generalization of Result \ref{result-3}] \label{result-5} Under Assumption \ref{assumption-4}, it follows that     
\begin{align*}
&\mathrm{logit}(\mathrm{Pr}(Y=1|A,Z,W)) =\beta_0^* +\beta_a^*A + \beta_u^*S + \tilde{\beta}_wW + \beta_{uw}^*SW , \\
&\text{where }
S = \mathrm{logit}(\mathrm{Pr}(W=1|A,Z,Y=1)), 
\ \beta_0^*= \tilde{\beta_0} - 
\beta_u (\frac{\tilde{\alpha}_0+\tilde{\alpha}_y}{\alpha_u+\alpha_{uy}}) - 
\beta_{uw} (\frac{\tilde{\alpha}_0+\tilde{\alpha}_y}{\alpha_u+\alpha_{uy}}), 
\\
& \quad \quad \ \
\beta_a^*=\beta_a, 
\ \beta_u^*=\frac{\beta_u}{\alpha_u+\alpha_{uy}}, 
\ \beta_{uy}^*=\frac{\beta_{uw}}{\alpha_u+\alpha_{uy}}.
\end{align*}
\end{result} 
Note $S=\mathrm{logit}(\mathrm{Pr}(W=1|A,Z,Y=1))$ from Result \ref{result-5} can be fitted via the first-stage logistic regression model with $Y-(A,Z)$ interactions incorporated, and Result \ref{result-5} requires the second-stage logistic regression model to include the term $SW$, which encodes for interactions $W-(A,Z)$, accounting interactions $(Y,W)-U$ in \eqref{lili-it1} and \eqref{lili-it2}.

\begin{proof} 
\\ \\
Due to the similarity with the proof in Section \ref{linot-linot}, we only provide minimal algebraic details.

Based on the odds ratio functions, we establish the following relationships
\begin{align*}
	OR(Y,W|A,Z,U) &= exp\{ YW(\beta_w + \beta_{uw}U) \} = exp\{ YW(\alpha_y + \alpha_{uy}U) \} \implies \beta_w=\alpha_y \text{ and } \beta_{uw}=\alpha_{uy}
\end{align*}

Using the odds ratio parametrization in \eqref{eq-odds ratio parametrization}, we find 
\begin{align*}
	&f(W,U|A,Z,Y=0) \\
	&\propto  f(W|A,Z,Y=0,U=0) f(U|A,Z,Y=0,W=0) OR(U,W|Y=0,A,Z) \\
	&= f(W|A,Z,Y=0,U=0) f(U|A,Z,Y=0,W=0) exp\{\alpha_u UW\}
 \end{align*}
By marginalizing $f(W,U|A,Z,Y=0)$ with respect to $U$, we have 
\begin{align*}
 &f(W|A,Z,Y=0)\\
 &\propto \int_\mathcal{U} f(W|A,Z,Y=0,U=0)  f(U=u|A,Z,Y=0,W=0) exp\{\alpha_u uW\} du \\
 &= f(W|A,Z,Y=0,U=0) exp\{\alpha_u Wm(A,Z)\} E[exp\{\alpha_uW \epsilon\}|Y=0,W=0] \\
 &\propto exp\{W(\tilde{\alpha}_0 + \alpha_u m(A,Z))\}  
\end{align*}
The first equality holds from the previous result.
The second equality holds from \eqref{lili-it-ass}
Therefore, we establish
\begin{align*}
f(W|A,Z,Y=0) \propto exp\{W(\tilde{\alpha}_0 + \alpha_u m(A,Z))\}   \numberthis \label{lili-it-tmp3}
\end{align*}
where $\tilde{\alpha}_0 = \alpha_0+\mathrm{log}\mathbf{M}_\epsilon(\alpha_u)$. 
From analogous algebra, we obtain the following result for $f(Y,U|A,Z,W=0)$:
\begin{align}
	&f(Y|A,Z,W=0) \propto exp\{Y(\tilde{\beta}_0 + \beta_aA+\beta_um(A,Z))\}   \numberthis  \label{lili-it-tmp4} 
\end{align}
where $\tilde{\beta}_0=\beta_0+\mathrm{log}\mathbf{M}_\epsilon(\beta_u)$. 
Combining \eqref{lili-it-tmp3}  and \eqref{lili-it-tmp4}, we obtain the following probability mass functions:
\begin{align}
	&f(W|A,Z,Y=0) = \frac{exp\{W(\tilde{\alpha}_0 + \alpha_u m(A,Z)) \}}{1+exp\{\tilde{\alpha}_0 + \alpha_u m(A,Z)\}} \numberthis \label{lili-it3}  \\
	&f(Y|A,Z,W=0) = \frac{exp\{Y(\tilde{\beta}_0 + \beta_aA + \beta_u m(A,Z))\}}{1+exp\{\tilde{\beta}_0 + \beta_aA + \beta_u m(A,Z)\}} \numberthis \label{lili-it4} 
\end{align}
Using the Bayes' rule, we can establish
\begin{align} \label{eq-bayes2}
	f(U|A,Z,Y=1,W=0)&= exp\{ \beta_0 -\tilde{\beta}_0 +\beta_u(U-m(A,Z))\}f(U|A,Z,Y=0,W=0)
\end{align}
Combining the above result with \eqref{eq-odds ratio parametrization} and $OR(U,W|A,Z,Y) = exp\{ \alpha_uUW + \alpha_{uy}UWY\}$, an alternative representation of $f(W,U|A,Z,Y=1)$ is given below
\begin{align*}
	&f(W,U|A,Z,Y=1) \\
	&\propto  f(W|A,Z,Y=1,U=0) f(U|A,Z,Y=1,W=0) OR(U,W|Y=1,A,Z) \\
	&= f(W|A,Z,Y=1,U=0) f(U|A,Z,Y=1,W=0) exp\{ (\alpha_u + \alpha_{uy}) UW \}
\end{align*}
By marginalizing the above representation with respect to $U$, we have 
\begin{align*}
	&f(W|A,Z,Y=1) \\
	&\propto \int_\mathcal{U} f(W|A,Z,Y=0,U=0)  f(U=u|A,Z,Y=1,W=0) exp\{(\alpha_u + \alpha_{uy}) uW\} du
 \end{align*}
 When $W=0$, we have $exp\{\alpha_uUW\}=1$, which implies 
 \begin{align*}
	&f(W=0|A,Z,Y=1) \propto 1 \cdot \int_\mathcal{U} f(U=u|A,Z,Y=1,W=0) du=1
 \end{align*}
 When $W=1$, we have $exp\{\alpha_uUW\}=\exp\{(\alpha_u + \alpha_{uy})U\}$, which implies
 \begin{align*}     
	&f(W=1|A,Z,Y=1) \\
	&\propto \int_\mathcal{U} f(W=1|A,Z,Y=1,U=0) exp\{(\alpha_u + \alpha_{uy}) u\} f(U=u|A,Z,Y=1,W=0) du\\
	&\propto exp\{\alpha_0+\alpha_y \} \int_\mathcal{U} exp\{(\alpha_u + \alpha_{uy}) u\} exp\{ \beta_0 -\tilde{\beta}_0 +\beta_u(u-m(A,Z))\}f(U=u|A,Z,Y=0,W=0) du
	\\
	&\propto exp\{\alpha_0+\alpha_y \}
	E[exp\{(\alpha_u + \alpha_{uy}) U\} exp\{ \beta_0 -\tilde{\beta}_0 +\beta_u(U-m(A,Z))\}|A,Z,Y=0,W=0]
	\\
	&= exp\{\alpha_0+\alpha_y \}
	E[exp\{(\alpha_u + \alpha_{uy}) (m(A,Z)+\epsilon) +  (\beta_0 -\tilde{\beta}_0 +\beta_u\epsilon )\}|Y=0,W=0] \\
	&= exp\{\alpha_0+\alpha_y \} exp\{(\alpha_u + \alpha_{uy}) m(A,Z)\} exp\{\beta_0 -\tilde{\beta}_0\}
	E[exp\{ (\alpha_u + \alpha_{uy} +\beta_u )\epsilon \}|Y=0,W=0] 
	\\
	&= exp\{\alpha_0+\alpha_y +(\alpha_u + \alpha_{uy}) m(A,Z) \} 
	exp\{ -\mathrm{log}(\mathbf{M}_\epsilon (\beta_u))\} exp\{ \mathrm{log}(\mathbf{M}_\epsilon (\alpha_u+\alpha_{uy} +\beta_u))\}\\
	&=exp\{ 
	\alpha_0 + \mathrm{log}(\mathbf{M}_\epsilon (\alpha_u)) +
	\alpha_y +
	\mathrm{log}
        \left(
        \frac
	{\mathbf{M}_\epsilon (\alpha_u+\alpha_{uy} +\beta_u)}
	{\mathbf{M}_\epsilon (\alpha_u)\mathbf{M}_\epsilon (\beta_u)}
        \right) + 
	(\alpha_u+\alpha_{uY}) m(A,Z)\}  
	\\
	&=exp\{ \tilde{\alpha}_0 + \tilde{\alpha}_y+  (\alpha_u+\alpha_{uy}) m(A,Z) \}
\end{align*}
The second line holds from \eqref{eq-odds ratio parametrization} and $OR(U,W|A,Z,Y) = exp\{ (\alpha_u + \alpha_{uy}) UW\} = \exp\{(\alpha_u + \alpha_{uy}) U\}$ when $W=1$. The third line holds from \eqref{lili-it-ass} and \eqref{eq-bayes2}. The fourth line is satisfied by the definition of the conditional expectation. The fifth line is satisfied from the conditional independence assumed in \eqref{lili-it-ass}. The sixth, seventh, and eighth lines are trivial from the definition of the MGF of $\mathbf{M}_\epsilon$. The last line is satisfied where $\tilde{\alpha}_0$ and $\tilde{\alpha}_y$ are defined as 
\begin{align*}
    &
    \tilde{\alpha}_0=\alpha_0 + \log(\mathbf{M}_\epsilon(\alpha_u)) \ , \quad  
    \tilde{\alpha}_y = \alpha_y + \mathrm{log}
    \left(
        \frac
	{\mathbf{M}_\epsilon (\alpha_u+\alpha_{uy} +\beta_u)}
	{\mathbf{M}_\epsilon (\alpha_u)\mathbf{M}_\epsilon (\beta_u)}
    \right)
\end{align*}
Consequently, we have
\begin{align*}
	f(W|A,Z,Y=1) \propto exp\{ W(\tilde{\alpha}_0 + \tilde{\alpha}_y+  (\alpha_u+\alpha_{uy}) m(A,Z)) \} \numberthis \label{lili-it-tmp3w}
\end{align*}
From analogous algebra, we find that $f(W|A,Z,Y=0)$ and $f(W|A,Z,Y=1)$ have the same form except for the intercept term and the slope term for $t(A,Z)$. 
Therefore, we can combine \eqref{lili-it-tmp3} and \eqref{lili-it-tmp3w}, and write
\begin{align*}
	&f(W|A,Z,Y)  \propto exp\{W(\tilde{\alpha}_0 + \alpha_um(A,Z) +  \alpha_{uy}m(A,Z)Y + \tilde{\alpha}_yY)\}  \numberthis \label{lili-it-final1} 
	\\
	&\text{where $\tilde{\alpha}_y= \alpha_y+
	\mathrm{log}
        \left(
        \frac{
	\mathbf{M}_\epsilon (\alpha_u+\alpha_{uy} +\beta_u)}{
	\mathbf{M}_\epsilon (\alpha_u) \mathbf{M}_\epsilon (\beta_u)}
        \right)
        $}
\end{align*}
By applying similar steps to $f(Y,U|A,Z,W=1)$, we obtain the equation below:
\begin{align*}
	&f(Y|A,Z,W) \propto exp\{Y(\tilde{\beta}_0 + \beta_aA + \beta_um(A,Z) +\beta_{uw} m(A,Z)W + \tilde{\beta}_wW)\} \numberthis \label{lili-it-final2}
	\\
	&\text{where $\tilde{\beta}_w=\beta_w+\mathrm{log}
        \left(
        \frac{
	\mathbf{M}_\epsilon (\alpha_u+\beta_{uw} +\beta_u)}{
	\mathbf{M}_\epsilon (\alpha_u)\mathbf{M}_\epsilon (\beta_u)}
        \right)$} 
\end{align*}

Combining \eqref{lili-it-final1}  and \eqref{lili-it-final2}, we get the following probability mass functions:
\begin{align}
	&f(W|A,Z,Y) = \frac{exp\{W(\tilde{\alpha}_0 + \alpha_u m(A,Z) + \alpha_{uy}m(A,Z)Y + \tilde{\alpha}_y Y)\}}{1+exp\{\tilde{\alpha}_0 + \alpha_u m(A,Z) + \alpha_{uy}m(A,Z)Y +\tilde{\alpha}_y Y\}} \numberthis \label{lili-dup-it3} 
	 \\
	&f(Y|A,Z,W) = \frac{exp\{Y(\tilde{\beta}_0 + \beta_aA + \beta_u m(A,Z) + \beta_{uw} m(A,Z)W+ \tilde{\beta}_wW)\}}{1+exp\{\tilde{\beta}_0 + \beta_aA + \beta_u m(A,Z) + \beta_{uw} m(A,Z)W+ \tilde{\beta}_wW\}} \numberthis \label{lili-dup-it4} \\
	&\text{where }  \tilde{\beta}_w=\tilde{\alpha}_y \notag
\end{align}
From \eqref{lili-dup-it3} and \eqref{lili-dup-it4}, we obtain the following two regression equations:
\begin{align*}
	&\mathrm{logit}(\mathrm{Pr}(W=1|A,Z,Y)) = \tilde{\alpha}_0 + \alpha_u m(A,Z) + \tilde{\alpha}_yY + \alpha_{uy}m(A,Z)Y \numberthis \label{lili-it5} \\
	&\mathrm{logit}(\mathrm{Pr}(Y=1|A,Z,W)) = \tilde{\beta}_0 + \beta_aA + \beta_u m(A,Z) + \tilde{\beta}_wW + \beta_{uw} m(A,Z)W \numberthis \label{lili-it6}
\end{align*}    

In this specification, we see $(Y,W)-U$ interactions can be incorporated by including $(Y,W)-(A,Z)$ interactions in the two-stage logistic regression procedure and Result \ref{result-5} is implied by equations \eqref{lili-it5} and \eqref{lili-it6}.
\end{proof}

Proximal control variable $S=\mathrm{logit}(\mathrm{Pr}(W=1|A,Z,Y=1))$ is a linear transformation of $E[U|A,Z,W=0,Y=0]=m(A,Z)$, i.e.
$$S=\mathrm{logit}(\mathrm{Pr}(W=1|A,Z,Y=1)) = (\tilde{\alpha}_0+\tilde{\alpha}_y)+(\alpha_u+\alpha_{uy})m(A,Z) \propto_L m(A,Z)$$
The procedure below shows how $S$ and $\beta_a$ can be estimated. Note that Procedure \ref{proc-4} is slightly different from Procedure \ref{proc-3}, as Procedure \ref{proc-4} adds another term $\mathrm{logit}(\mathrm{Pr}(W=1|A,Z,Y=1))W=SW$ in the second-stage regression.
\\
\begin{proc} \label{proc-4} Logit $Y$ Link, Logit $W$ Link Case with Interaction Terms
\begin{enumerate}
  \item Specify a linear regression function for $\mathrm{logit}(\mathrm{Pr}(W=1|A,Z,Y))$. Say $W \sim A+Z+Y+AZ+AY+ZY+AZY$.
  \item Perform the first-stage logistic regression $W\sim A+Z+Y+AZ+AY+ZY+AZY$.
  \item Compute $\hat{S}= \mathrm{logit}(\widehat{\mathrm{Pr}}(W=1|A,Z,Y=1))=(\hat{\alpha}_0^*+ \tilde{\alpha}_y) + (\hat{\alpha}_a^*+\hat{\alpha}_{ay}^*)A + (\hat{\alpha}_z^*+\hat{\alpha}_{zy}^*)Z+(\hat{\alpha}_{az}^*+\hat{\alpha}_{azy}^*)AZ$ using the estimated coefficients from the first-stage regression.
  \item Perform the second-stage logistic regression $Y \sim A + \hat{S} + W + \hat{S}W$ to obtain $\hat{\beta}_a$.
\end{enumerate}    
\end{proc}
\newpage

\subsubsection{Extension to Polytomous Regression}
\label{lip-lip}
In this section, we present details on the multinomial $Y$ and multinomial $W$ case, which corresponds to the gray cell in the table below.
\begin{figure}[H]
\centering
\includegraphics[width=1\textwidth]{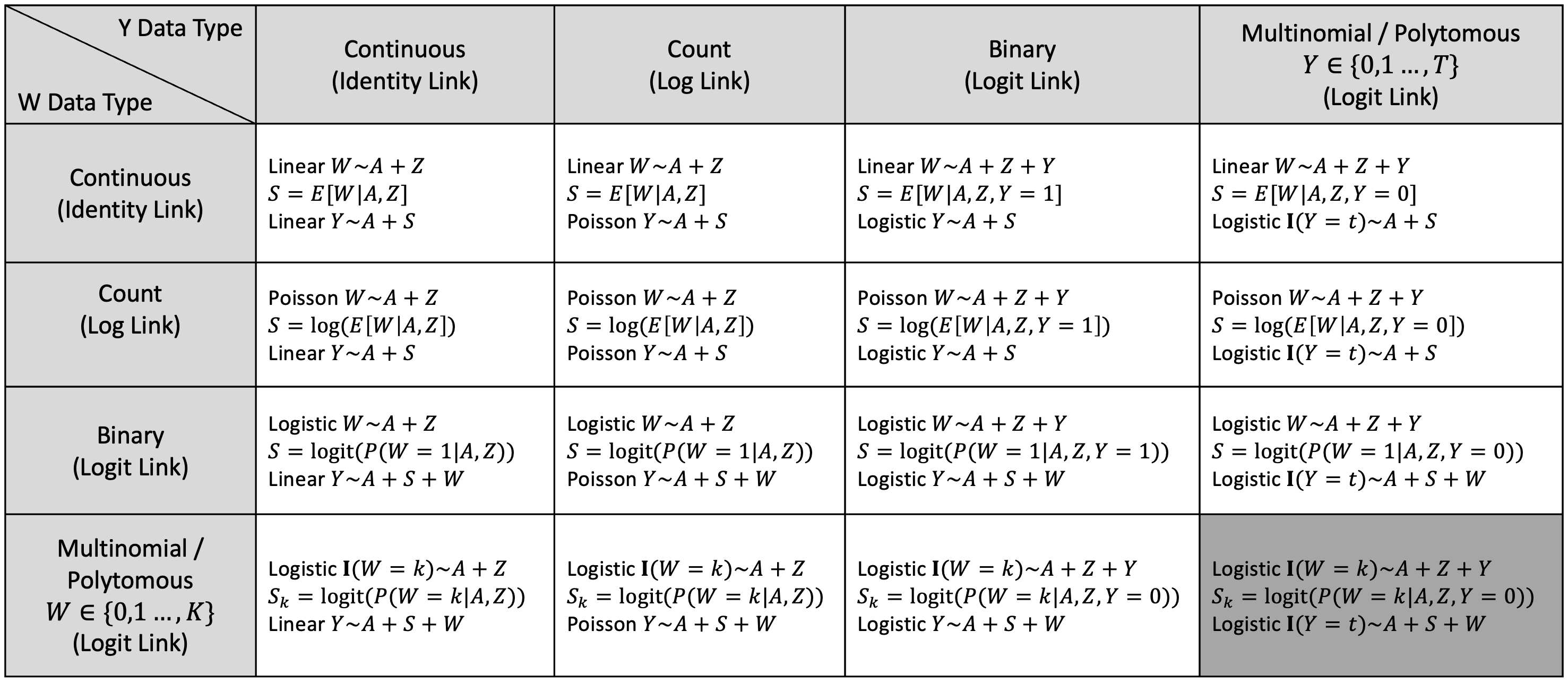}
\label{fig:uni_5}
\end{figure}
In the main text, $U$ is assumed to be a scalar throughout. Here, we consider a more general scenario with multidimensional $U$. We present the derivation for the logit links case in which $W$ has $K+1$ categories $(0,\ldots,K)$ (which is of dimension $K$) and $Y$ has $T+1$ categories $(0,\ldots,T)$. We assume $Z$ has a dimension of $Q$ and $U$ spans $R^V$ dimensions with $V \leq min(K,Q)$ so that the rank condition for identification is satisfied \cite{miao2018identifying}. We set $0$ as the reference level for $W$ and $Y$ and proceed with the following parameterization:
\begin{align*}
	&\mathrm{logit} (\mathrm{Pr}(W=k))
	 = \mathrm{log} \left( 
	 \frac{\mathrm{Pr}(W=k)}{\mathrm{Pr}(W=0)}
	 \right)
	 \text{ and }
	 \mathrm{logit} (\mathrm{Pr}(Y=t))
	 = \mathrm{log} \left( 
	 \frac{\mathrm{Pr}(Y=t)}{\mathrm{Pr}(Y=0)}
	 \right),\\
        &\textrm{for $k=1,\ldots,K$ and $t=1,\ldots,T$.}
\end{align*}

\begin{assumption}\label{assumption-poly}
\begin{align*}
& \mathrm{logit} (\mathrm{Pr}(Y=t|A,Z,W,U)) = \beta_{0t} + \beta_{at}A  + \beta_{ut}' U + \sum_{k=1}^K \beta_{wtk} I(W=k)) \numberthis \label{lili-p1} \\
& \mathrm{logit} (\mathrm{Pr}(W=k|A,Z,Y,U)) = \alpha_{0k}  + \alpha_{uk}' U +
\sum_{t=1}^T  \alpha_{ykt} I(Y=t)  \numberthis \label{lili-p2} \\
	&U|A,Z,Y=0,W=0 \sim \underbrace{E[U|A,Z,Y=0,W=0]}_{=m(A,Z)} + \epsilon; \ E[\epsilon]=0; \ \epsilon \indep (A,Z)|Y=0,W=0; \numberthis \label{aasp3} \\
	&\textrm{$\mathbf{M}_\epsilon(*)$, the MGF of $\epsilon|Y=0,W=0$ exists for any real value *.}\\
	& \text{\eqref{lili-p1} and \eqref{lili-p2} $\implies$ } \beta_{wtk} = \alpha_{ykt}
\end{align*}
\end{assumption}
We note that $\beta_{ut},\alpha_{uk},m(A,Z)\in R^V$ of the same dimension of the vector $U=(U_1,\ldots,U_V)'$ .

We now prove the general version of Result \ref{result-4} under Assumption \ref{assumption-poly}.

\begin{proof}
\\ \\
Due to the similarity with the proof in Section \ref{linot-linot}, we again only provide minimal algebraic details.

Based on the odds ratio functions, we establish the following relationships:
\begin{align*}
	OR(W=k,Y=t|A,Z,U)=exp\{ \beta_{wtk} \} = exp\{ \alpha_{ykt} \} \implies \beta_{wtk} = \alpha_{ykt}
\end{align*}

Using the odds ratio parametrization in \eqref{eq-odds ratio parametrization}, we find 
\begin{align*}
	&f(W=k,U|A,Z,Y) \\
	&\propto f(W=k|A,Z,Y,U= \underbar0)f(U|A,Z,Y,W=0) OR(W=k,U|A,Z,Y) \\
	&\propto exp\{\alpha_{0k}+ \sum_{t=1}^T\alpha_{ykt}I(Y=t)\}f(U|A,Z,Y,W=0) exp\{ \alpha_{uk}' U \}
\end{align*}
By marginalizing $f(W,U|A,Z,Y)$ with respect to $U$, we have 
\begin{align*}
        &f(W=k|A,Z,Y) \\
	&\propto exp\{\alpha_{0k}+\sum_{t=1}^T\alpha_{ykt}I(Y=t)\} \int_\mathcal{U} f(U=\underbar u|A,Z,Y,W=0) exp\{\alpha_{uk}'\underbar u\} d \underbar u \\
 	&= exp\{ 
	\alpha_{0k}+\mathrm{log}(\mathbf{M}_\epsilon (\alpha_{uk}'))
	 + \alpha_{uk}'m(A,Z) + \sum_{t=1}^T 
	\left(
        \alpha_{ykt}+\mathrm{log}
        \left(
        \frac{
	\mathbf{M}_\epsilon (\alpha_{uk}'+\beta_{ut}')}{
	\mathbf{M}_\epsilon (\alpha_{uk}')\mathbf{M}_\epsilon (\beta_{ut}')}
        \right)
        \right)
        I(Y=t)
	\} \\
	&= exp\{ 
	\tilde{\alpha}_{0k}
	 + \alpha_{uk}'m(A,Z) + \sum_{t=1}^T  \tilde{\alpha}_{ykt}I(Y=t)
	\} 
	\numberthis \label{lili-p-tmp1} \\
        &\textrm{where } 
        \tilde{\alpha}_{0k} = \alpha_{0k}+\mathrm{log}(\mathbf{M}_\epsilon (\alpha_{uk}')), \ 
        \tilde{\alpha}_{ykt} = \alpha_{ykt}+\mathrm{log}
        \left(
        \frac{
	\mathbf{M}_\epsilon (\alpha_{uk}'+\beta_{ut}')}{
	\mathbf{M}_\epsilon (\alpha_{uk}')\mathbf{M}_\epsilon (\beta_{ut}')}
        \right)
\end{align*}
Following analogous algebraic steps as in the binary $(Y, W)$ case, the transition from the second line to the third line becomes evident. Note the complete proof entails the representation from Bayes' rule applied to $f(W=k,U|A,Z,Y=t)$ for each $t \in \{1,\ldots,T\}$, which is omitted here.

From analogous algebra, we obtain the following result for $f(Y,U|A,Z,W)$
\begin{align*}
	&f(Y=t|A,Z,W) \\
	&\propto exp\{
	\beta_{0t} +\mathrm{log}(\mathbf{M}_\epsilon (\beta_{ut}'))
	+ \beta_{at}A + \beta_{ut}'m(A,Z) +
	\sum_{k=1}^K 
	\left(
        \beta_{wtk}+\mathrm{log}
        \left(
        \frac{
	\mathbf{M}_\epsilon (\alpha_{uk}'+\beta_{ut}')}{
	\mathbf{M}_\epsilon (\alpha_{uk}')\mathbf{M}_\epsilon (\beta_{ut}')}
        \right)
        \right)
	I(W=k)        
	\} \\
	&\propto exp\{
	\tilde{\beta}_{0t}
	+ \beta_{at}A + \beta_{ut}'m(A,Z) +
	\sum_{k=1}^K \tilde{\beta}_{wtk} I(W=k)
	\}
        \numberthis \label{lili-p-tmp2} \\
        &\textrm{where } 
        \tilde{\beta}_{0t} = \beta_{0t} +\mathrm{log}(\mathbf{M}_\epsilon (\beta_{ut}')), \
        \tilde{\beta}_{wtk} = \beta_{wtk}+\mathrm{log}
        \left(\frac{
	\mathbf{M}_\epsilon (\alpha_{uk}'+\beta_{ut}')}{
	\mathbf{M}_\epsilon (\alpha_{uk}')\mathbf{M}_\epsilon (\beta_{ut}')}
        \right)
\end{align*}
Combining \eqref{lili-p-tmp1} and \eqref{lili-p-tmp2}, we get the following probability mass functions
\begin{align*}
	&f(W=k|A,Z,Y) = \frac{exp\{
	\tilde{\alpha}_{0k}+\alpha_{uk}'m(A,Z) + \sum_{t=1}^T \tilde{\alpha}_{ykt} I(Y=t)
	\}}{
	1+ \sum_{k=1}^K exp\{\tilde{\alpha}_{0k}+\alpha_{uk}'m(A,Z) + \sum_{t=1}^T \tilde{\alpha}_{ykt} I(Y=t)
	\}
	} 
	\numberthis \label{lili-p3}\\
	&f(Y=t|A,Z,W)=\frac{exp\{
	\tilde{\beta}_{0t} + \beta_{at}A + \beta_{ut}'m(A,Z) + \sum_{k=1}^K \tilde{\beta}_{wtk}I(W=k)
	\}}{
	1+ \sum_{t=1}^T exp\{
		\tilde{\beta}_{0t} + \beta_{at}A + \beta_{ut}'m(A,Z) + \sum_{k=1}^K \tilde{\beta}_{wtk}I(W=k)
	\}
	}
	\numberthis \label{lili-p4}
	\\
	&\text{where }  \tilde{\beta}_{wtk}=\tilde{\alpha}_{ykt} \notag
\end{align*}
From \eqref{lili-p3} and \eqref{lili-p4}, we obtain the following two regression equations:
\begin{align*}
	&\mathrm{logit}(\mathrm{Pr}(W=k|A,Z,Y)) = \tilde{\alpha}_{0k}+\alpha_{uk}'m(A,Z) + \sum_{t=1}^T \tilde{\alpha}_{ykt} I(Y=t)  \numberthis \label{lili-p5} \\
	&\mathrm{logit}(\mathrm{Pr}(Y=t|A,Z,W)) = 	\tilde{\beta}_{0t} + \beta_{at}A + \beta_{ut}'m(A,Z) + \sum_{k=1}^K \tilde{\beta}_{wtk}I(W=k) \numberthis \label{lili-p6}
\end{align*}
Equations \eqref{lili-p5} and \eqref{lili-p6} imply a more general version of Result \ref{result-4} of the main paper.
\begin{align*}
	&\mathrm{logit} (\mathrm{Pr}(Y=t|A,Z,W))  \\
	&= \beta_{0t}^* + \beta_{at}^*A+ \sum_{k=1}^K\beta_{utk}^{*}\mathrm{logit} (\mathrm{Pr}(W=k|A,Z,Y=0))  +  \sum_{k=1}^K \tilde{\beta}_{wtk}I(W=k) \ \textrm{, where $\beta_{at}^*=\beta_{at}$.}
\end{align*}

\end{proof} 
Proximal control variable $S$ is a linear transformation of $E[U|A,Z,W=0,Y=0]=m(A,Z)$ where each $S_k$ is a linear combination of the element $m_v(A,Z)$ of $m(A,Z)$, i.e.
\begin{align*}
	&S = (S_1,\ldots,S_K) \\
	&m(A,Z) = (m_1(A,Z),\ldots,m_V(A,Z))\\
	&S_k = \mathrm{logit}(\mathrm{Pr}(W=k|A,Z,Y=0)) = \tilde{\alpha}_{0k}+\alpha_{uk}^{*'}m(A,Z) \propto_L m(A,Z) \\
	&S= C m(A,Z), \ where \ C \in R^{K\times V} \ of\ rank\ V
\end{align*}
Note that when \((W, Z)\) has a higher dimension than \(U\), i.e., \(\min(T, K) > V\), the rows of \(C\) will not be linearly independent, leading to collinearity among the elements of \(S\). This collinearity will be reflected in the second-stage regression. Therefore, if the estimated variance of the \(S\) coefficients in the second-stage regression is excessively large, one possible explanation is that \(U\) may be of significantly lower dimension than $W$ and $Z$.
\\
\\
The procedure below shows how $S$ and $\beta_a$ can be estimated.
\\
\begin{proc} \label{proc-5} Polytomous-Logit $Y$ Link, Polytomous-Logit $W$ Link Case
    \begin{enumerate}
  \item Specify a linear regression function for $\mathrm{logit}(\mathrm{Pr}(W=k|A,Z,Y))$. Say $W \sim A+Z+Y$ for all strata of $W$.
  \item Perform the first-stage polytomous logistic regression $W\sim A+Z+Y$.
  \item Compute $\hat{S} =  \left( \mathrm{logit}(\mathrm{Pr}(W=1|A,Z,Y=0)), \ldots ,\mathrm{logit}(\mathrm{Pr}(W=K|A,Z,Y=0)) \right)$ using the estimated coefficients from the first-stage regression.
  \item Perform the second-stage polytomous logistic regression $Y \sim A + \hat{S} + W$ to obtain $\hat{\beta}_{at}$.
\end{enumerate}
Similarly, following steps in Section \ref{lit-lit}, we can have interaction terms involved in regression function specification. Moreover, since the proof above does not require $K=T$, similar procedures apply to the cases where $(Y,W)$ are binary and multinomial (more than two categories), respectively, which correspond to the two gray cells of the table below. Details are omitted.
\end{proc}
\begin{figure}[H]
\centering
\includegraphics[width=1\textwidth]{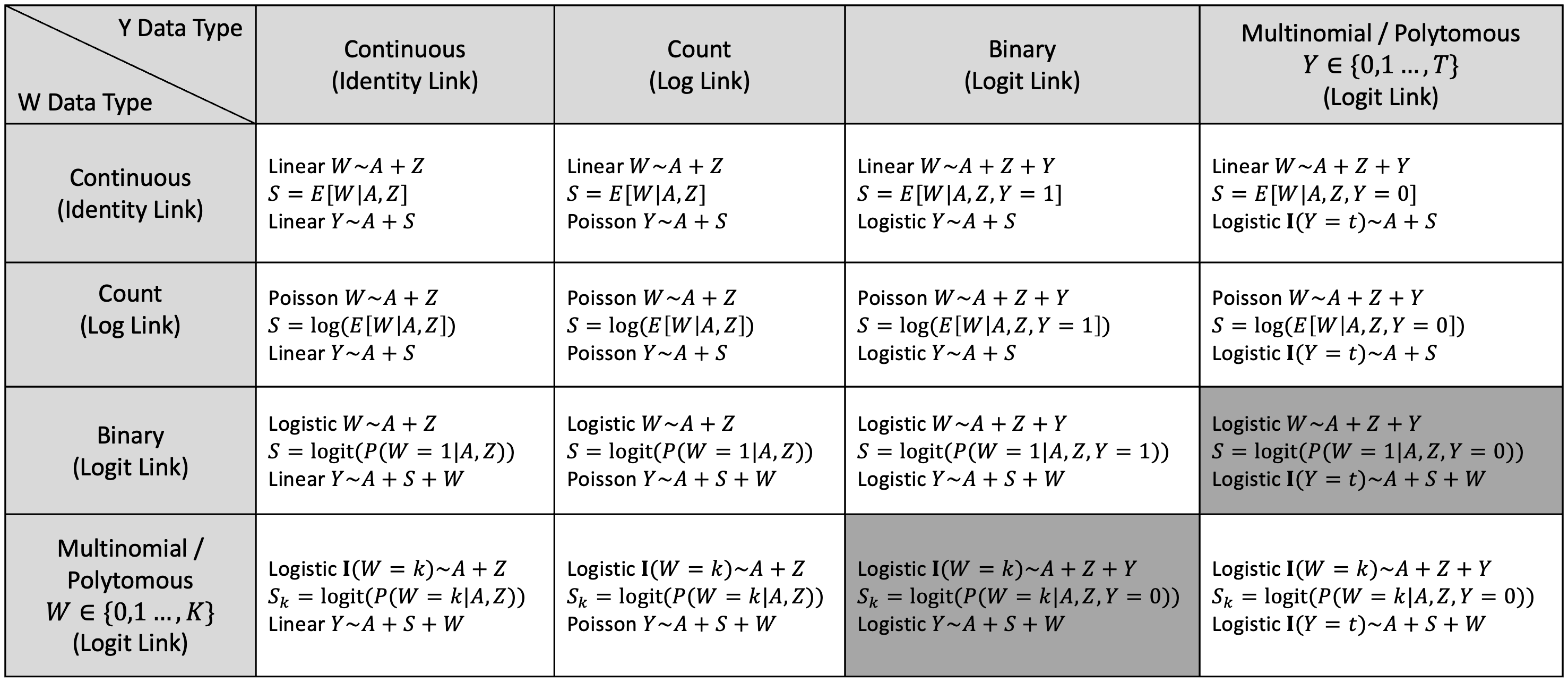}
\label{fig:SPEC3}
\end{figure}
\newpage

\subsection{
Identity and Log Links Combination}
In this section, we present details on the continuous $Y$ and count $W$ and count $Y$ and continuous $Y$ case, which correspond to the gray cells in the table below.
\begin{figure}[H]
\centering
\includegraphics[width=1\textwidth]{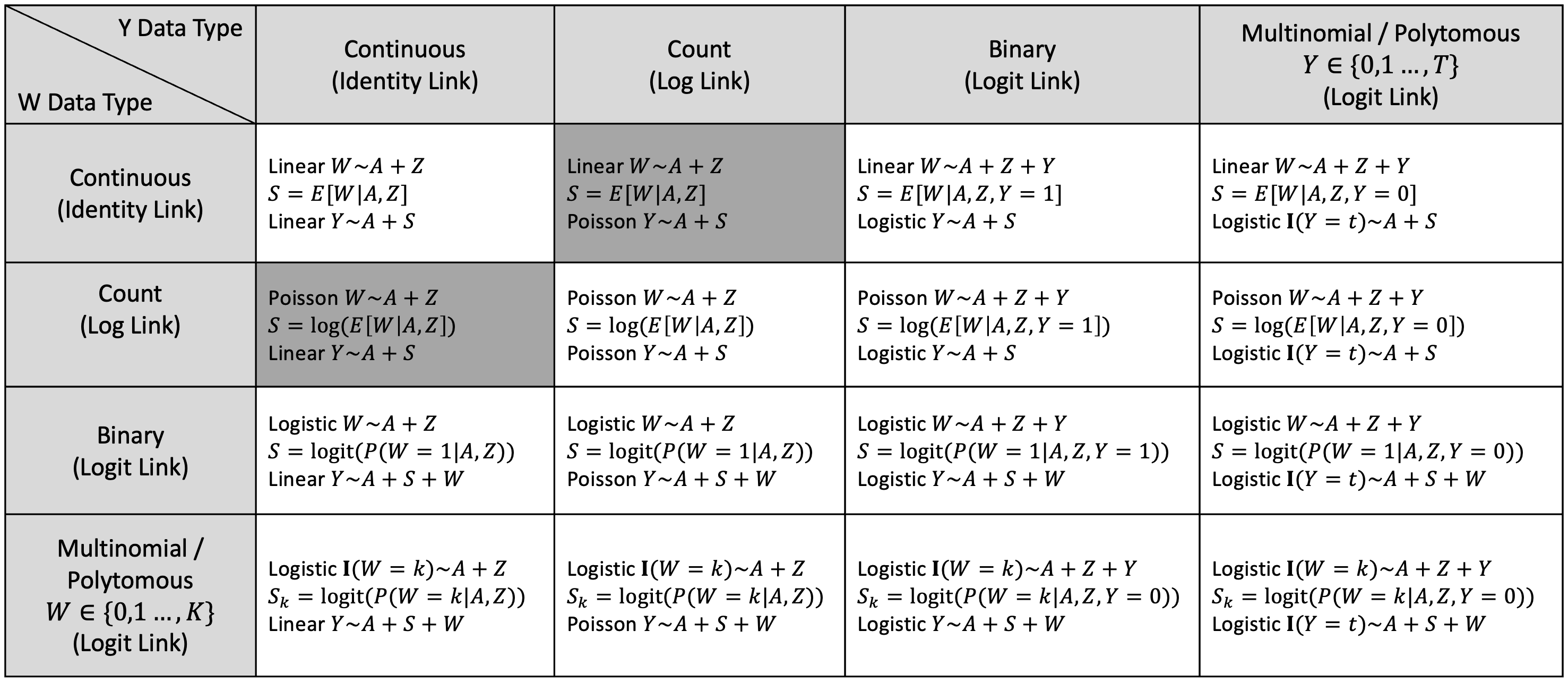}
\label{fig:uni_6}
\end{figure}
\subsubsection{Log $Y$ Link and Identity $W$ Link} 

We begin by presenting the derivation for the case ($Y$-Log, $W$-Identity) with the following model:

\begin{assumption}\label{assumption-logY-identityW}
\begin{align*}
	&\mathrm{log}(E[Y|A,Z,U]) = \beta_0+\beta_aA+\beta_uU  \numberthis \label{lo-id1} \\
	&E[W|A,Z,U] = \alpha_0 + \alpha_uU \numberthis \label{lo-id2}  \\
	&U|A,Z \sim E[U|A,Z] + \epsilon; \ E[\epsilon] = 0; \ \epsilon \indep (A,Z); \numberthis \label{lo-id-ass}  \\
        &\textrm{$\mathbf{M}_\epsilon(*)$, the MGF of $\epsilon$ exists for any real value *.}
\end{align*}
\end{assumption}

\begin{result} \label{result-6} Under Assumption \ref{assumption-logY-identityW}, it follows that     
\begin{align*}
	&\mathrm{log}(E[Y|A,Z]) = \beta_0^* + \beta_a^*A + \beta_u^*S, \\
	&\text{where } S=E[W|A,Z], \ \beta_a^* = \beta_a.
\end{align*}
\end{result} 
We now prove Result \ref{result-6} under Assumption \ref{assumption-logY-identityW}.

\begin{proof} 
\\ \\
Due to the shared collapsibility of identity and log links, iterating similar steps in Sections \ref{id-id} and \ref{lo-lo} leads to the answer. We only provide minimal algebraic details.
\\ \\
By taking the expectation of regression equations \eqref{lo-id1} and \eqref{lo-id2} with respect to $U$ conditional on $(A,Z)$, we obtain the following two regression equations which imply Result \ref{result-6}:
\begin{align*}
	&E[W|A,Z]= \alpha_0 + \alpha_u E[U|A,Z] \\
	&\mathrm{log}(E[Y|A,Z]) = \tilde{\beta}_0 + \beta_aA + \beta_u E[U|A,Z] \\
        &\textrm{where } \tilde{\beta}_0= \beta_0 + \mathrm{log}\mathbf{M}_\epsilon(\beta_u)
\end{align*}
\end{proof}
Proximal control variable $S$ is a linear transformation of $E[U|A,Z]$, i.e.
$$S=E[W|A,Z] \propto_L E[U|A,Z]$$
The procedure below shows how $S$ and $\beta_a$ can be estimated.
\\
\begin{proc} \label{proc-6} Log $Y$ Link, Identity $W$ Link Case
\begin{enumerate}
  \item Specify a linear regression function for $E[W|A,Z]$. Say $W\sim A+Z$.
  \item Perform the first-stage linear regression $W\sim A+Z$.
  \item Compute $\hat{S} =\hat{E}[W|A,Z]= \hat{\alpha}_0^* + \hat{\alpha}_a^*A + \hat{\alpha}_z^*Z$ using the estimated coefficients from the first-stage regression.
  \item Perform the second-stage Poisson regression $Y \sim A + \hat{S} $ to obtain $\hat{\beta}_a$.
\end{enumerate}
\end{proc}

\subsubsection{Identity $Y$ Link and Log $W$ Link}
The case ($Y$-Identity, $W$-Log) exhibits a symmetrical setting in both the model and the proof. We give the assumption, result and procedure of estimation directly.  

\begin{assumption}\label{assumption-identityY-logW}
\begin{align*}
	&E[Y|A,Z,U] = \beta_0+\beta_aA+\beta_uU \\
	&\mathrm{log}(E[W|A,Z,U]) = \alpha_0 + \alpha_uU \\
	&U|A,Z \sim E[U|A,Z] + \epsilon; \ E[\epsilon] = 0; \ \epsilon \indep (A,Z); \\
        &\textrm{$\mathbf{M}_\epsilon(*)$, the MGF of $\epsilon$ exists for any real value *.}
\end{align*}
\end{assumption}
	
\begin{result}[Symmetrical to Result \ref{result-6}] \label{result-7} Under Assumption \ref{assumption-identityY-logW}, it follows that     
\begin{align*}
	&E[Y|A,Z] = \beta_0^* + \beta_a^*A + \beta_u^*S, \\
	&\text{where } S=\mathrm{log}(E[W|A,Z]), \ \beta_a^* = \beta_a.
\end{align*}
\end{result}
Proximal control variable $S$ is a linear transformation of $E[U|A,Z]$, i.e.
$$S=\mathrm{log}(E[W|A,Z]) \propto_L E[U|A,Z]$$
The procedure below shows how $S$ and $\beta_a$ can be estimated.
\\
\begin{proc} \label{proc-7} Identity $Y$ Link, Log $W$ Link Case
    \begin{enumerate}
  \item Specify a linear regression function for $\mathrm{log}(E[W|A,Z])$. Say $W\sim A+Z$.
  \item Perform the first-stage Poisson regression $W\sim A+Z$.
  \item Compute $ \hat{S} = \mathrm{log}(\hat{E}[W|A,Z])= \hat{\alpha}_0^* + \hat{\alpha}_a^*A + \hat{\alpha}_z^*Z$ using the estimated coefficients from the first-stage regression.
  \item Perform the second-stage linear regression $Y \sim A + \hat{S}$ to obtain $\hat{\beta}_a$.
\end{enumerate}
\end{proc}
\newpage
\subsection{Identity and Logit Links Combination}
\label{id-li}

In this section, we present details on the continuous $Y$ and binary $W$ and binary $Y$ and continuous $Y$ case, which corresponds to the gray cells in the table below.
\begin{figure}[H]
\centering
\includegraphics[width=1\textwidth]{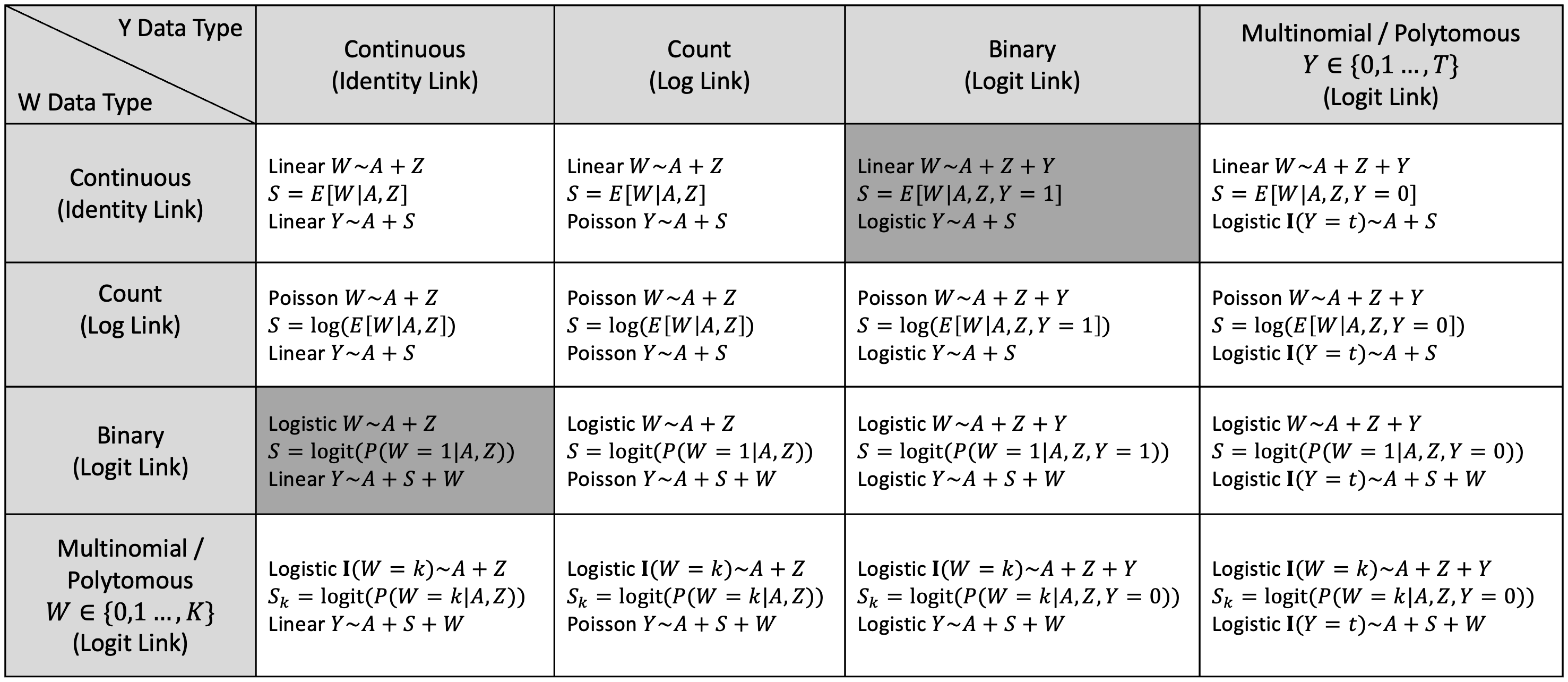}
\label{fig:uni_7}
\end{figure}
We again need compatible assumptions due to the property of the logit link. The single-logit case can be seen as the reduced form of what we present in Section \ref{li-li} for the double-logit link case. We borrow the same idea from Section \ref{li-li} and modify our assumptions accordingly. 

\subsubsection{Logit $Y$ Link and Identity $W$ Link Combination with No Interaction Term} 
\label{id-li-noint}
We first investigate the case ($Y$-Logit, $W$-Identity) with no $Y-U$ interaction. A more general model specification which allows for $Y-U$ interactions is detailed in Section \ref{id-li-int}. We consider the following model:
\begin{assumption}\label{assumption-logitY-identityW}
\begin{align*}
	&\mathrm{logit}(\mathrm{Pr}(Y=1|A,Z,U)) = \beta_0+\beta_aA+\beta_uU \numberthis \label{ali1} \\
	&E[W|A,Z,Y,U] = \alpha_0 + \alpha_uU + \alpha_y Y \numberthis \label{ali2} \\
	&U|A,Z,Y=0 \sim \underbrace{E[U|A,Z,Y=0]}_{=t(A,Z)} + \epsilon; \ E[\epsilon]=0; \ \epsilon \indep (A,Z)|Y=0; \numberthis \label{alia3} \\
	&\textrm{$\mathbf{M}_\epsilon(*)$, the MGF of $\epsilon|Y=0$ exists for any real value *.}
\end{align*}
\end{assumption}

\begin{result} \label{result-8} Under Assumption \ref{assumption-logitY-identityW}, it follows that     
\begin{align*}
&\mathrm{logit}(\mathrm{Pr}(Y=1|A,Z)) = \beta_0^* + \beta_a^*A + \beta_u^* S \\
&\text{where $S=E[W|A,Z,Y=1], \ \beta_a^*=\beta_a$.}
\end{align*}
\end{result} 

\begin{proof}
We establish the following results based on similar algebra in Section \ref{li-li}:
\begin{align*}
	f(Y,U|A,Z) &\propto f(Y|A,Z,U=0)f(U|A,Z,Y=0)OR(Y,U|A,Z) \\
	&=f(Y|A,Z,U=0) exp\{\beta_uYU\}f(U|A,Z,Y=0) \\
	f(Y|A,Z) 
	&\propto \int_{\mathcal{U}} f(Y|A,Z,U=0) exp\{\beta_uYu\}f(U=u|A,Z)du\\
	&=f(Y|A,Z,U=0)E[exp\{\beta_uYU\}|U=0,A,Z)]\\
	&=f(Y|A,Z,U=0)E[exp\{\beta_uY (t(A,Z)+\epsilon) \}|Y=0,A,Z)]\\
	&=f(Y|A,Z,U=0)exp\{\beta_uY t(A,Z)\} E[exp\{\beta_uY \epsilon \}|Y=0,A,Z)]\\	
	&=f(Y|A,Z,U=0)exp\{\beta_uY t(A,Z)\} exp\{\mathrm{log}\mathbf{M}_\epsilon (\beta_uY)\} \\
	&\propto exp\{Y(\beta_0 + \beta_aA+\beta_ut(A,Z) + \mathrm{log}\mathbf{M}_\epsilon (\beta_uY))\}	\\
	&= exp\{Y(\tilde{\beta}_0 + \beta_aA+\beta_ut(A,Z))\} \\
	f(Y|A,Z) &= \frac{exp\{Y(\tilde{\beta}_0 + \beta_aA+\beta_ut(A,Z))\}}{1+exp\{\tilde{\beta}_0 + \beta_aA+\beta_ut(A,Z)\}} \numberthis \label{ali4}
        \\ 
	&\textrm{where $\tilde{\beta}_0=\beta_0+ \mathrm{log}\mathbf{M}_\epsilon (\beta_u)$} 
\end{align*}
Taking the conditional expectation of \eqref{ali2} with respect to $U$ conditional on $(A,Z)$ in the stratum $(Y=0)$, we have:
\begin{align*}
	E[W|A,Z,Y=0] = \alpha_0 + \alpha_u E[U|A,Z,Y=0] = \alpha_0 + \alpha_u t(A,Z) \label{aid-lointermediate3} \numberthis
\end{align*}

From Bayes' rule, we have
\begin{align*}
	f(U|A,Z,Y=0) & = \frac{f(Y=0|U,A,Z)f(U|A,Z)}{f(Y=0|A,Z)} \\
	f(U|A,Z,Y=1) & = \frac{f(Y=1|U,A,Z)f(U|A,Z)}{f(Y=1|A,Z)} \\
	\frac{f(U|A,Z,Y=1)}{f(U|A,Z,Y=0)} & = \frac{f(Y=1|U,A,Z)f(Y=0|A,Z)}{f(Y=0|U,A,Z)f(Y=1|A,Z)} \\
	&=\frac{exp\{ \beta_0+\beta_aA+\beta_uU\}  }
	{exp\{\tilde{\beta}_0+\beta_aA+\beta_ut(A,Z)\}} \\
    & = exp\{ \beta_0 -\tilde{\beta}_0 +\beta_u(U-t(A,Z))\}
\end{align*}
which implies
\begin{align*}
    f(U|A,Z,Y=1) = exp\{ \beta_0 -\tilde{\beta}_0 +\beta_u(U-t(A,Z))\}f(U|A,Z,Y=0) \numberthis \label{eq-bayes-5}
\end{align*}
With the above representation, we find an alternative representation of $E[W|A,Z,Y=1]$ as follows:
\begin{align*}
    &E[U|A,Z,Y=1] \\
    & =\int_{\mathcal{U}} u f(U=u|A,Z,Y=1)du \\
    &=\int_{\mathcal{U}} u \cdot  exp\{\beta_0-\tilde{\beta}_0 + \beta_u(u-E[U|A,Z,Y=0])\}f(U=u|A,Z,Y=0) du \\
    &=E[ (E[U|A,Z,Y=0]+\epsilon)   exp\{ \beta_0-\tilde{\beta}_0+\beta_u (E[U|A,Z,Y=0] +\epsilon-E[U|A,Z,Y=0]) \}   | A,Z,Y=0)] \\
    &=exp\{\beta_0-\tilde{\beta}_0\} E[(t(A,Z) + \epsilon)exp\{\beta_u \epsilon \}|Y=0] \\
    &=exp\{\beta_0-\tilde{\beta}_0\} E[t(A,Z) exp\{\beta_u \epsilon \} |Y=0] + exp\{\beta_0-\tilde{\beta}_0\} E[\epsilon exp\{\beta_u \epsilon \}|Y=0] \\
    & = t(A,Z) + c \numberthis \label{eq-5-1}
\end{align*}
The second line is trivial from the definition of the conditional expectation.
The third line holds from \eqref{eq-bayes-5}.
The fourth line holds from the definition of conditional expectation.
The fifth line holds from the definition of $t(A,Z)$ and $\epsilon \indep (A,Z)) | Y=0$, which are assumed in \eqref{alia3}.
The sixth line holds where with a constant $c=exp\{\beta_0-\tilde{\beta}_0\} E[\epsilon exp\{\beta_u \epsilon \}|Y=0]$.
The last line holds as follows:
 \begin{align*} 
&exp\{\beta_0-\tilde{\beta}_0\} E[t(A,Z) exp\{\beta_u \epsilon \}|Y=0] \\
&=exp\{\beta_0-\tilde{\beta}_0\} t(A,Z) E[exp\{\beta_u \epsilon \}|Y=0] \\
&=exp\{\beta_0-\tilde{\beta}_0\} t(A,Z) \mathbf{M}_\epsilon (\beta_u) \\
&\stackrel{(a)}{=}exp\{\beta_0-\tilde{\beta}_0\} t(A,Z) exp\{\tilde{\beta}-\beta_0 \}  \\
&=t(A,Z) 
\end{align*}
where equality (a) is from $\tilde{\beta}_0=\beta_0+\mathrm{log}(\mathbf{M}_\epsilon (\beta_u))$.

Taking the conditional expectation of \eqref{ali2} with respect to $U$ conditional on $(A,Z)$ in the stratum $(Y=1)$, we have:
\begin{align*}
	E[W|A,Z,Y=1] = \alpha_0 + \alpha_u E[U|A,Z,Y=1] + \alpha_y = \alpha_0 + \alpha_u t(A,Z) + \alpha_y + \alpha_u c \label{aid-lointermediate4} \numberthis
\end{align*}
where the second equality holds from \eqref{eq-5-1}.
We can find that $E[W|A,Z,Y=0]$ and $E[W|A,Z,Y=1]$ have the same form except for the intercept term. 
Therefore, we can combine \eqref{aid-lointermediate3} and \eqref{aid-lointermediate4}, and write
\begin{align*}
&E[W|A,Z,Y]  =\alpha_0 + \alpha_uE[U|A,Z,Y] +\alpha_yY = \alpha_0 + \alpha_u t(A,Z) + \tilde{\alpha}_y Y, \\
&\textrm{where } \tilde{\alpha}_y = \alpha_y+\alpha_u c.
\end{align*}
Consequently, we obtain the following two regression equations which imply Result \ref{result-8}:
\begin{align*}
	&E[W|A,Z,Y] = \alpha_0 + \alpha_u t(A,Z) + \tilde{\alpha}_y Y  \\
	&\mathrm{logit}(\mathrm{Pr}(Y=1|A,Z)) = \tilde{\beta}_0 + \beta_aA + \beta_ut(A,Z)
\end{align*}
\end{proof} 

Proximal control variable $S$ is a linear transformation of $E[U|A,Z,Y=0]=t(A,Z)$, i.e.
$$S= E[W|A,Z,Y=1] =\alpha_0 + \tilde{\alpha}_y + \alpha_u t(A,Z) \propto_L t(A,Z)$$
The procedure below shows how $S$ and $\beta_a$ can be estimated.
\begin{proc} \label{proc-8} Logit $Y$ Link, Identity $W$ Link Case with No Interaction Term
    \begin{enumerate}
  \item Specify a linear regression function for $E[W|A,Z,Y]$. Say $W \sim A+Z+Y$.
  \item Perform the first-stage linear regression $W\sim A+Z+Y$.
  \item Compute $\hat{S} = \hat{E}[W|A,Z,Y=1] = \hat{\alpha}_0^* + \hat{\alpha}_a^*A + \hat{\alpha}_z^*Z + \hat{\alpha}_y^*$ using the estimated coefficients from the first-stage regression.
  \item Perform the second-stage logistic regression $Y \sim A +\hat{S}$ to obtain $\hat{\beta}_a$.
\end{enumerate}
\end{proc}
The same procedure applies to DAG (1) in Figure \ref{fig:1} for the same reason explained in Section \ref{li-li}.
\newpage

\subsubsection{Logit $Y$ Link and Identity $W$ Link Combination with Interaction Terms}
\label{id-li-int}
We consider the ($Y$-Logit, $W$-Identity) case with an alternative model specification below with $Y-U$ interaction, which is analogous to that in Section \ref{lit-lit}: 
\begin{assumption} \label{assumption-logitY-identityW-int}
\begin{align*}
	&\mathrm{logit}(\mathrm{Pr}(Y=1|A,Z,U)) = \beta_0+\beta_aA+\beta_uU \numberthis \label{alit1} \\
	&E[W|A,Z,Y,U] = \alpha_0 + \alpha_uU + \alpha_y Y + \alpha_{uy}UY \numberthis \label{alit2} \\
	&U|A,Z,Y=0 \sim \underbrace{E[U|A,Z,Y=0]}_{=t(A,Z)} + \epsilon; \ E[\epsilon]=0; \ \epsilon \indep (A,Z)|Y=0; \numberthis \label{aliat3} \\
	&\textrm{$\mathbf{M}_\epsilon(*)$, the MGF of $\epsilon|Y=0$ exists for any real value *.}
\end{align*}
\end{assumption} 
The model encodes an interaction between $U$ and $Y$ in the structural linear model for $W$. The following result modifies the two-stage regression procedure to account for such interactions.
\begin{result} \label{result-9} Under Assumption \ref{assumption-logitY-identityW-int}, it follows that     
\begin{align*}
&\mathrm{logit}(\mathrm{Pr}(Y=1|A,Z)) = \beta_0^* + \beta_a^*A + \beta_u^* S, \\
&\text{where $S=E[W|A,Z,Y=1], \ \beta_a^*=\beta_a$.}
\end{align*}
\end{result} 
Note $S=E[W|A,Z,Y=1]$ from Result \ref{result-9} can be fitted via the first-stage linear regression model with $Y-(A,Z)$ interactions incorporated, accounting for interaction $Y-(A,Z)$ in \eqref{alit2}.
\\ \\
\begin{proof}
\\ \\
Since the proof is similar to that presented in Section \ref{id-li-noint}, we only provide minimal algebraic details. Note that $f(Y|A,Z)$ has the same form as below:
\begin{align*}
	f(Y|A,Z) &= \frac{exp\{Y(\tilde{\beta}_0 + \beta_aA+\beta_ut(A,Z))\}}{1+exp\{\tilde{\beta}_0 + \beta_aA+\beta_ut(A,Z)\}},  
	\numberthis \label{alit4}
\end{align*}
where $\tilde{\beta}_0=\beta_0+ \mathrm{log}\mathbf{M}_\epsilon (\beta_u)$. 
For the expectation of regression equations \eqref{alit2} for $U$ conditional on $(A,Z)$ in the stratum $(Y=0)$, we have
\begin{align*}
	E[W|A,Z,Y=0] = \alpha_0 + \alpha_u E[U|A,Z,Y=0] = \alpha_0 + \alpha_u t(A,Z) \numberthis \label{alitintermediate1}
\end{align*}
The same equation is obtained from the Bayes' rule:
\begin{align*}
	&f(U|A,Z,Y=1) = exp\{ \beta_0 -\tilde{\beta}_0 +\beta_u(U-t(A,Z))\}f(U|A,Z,Y=0)
\end{align*}
With the above representation, we can find an alternative representation of $E[W|A,Z,Y=1]$ by following similar algebra in Section \ref{id-li-noint}
\begin{align*}
	&E[U|A,Z,Y=1] =\int_{\mathcal{U}}u f(U=u|A,Z,Y=1)du 
    =  t(A,Z) + c
\end{align*}
where the constant $c$ is the same as the one in \eqref{eq-5-1}. Therefore, we have 
\begin{align*}
E[W|A,Z,Y=1] = \alpha_0 + (\alpha_u + \alpha_{uy}) t(A,Z)  + (\alpha_u+\alpha_{uy})c \numberthis \label{alitintermediate2}
\end{align*}
We find that $E[W|A,Z,Y=0]$ and $E[W|A,Z,Y=1]$ have the same form except for the intercept term and the slope term for $t(A,Z)$. 
Therefore, we can combine \eqref{alitintermediate1} and \eqref{alitintermediate2}, and write
\begin{align*}
	&E[W|A,Z,Y]  
	=\alpha_0 + \alpha_ut(A,Z) + \tilde{\alpha}_y Y + \alpha_{uy}t(A,Z)Y \\
        &\textrm{where $\tilde{\alpha}_y = (\alpha_u+\alpha_{uy})c $}
\end{align*} 
Consequently, we obtain the following two regression equations which imply Result \ref{result-9}:
\begin{align*}
	&E[W|A,Z,Y] =  \alpha_0 + \alpha_ut(A,Z) + \tilde{\alpha}_y Y  + \alpha_{uy}t(A,Z)Y \\
	&\mathrm{logit}(\mathrm{Pr}(Y=1|A,Z))  = \tilde{\beta}_0 + \beta_aA+\beta_ut(A,Z)
\end{align*}
In this specification, we see $Y-U$ interactions can be incorporated by including $Y-(A,Z)$ interactions in the first-stage linear regression in the estimation procedure.
\end{proof}

Proximal control variable $S$ is a linear transformation of $E[U|A,Z,Y=1]=t(A,Z)$, i.e.
$$S= E[W|A,Z,Y=1] = (\alpha_0+\tilde{\alpha}_y) + (\alpha_u+\alpha_{uy}) t(A,Z) \propto_L t(A,Z)$$
The procedure below shows how $S$ and $\beta_a$ can be estimated.
\begin{proc} \label{proc-9} Logit $Y$ Link, Identity $W$ Link Case with Interaction Terms
    \begin{enumerate}
  \item Specify a linear regression function for $E[W|A,Z,Y]$. Say $W \sim A+Z+Y+AZ+AY+ZY+AZY$.
  \item Perform the first-stage linear regression $W \sim A+Z+Y+AZ+AY+ZY+AZY$.
  \item Compute $\hat{S} = \hat{E}[W|A,Z,Y=1] = (\hat{\alpha}_0^*+ \tilde{\alpha}_y) + (\hat{\alpha}_a^*+\hat{\alpha}_{ay}^*)A + (\hat{\alpha}_z^*+\hat{\alpha}_{zy}^*)Z+(\hat{\alpha}_{az}^*+\hat{\alpha}_{azy}^*)AZ$ using the estimated coefficients from the first-stage regression.
  \item Perform the second-stage logistic regression $Y \sim A + \hat{S}$ to obtain $\hat{\beta}_a$.
\end{enumerate}
\end{proc}



\subsubsection{Identity $Y$ Link and Logit $W$ Link Combination with No Interaction Term}
\label{id-li-sym-noint}
The case ($Y$-Identity, $W$-Logit) with no $W-U$ interaction is symmetrical to the case in Section \ref{id-li-noint}. A more general model specification which allows for $W-U$ interactions is detailed in Section \ref{id-li-sym-int}. We consider the following model:

\begin{assumption}\label{assumption-identityY-logitW}
    \begin{align*}
	&E[Y|A,Z,W,U] = \beta_0+\beta_aA+\beta_uU + \beta_wW  \\
	&\mathrm{logit}(\mathrm{Pr}(W=1|A,Z,U)) = \alpha_0 + \alpha_uU \\
	&U|A,Z,W=0 \sim \underbrace{E[U|A,Z,W=0]}_{=h(A,Z)} + \epsilon; \ E[\epsilon]=0; \ \epsilon \indep (A,Z)|W=0; \\
	&\textrm{$\mathbf{M}_\epsilon(*)$, the MGF of $\epsilon|W=0$ exists for any real value *.}
\end{align*}
\end{assumption}

\begin{result}[Symmetrical to Result \ref{result-8}] \label{result-10} Under Assumption \ref{assumption-identityY-logitW}, it follows that     
\begin{align*}
&E[Y|A,Z,W] = \beta_0^* + \beta_a^*A + \beta_u^* S + \beta_w^* W, \\
&\text{where $S=\mathrm{logit}(\mathrm{Pr}(W=1|A,Z)), \ \beta_a^*=\beta_a$.}
\end{align*}
\end{result} 

\begin{proof} \\ \\
Based on similar algebra in \eqref{id-li-noint}, we obtain the following two regression equations which imply Result \ref{result-10}:
\begin{align*} 
	&\mathrm{logit}(\mathrm{Pr}(W=1|A,Z)) = \tilde{\alpha}_0 + \alpha_uh(A,Z)  \\
	&E[Y|A,Z,W] = \beta_0 + \beta_aA + \beta_u h(A,Z) + \tilde{\beta}_wW
\end{align*}    
\end{proof}

Proximal control variable $S$ is is a linear transformation of $E[U|A,Z,W=0]=h(A,Z)$, i.e., 
$$S= \mathrm{logit}(\mathrm{Pr}(W=1|A,Z)) =\tilde{\alpha}_0+\alpha_u h(A,Z) \propto_L h(A,Z)$$
The procedure below shows how $S$ and $\beta_a$ can be estimated.
\begin{proc} \label{proc-10} Identity $Y$ Link, Logit $W$ Link Case with No Interaction Term
    \begin{enumerate}
  \item Specify a linear regression function for $\mathrm{logit}(\mathrm{Pr}(W=1|A,Z))$. Say $W \sim A+Z$.
  \item Perform the first-stage logistic linear regression $W\sim A+Z$.
  \item Compute $\hat{S}=\mathrm{logit}(\widehat{\mathrm{Pr}}(W=1|A,Z))= \hat{\alpha}_0^* + \hat{\alpha}_a^*A + \hat{\alpha}_z^*Z$ using the estimated coefficients from the first-stage regression.
  \item Perform the second-stage linear regression $Y \sim A + \hat{S} + W$ to obtain $\hat{\beta}_a$.
\end{enumerate}
\end{proc}

The same procedure applies to DAG (1) in Figure \ref{fig:1} for the same reason explained in Section \ref{li-li}. \\ \\
The above model specification excludes $W-(A,Z)$ interactions in the estimation procedure. The more general model specification is detailed in the next section.

\subsubsection{Identity $Y$ Link and Logit $W$ Link Combination with Interaction Terms}
\label{id-li-sym-int}
The case ($Y$-Identity, $W$-Logit) with $W-U$ interaction is symmetrical to the case in Section \ref{id-li-int}. Specifically, we consider the following model:
\begin{assumption} \label{assumption-identityY-logitW-int}
    \begin{align*}
	&E[Y|A,Z,W,U] = \beta_0+\beta_aA+\beta_uU + \beta_wW + \beta_{uw}UW \numberthis \label{id-lo-sym-1} \\
	&\mathrm{logit}(\mathrm{Pr}(W=1|A,Z,U)) = \alpha_0 + \alpha_uU  \\
	&U|A,Z,W=0 \sim \underbrace{E[U|A,Z,W=0]}_{=h(A,Z)} + \epsilon; \ E[\epsilon]=0; \ \epsilon \indep (A,Z)|W=0; \\
	&\textrm{$\mathbf{M}_\epsilon(*)$, the MGF of $\epsilon|W=0$ exists for any real value *.}
    \end{align*}
\end{assumption}
The model encodes an interaction between $U$ and $W$ in the structural linear model for $Y$. The following result modifies the two-stage regression procedure to account for such interactions.
\begin{result}[Symmetrical to Result \ref{result-9}] \label{result-11} Under Assumption \ref{assumption-identityY-logitW-int}, it follows that     
\begin{align*}
&E[Y|A,Z,W] = \beta_0^* + \beta_a^*A + \beta_u^* S + \beta_w^* W + \beta_{uw}^*SW, \\
&\text{where $S=\mathrm{logit}(\mathrm{Pr}(W=1|A,Z)), \ \beta_a^*=\beta_a$.}
\end{align*}
\end{result}
Note Result \ref{result-11} requires the second-stage logistic regression model to include the term $SW$, which encodes for interactions $W-(A,Z)$, accounting interactions $W-U$ in \eqref{id-lo-sym-1}.

\begin{proof} \\ \\
Under the model allowing for interactions between $W-U$, we can show the following two regression equations which imply Result \ref{result-11} from some algebra analogous to that presented in Section \ref{id-li-int}.
\begin{align*}
    &\mathrm{logit}(\mathrm{Pr}(W=1|A,Z)) = \tilde{\alpha}_0 + \alpha_uh(A,Z)  \\
    &E[Y|A,Z,W] = \beta_0+\beta_aA+\beta_u h(A,Z) + \tilde{\beta}_w W + \beta_{uw} h(A,Z)W 
\end{align*}
In this specification, we see $W-U$ interactions can be incorporated by including $SW$ interaction in the second-stage linear regression in the estimation procedure.
\end{proof}

Proximal control variable $S$ is is a linear transformation of $E[U|A,Z,W=0]=h(A,Z)$, i.e.
$$S= \mathrm{logit}(\mathrm{Pr}(W=1|A,Z)) = \tilde{\alpha}_0+\alpha_u h(A,Z) \propto_L h(A,Z)$$
The procedure below shows how $S$ and $\beta_a$ can be estimated.

\begin{proc} \label{proc-11} Identity $Y$ Link, Logit $W$ Link Case with Interaction Terms
\begin{enumerate}
  \item Specify a linear regression function for $\mathrm{logit}(\mathrm{Pr}(W=1|A,Z))$. Say $W \sim A+Z$.
  \item Perform the first-stage logistic linear regression $W\sim A+Z$.
  \item Compute $\hat{S}=\mathrm{logit}(\widehat{\mathrm{Pr}}(W=1|A,Z))= \hat{\alpha}_0^* + \hat{\alpha}_a^*A + \hat{\alpha}_z^*Z$ using the estimated coefficients from the first-stage regression.
  \item Perform the second-stage linear regression $Y \sim A + \hat{S} + W + \hat{S}W$ to obtain $\hat{\beta}_a$.
\end{enumerate}
\end{proc}

By similar algebra steps, we can solve the case where the logit link is applied to the multinomial $Y$ or $W$. The procedures are given in the two gray cells of the table below. Details are omitted.

\begin{figure}[H]
\centering
\includegraphics[width=1\textwidth]{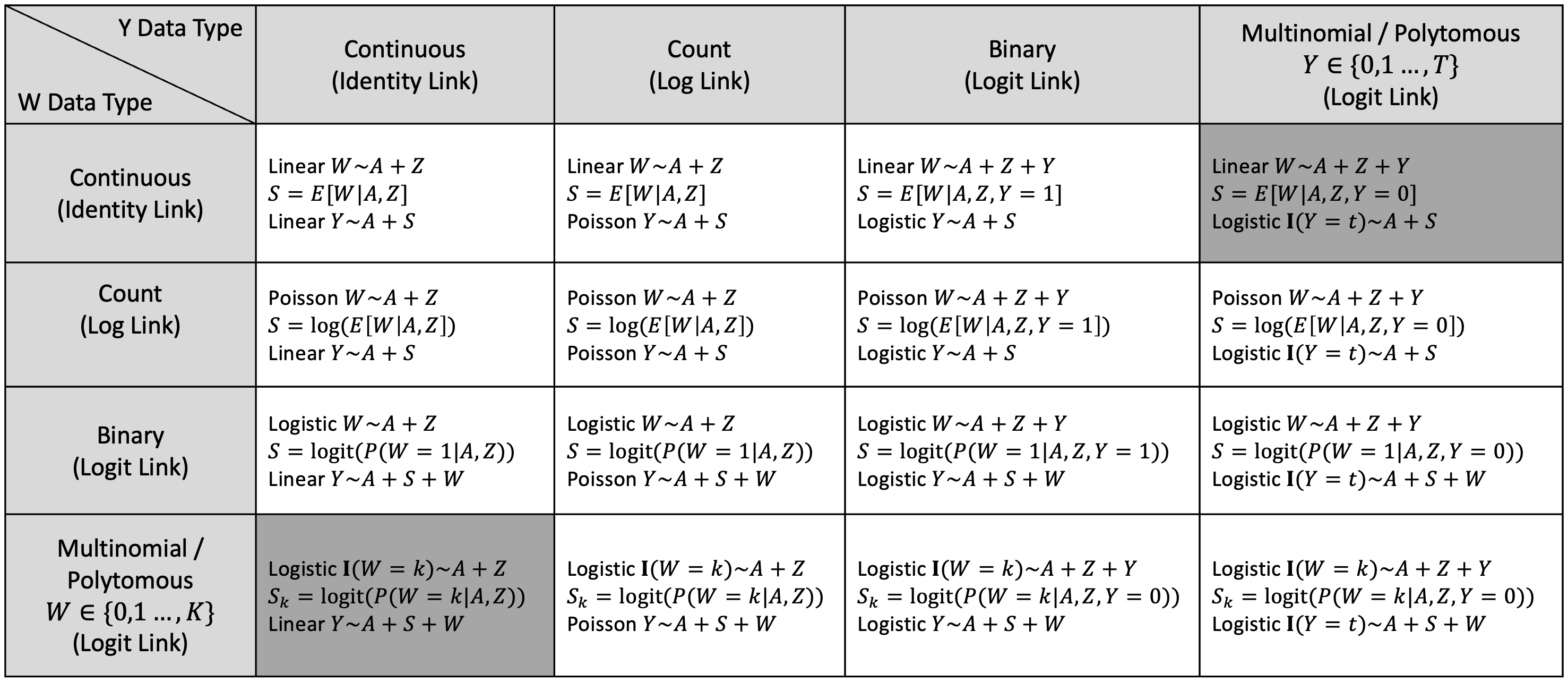}
\label{fig:SPEC1}
\end{figure}

\newpage

%


\subsection{Log and Logit Links Combination}
\label{lo-li}

In this section, we present details on the count $Y$ and binary $W$ and binary $Y$ and count $Y$ case, which corresponds to the gray cells in the table below.

\begin{figure}[H]
\centering
\includegraphics[width=1\textwidth]{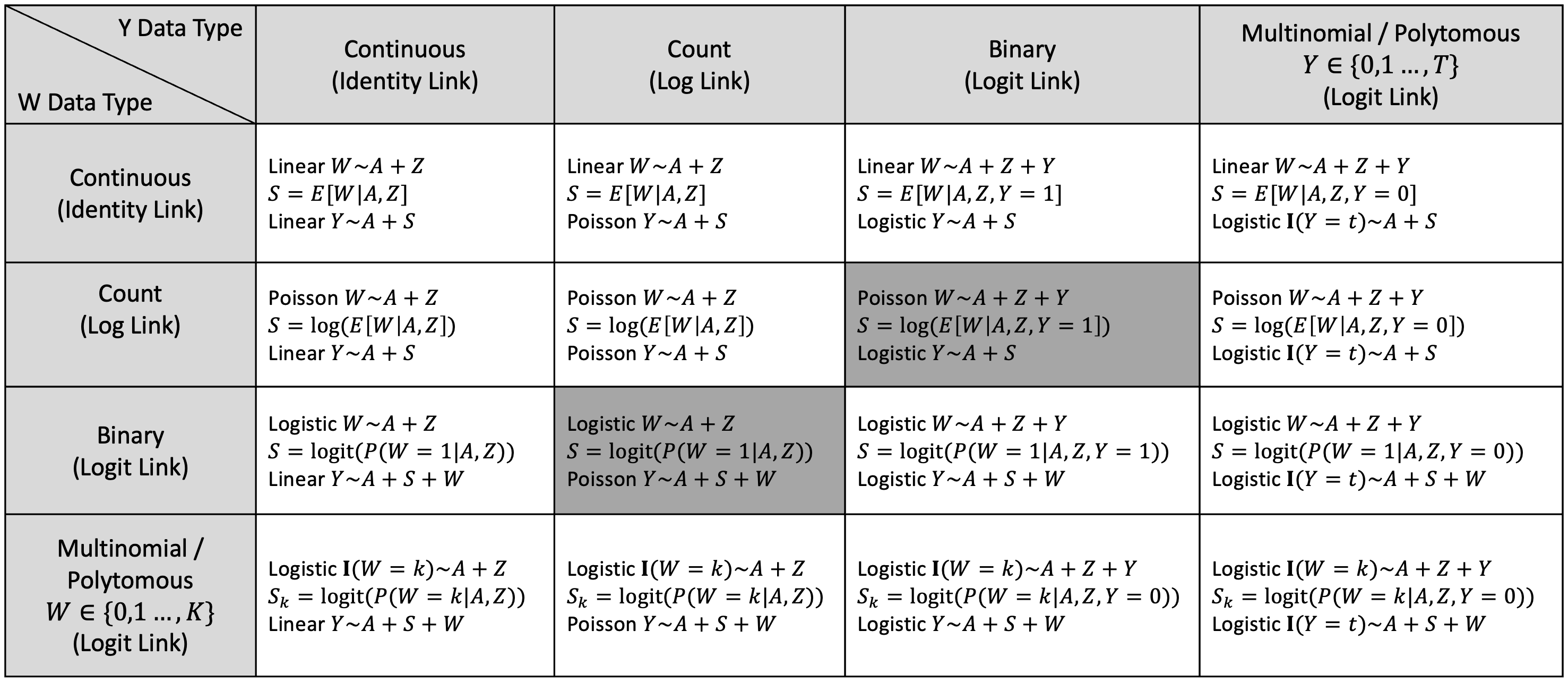}
\label{fig:uni_8}
\end{figure}

It is not entirely surprising that the steps followed in Section \ref{id-li} extend to the current setting, largely due to similar properties of the identity and the log links. The main difference is the need to replace the identity link from Section \ref{id-li} and the intercept terms absorbing constants from MGFs; a similar phenomenon has been discussed in Section \ref{lo-lo}. Due to algebraic similarities, we omit the details and directly state the results.

\subsubsection{Logit $Y$ Link and Log $W$ Link Combination with No Interaction Term}
\label{lo-li-noint}
We first investigate the case ($Y$-Logit, $W$-Log) with no $Y-U$ interaction. A more general model specification which allows for $Y-U$ interactions is detailed in Section \ref{lo-li-int}. We consider the following model:
\begin{assumption} \label{assumption-logitY-logW}
    \begin{align*}
	&\mathrm{logit}(\mathrm{Pr}(Y=1|A,Z,U)) = \beta_0+\beta_aA+\beta_uU  \\
	&\mathrm{log}(E[W|A,Z,Y,U]) = \alpha_0 + \alpha_uU + \alpha_y Y \\
	&U|A,Z,Y=0 \sim \underbrace{E[U|A,Z,Y=0]}_{=t(A,Z)} + \epsilon; \ E[\epsilon]=0; \ \epsilon \indep (A,Z)|Y=0;  \\
	&\textrm{$\mathbf{M}_\epsilon(*)$, the MGF of $\epsilon|Y=0$ exists for any real value *.}
\end{align*}
\end{assumption}

\begin{result} \label{result-12} Under Assumption \ref{assumption-logitY-logW}, it follows that     
\begin{align*}
&\mathrm{logit}(\mathrm{Pr}(Y=1|A,Z)) = \beta_0^* + \beta_a^*A + \beta_u^* S, \\
&\text{where $S=\mathrm{log}(E[W|A,Z,Y=1]), \ \beta_a^*=\beta_a$.}
\end{align*}
\end{result} 

\begin{proof}
\\ \\
Following algebra in Section \ref{lo-lo} and \ref{id-li-noint}, we obtain the following two regression equations which imply Result \ref{result-12}:
\begin{align*} 
	&\mathrm{log}(E[W|A,Z,Y]) = \tilde{\alpha}_0 + \alpha_u t(A,Z) + \tilde{\alpha}_yY  \\
	&\mathrm{logit}(\mathrm{Pr}(Y=1|A,Z)) = \tilde{\beta}_0 + \beta_aA + \beta_u t(A,Z)
\end{align*}
\end{proof}
Proximal control variable $S$ is a linear transformation of $E[U|A,Z,Y=0]=t(A,Z)$, i.e.
$$S= \mathrm{log}(E[W|A,Z,Y=1]) = \tilde{\alpha}_0 + \tilde{\alpha}_y + \alpha_u t(A,Z) \propto_L t(A,Z)$$
The procedure below shows how $S$ and $\beta_a$ can be estimated.
\\
\begin{proc} \label{proc-12} Logit $Y$ Link, Log $W$ Link Case with No Interaction Term
\begin{enumerate}
  \item Specify a linear regression function for $\mathrm{log}(E[W|A,Z,Y])$. Say $W \sim A+Z+Y$.
  \item Perform the first-stage Poisson regression $W\sim A+Z+Y$.
  \item Compute $\hat{S} = \mathrm{log}(\hat{E}[W|A,Z,Y=1]) = \hat{\alpha}_0^* + \hat{\alpha}_a^*A + \hat{\alpha}_z^*Z + \hat{\alpha}_y^*$ using the estimated coefficients from the first-stage regression.
  \item Perform the second-stage logistic regression $Y \sim A + \hat{S}$ to obtain $\hat{\beta}_a$.
\end{enumerate}
\end{proc}
The same procedure applies to DAG (1) in Figure \ref{fig:1} for the same reason explained in Section \ref{li-li}. 

\subsubsection{Logit $Y$ Link and Log $W$ Link Combination with Interaction Terms}
\label{lo-li-int}
We consider the ($Y$-Logit, $W$-Log) case with an alternative model specification below with $Y-U$ interaction, which is analogous to that in Section \ref{id-li-int}
\begin{assumption} \label{assumption-logitY-logW-int}
    \begin{align*}
	&\mathrm{logit}(\mathrm{Pr}(Y=1|A,Z,U)) = \beta_0+\beta_aA+\beta_uU \\
	&\mathrm{log}(E[W|A,Z,Y,U]) = \alpha_0 + \alpha_uU + \alpha_y Y + \alpha_{uy}UY \numberthis \label{lo-li-int-2} \\
	&U|A,Z,Y=0 \sim \underbrace{E[U|A,Z,Y=0]}_{=t(A,Z)} + \epsilon; \ E[\epsilon]=0; \ \epsilon \indep (A,Z)|Y=0; \\
	&\textrm{$\mathbf{M}_\epsilon(*)$, the MGF of $\epsilon|Y=0$ exists for any real value *.}
\end{align*}
\end{assumption}

The model encodes an interaction between $U$ and $Y$ in the structural log-linear model for $W$. The following result modifies the two-stage regression procedure to account for such interactions.

\begin{result} \label{result-13} Under Assumption \ref{assumption-logitY-logW-int}, it follows that     
\begin{align*}
&\mathrm{logit}(\mathrm{Pr}(Y=1|A,Z)) = \beta_0^* + \beta_a^*A + \beta_u^* S, \\
&\text{where $S=\mathrm{log}(E[W|A,Z,Y=1]), \ \beta_a^*=\beta_a$.}
\end{align*}
\end{result} 
Note $S=\mathrm{log}(E[W|A,Z,Y=1])$ from Result \ref{result-13} can be fitted via the first-stage Poisson regression model with $Y-(A,Z)$ interactions incorporated, accounting for interaction $Y-(A,Z)$ in \eqref{lo-li-int-2}.

\begin{proof} \\ \\
Under the model allowing for $Y-U$ interaction, we can show the following two regression equations which imply Result \ref{result-13} from some algebra analogous to that presented in Section \ref{id-li-int}.
\begin{align*} 
	&\mathrm{log}(E[W|A,Z,Y]) = \tilde{\alpha}_0 + \alpha_u t(A,Z) + \tilde{\alpha}_yY + \alpha_{uy}t(A,Z)Y\\
	&\mathrm{logit}(\mathrm{Pr}(Y=1|A,Z)) = \tilde{\beta}_0 + \beta_aA + \beta_u t(A,Z) 
\end{align*}
In this specification, we see $Y-U$ interactions can be incorporated by including $Y-(A,Z)$ interactions in the fist-stage Poisson regression in the estimation procedure.
\end{proof}

Proximal control variable $S$ is a linear transformation of $E[U|A,Z,Y=0]=t(A,Z)$, i.e.
$$S= \mathrm{log}(E[W|A,Z,Y=1]) = (\tilde{\alpha}_0 + \tilde{\alpha}_y) + (\alpha_u +  \tilde{\alpha}_y) t(A,Z)  \propto_L t(A,Z)$$
The procedure below shows how $S$ and $\beta_a$ can be estimated.

\begin{proc} \label{proc-13} Logit $Y$ Link, Log $W$ Link Case with Interaction Terms    
\end{proc}
\begin{enumerate}
  \item Specify a linear regression function for $\mathrm{log}(E[W|A,Z,Y])$. Say $W \sim A+Z+Y+AZ+AY+ZY+AZY$.
  \item Perform the first-stage Poisson regression $W \sim A+Z+Y+AZ+AY+ZY+AZY$.
  \item Compute $\hat{S} = \mathrm{log}(\hat{E}[W|A,Z,Y=1]) = (\hat{\alpha}_0^*+ \tilde{\alpha}_y) + (\hat{\alpha}_a^*+\hat{\alpha}_{ay}^*)A + (\hat{\alpha}_z^*+\hat{\alpha}_{zy}^*)Z+(\hat{\alpha}_{az}^*+\hat{\alpha}_{azy}^*)AZ$ using the estimated coefficients from the first-stage regression.
  \item Perform the second-stage logistic regression $Y \sim A + \hat{S}$ to obtain $\hat{\beta}_a$.
\end{enumerate}

\subsubsection{Log $Y$ Link and Logit $W$ Link Combination with No Interaction Term}
\label{lo-li-noint-sym}
The case ($Y$-Log, $W$-Logit) is with no $W-U$ interaction is symmetrical to the case in Section \ref{lo-li-noint}. A more general model specification which allows for $W-U$ interactions is detailed in Section \ref{lo-li-int-sym}. We consider the following model:

\begin{assumption}\label{assumption-logY-logitW}
\begin{align*}
	&\mathrm{log}(E[Y|A,Z,W,U]) = \beta_0+\beta_aA+\beta_uU + \beta_wW  \\
	&\mathrm{logit}(\mathrm{Pr}(W=1|A,Z,U)) = \alpha_0 + \alpha_uU \\
	&U|A,Z,W=0 \sim \underbrace{E[U|A,Z,W=0]}_{=h(A,Z)} + \epsilon; \ E[\epsilon]=0; \ \epsilon \indep (A,Z)|W=0;\\
	&\textrm{$\mathbf{M}_\epsilon(*)$, the MGF of $\epsilon|W=0$ exists for any real value *.}
\end{align*}
\end{assumption}

\begin{result}[Symmetrical to Result \ref{result-12}] \label{result-14} Under Assumption \ref{assumption-logY-logitW}, it follows that     
\begin{align*}
&\mathrm{log}(E[Y|A,Z,W]) = \beta_0^* + \beta_a^*A + \beta_u^* S + \beta_w^*W, \\
&\text{where $S=\mathrm{logit}(\mathrm{Pr}(W=1|A,Z)), \ \beta_a^*=\beta_a$.}
\end{align*}
\end{result}

\begin{proof} \\ \\
    Following algebra in Section \ref{id-li-sym-noint}, we obtain the following two regression equations:
\begin{align*} 
	&\mathrm{logit}(\mathrm{Pr}(W=1|A,Z)) = \tilde{\alpha}_0 + \alpha_uh(A,Z)  \\
	&\mathrm{log}(E[Y|A,Z,W]) = \beta_0 + \beta_aA + \beta_u h(A,Z) + \tilde{\beta}_wW 
\end{align*}
\end{proof} 

Proximal control variable $S$ is a linear transformation of $E[U|A,Z,W=0]=t(A,Z)$, i.e.
$$S= \mathrm{logit}(\mathrm{Pr}(W=1|A,Z)) =\alpha_0+\alpha_u h(A,Z) \propto_L h(A,Z)$$
The procedure below shows how $S$ and $\beta_a$ can be estimated.
\begin{proc} \label{proc-14} Log $Y$ Link, Logit $W$ Link Case with No Interaction Term
    \begin{enumerate}
  \item Specify a linear regression function for $\mathrm{logit}(\mathrm{Pr}(W=1|A,Z,Y))$. Say $W \sim A+Z$.
  \item Perform the first-stage logistic regression $W\sim A+Z$.
  \item Compute $\hat{S} = \mathrm{logit}(\widehat{\mathrm{Pr}}(W=1|A,Z)) = \hat{\alpha}_0^* + \hat{\alpha}_a^*A + \hat{\alpha}_z^*Z$ using the estimated coefficients from the first-stage regression.
  \item Perform the second-stage Poisson regression $Y \sim A + \hat{S} +W$ to obtain $\hat{\beta}_a$.
\end{enumerate}
\end{proc}

The same procedure applies to DAG (1) in Figure \ref{fig:1} for the same reason explained in Section \ref{li-li}. 

\newpage

\subsubsection{Log $Y$ Link and Logit $W$ Link Combination with Interaction Terms}
\label{lo-li-int-sym}

The case ($Y$-Log, $W$-Logit) with $W-U$ interaction is symmetrical to the case in Section \ref{lo-li-int}. Specifically, we consider the following model:

\begin{assumption} \label{assumption-logY-logitW-int}
\begin{align*}
	&\mathrm{log}(E[Y|A,Z,W,U]) = \beta_0+\beta_aA+\beta_uU + \beta_wW + \beta_{uw}UW \numberthis \label{lo-li-sym-int} \\
	&\mathrm{logit}(\mathrm{Pr}(W=1|A,Z,U)) = \alpha_0 + \alpha_uU  \\
	&U|A,Z,W=0 \sim \underbrace{E[U|A,Z,W=0]}_{=h(A,Z)} + \epsilon; \ E[\epsilon]=0; \ \epsilon \indep (A,Z)|W=0; \\
	&\textrm{$\mathbf{M}_\epsilon(*)$, the MGF of $\epsilon|W=0$ exists for any real value *.}
\end{align*}
\end{assumption}

The model encodes an interaction between $U$ and $W$ in the structural log-linear model for $Y$. The following result modifies the two-stage regression procedure to account for such interactions.

\begin{result}[Symmetrical to Result \ref{result-13}] \label{result-15} Under Assumption \ref{assumption-logY-logitW-int}, it follows that     
\begin{align*}
&\mathrm{log}(E[Y|A,Z,W]) = \beta_0^* + \beta_a^*A + \beta_u^* S + \beta_w^*W + \beta_{uw}^*SW, \\
&\text{where $S=\mathrm{logit}(\mathrm{Pr}(W=1|A,Z)), \ \beta_a^*=\beta_a$.}
\end{align*}
\end{result} 

Note Result \ref{result-15} requires the second-stage Poisson regression model to include the term $SW$, which encodes for interactions $W-(A,Z)$, accounting interactions $W-U$ in \eqref{lo-li-sym-int}.

\begin{proof} 
\\ \\
Under the model allowing for interactions between $W-U$, we can show the following result from some algebra analogous to that presented in Section \ref{id-li-sym-int}:
\begin{align*} 
	&\mathrm{logit}(\mathrm{Pr}(W=1|A,Z)) = \tilde{\alpha}_0 + \alpha_uh(A,Z)  \\
	&\mathrm{log}(E[Y|A,Z,W]) = \tilde{\beta}_0+\beta_aA+\beta_uh(A,Z) + \tilde{\beta}_wW + \beta_{uw}h(A,Z)W
\end{align*}
In this specification, we see $W-U$ interactions can be incorporated by including $SW$ interaction in the second-stage Poisson regression in the estimation procedure.
\end{proof}
Proximal control variable $S$ is a linear transformation of $E[U|A,Z,W=0]=t(A,Z)$, i.e.
$$S= \mathrm{logit}(\mathrm{Pr}(W=1|A,Z)) =\alpha_0+\alpha_u h(A,Z) \propto_L h(A,Z)$$
The procedure below shows how $S$ and $\beta_a$ can be estimated.

\begin{proc} \label{proc-15} Log $Y$ Link, Logit $W$ Link Case with Interaction Terms
\begin{enumerate}
  \item Specify a linear regression function for $\mathrm{logit}(\mathrm{Pr}(W=1|A,Z,Y))$. Say $W \sim A+Z$.
  \item Perform the first-stage logistic regression $W\sim A+Z$.
  \item Compute $\hat{S} = \mathrm{logit}(\widehat{\mathrm{Pr}}(W=1|A,Z)) = \hat{\alpha}_0^* + \hat{\alpha}_a^*A + \hat{\alpha}_z^*Z$ using the estimated coefficients from the first-stage regression.
  \item Perform the second-stage Poisson regression $Y \sim A + \hat{S} + W + \hat{S}W$ to obtain $\hat{\beta}_a$.
\end{enumerate}
\end{proc}

By similar algebra steps, we can solve the case where the logit link is applied to the multinomial $Y$ or $W$. The procedures are given in the two gray cells of the table below. Details are omitted.

\begin{figure}[H]
\centering
\includegraphics[width=1\textwidth]{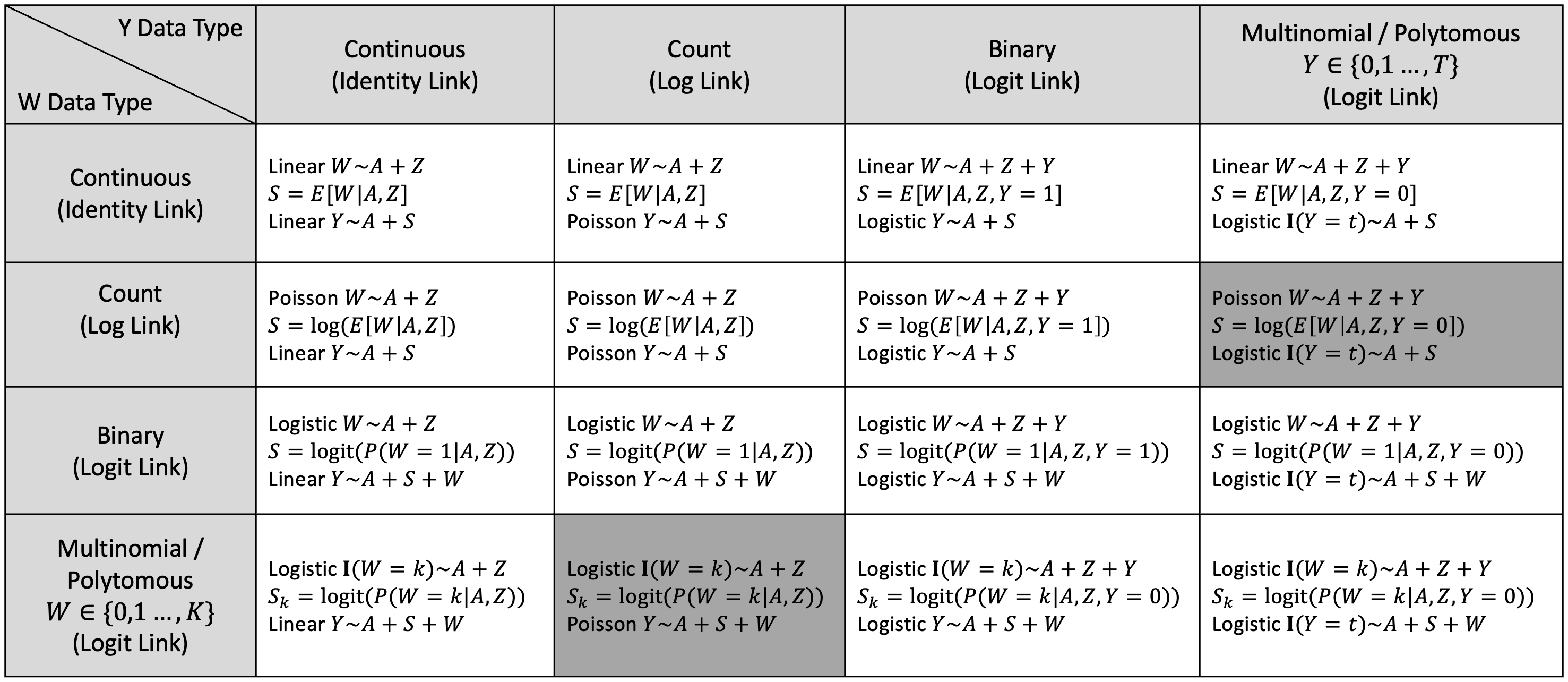}
\label{fig:SPEC1}
\end{figure}

\newpage
\subsection{Simulation Study}
\label{ss}
In this section, we evaluate the finite sample performance of the proposed estimator for binary $(Y,W)$ data pair with logit links. From Section \ref{linot-linot} , the following assumption are made:
\begin{align*}
	&\mathrm{logit}(\mathrm{Pr}(Y=1|A,Z,W,U)) = \beta_0+\beta_aA+\beta_uU+\beta_w W \numberthis \label{local1} \ \\
	&\mathrm{logit}(\mathrm{Pr}(W=1|A,Z,Y,U)) = \alpha_0+\alpha_uU+\alpha_y Y \numberthis \label{local2}  \\
	&U|A,Z,Y=0,W=0 \sim \underbrace{E[U|A,Z,Y=0,W=0]}_{=m(A,Z)} + \epsilon; \ E[\epsilon]=0; \ \epsilon \indep (A,Z)|Y=0,W=0; \\
	&\textrm{$\mathbf{M}_\epsilon(*)$, the MGF of $\epsilon|Y=0,W=0$ exists for any real value *.}\\
	&\text{\eqref{local1} and \eqref{local2} $\implies$ }\beta_w=\alpha_y
\end{align*}
Before embarking on the simulation, two useful results are derived. Let $f_\epsilon$ denote the probability function of $\epsilon$. By odds ratio parametrization, we find:
\begin{align*}
	&f(U|A,Z) \\
	&= \sum_{w=0,1} \sum_{y=0,1} f(U,Y=y,W=w|A,Z) \\
	&\propto \sum_{w=0,1} \sum_{y=0,1} f(U|A,Z,Y=0,W=0)f(Y=y|A,Z,W=0,U=0)f(W=w|A,Z,Y=0,U=0) \\
	& \quad \quad \cdot exp\{\beta_uyU\}\cdot exp\{\alpha_uwU\}\cdot exp\{\beta_wyw\} \\
	&=  \sum_{w=0,1} f_\epsilon(U-m(A,Z))f(W=w|A,Z,Y=0,U=0) \{ \frac{1+exp\{\beta_0+\beta_aA +\beta_uU\} exp\{\beta_w w\}}{1+exp\{\beta_0+\beta_aA\}}\}\\
	&=  f_\epsilon(U-m(A,Z)) \{ 
	\frac{1+
	exp\{\beta_0+\beta_aA +\beta_uU\} +
		(1+exp\{\beta_0+\beta_aA +\beta_uU\}) \exp\{\beta_w\} \exp\{\alpha_0 +\alpha_u U\}
	}{(1+exp\{\beta_0+\beta_aA\})(1+exp\{\alpha_0\})}                                                                         
	\} \numberthis \label{gen1}
\end{align*}
The third line holds from the following property of the odds ratio function. Specifically, the joint density function of $(A,B,C)$ given $D$ is represented as 
\begin{align*}
    &f(A,B,C|D) = \frac{f(A|B=0,C=0,D)\cdot f(B|A=0,C=0,D)\cdot f(C|A=0,B=0,D)}{\text{Normalizing Constant}} \notag \\
	&\quad \quad\quad\quad\quad\quad\quad\quad \cdot OR(A,B|C=0,D) \cdot OR(A,C|B=0,D) \cdot OR(B,C|A=0,D) \cdot T (A,B,C|D) \\
	&\text{where } \ T(A,B,C|D)= \frac{OR(A,B|C,D)}{OR(A,B|C=0,D)} = \frac{OR(A,C|B,D)}{OR(A,C|B=0,D)} =\frac{OR(B,C|A,D)}{OR(B,C|A=0,D)}
\end{align*}
This result implies that for an arbitrary error distribution $f_\epsilon$ in the stratum $(Y=0,W=0)$, the error distribution is indirectly induced for other three strata $(Y=1,W=0)$, $(Y=0,W=1)$ and $(Y=1,W=1)$ while being compatible with the restrictions imposed by regression equations \eqref{local1} and \eqref{local2}. This also reveals the reason for constraining assumptions for $\epsilon$ within the baseline stratum $(Y=0,W=0)$. With this result, we can generate $U$ from $(A,Z)$ through $f(U|A,Z)$ via the accept-reject sampling algorithm.
\\ \\
Again, by odds ratio parametrization, we find:
\begin{align*}
    f(Y,W|U,A,Z)  
	&\propto f(Y|A,Z,W=0,U)f(W|A,Z,Y=0,U) OR(Y,W|A,Z,U) \\
	&\propto exp\{Y(\beta_0+\beta_aA+\beta_uU)\} exp\{W(\alpha_0+\alpha_uU)\} exp\{\beta_w YW\} \numberthis \label{gen2}
\end{align*}
This result allows for generating $(Y,W)$ from $(U,A,Z) $ through $f(Y,W|U,A,Z)$ via a multinomial distribution (which encodes the bivariate Bernoulli distribution using 4 levels). 
\\ \\
Then, we consider the following data generation process with pre-specified parameters and error distribution.
\begin{enumerate}
    \item Generate $A$ and $Z$ independently from $N(0,0.5)$
    \item Generate $U|A,Z$ from $f(U|A,Z)$ presented in \eqref{gen1}, where we let $m(A,Z) = E[U|A,Z,Y=0,W=0] = -0.4 + 0.8A + 1.2Z - AZ$ and $f_\epsilon$ takes Logistic$(\mu=0,s=0.3)$, using the accept-reject sampling algorithm.
    \item Generating $(Y,W)$ from $f(Y,W|U,A,Z)$ presented in \eqref{gen2} with $\beta_0 = -1.4$, $\beta_a = 1.2$, $\beta_u = -0.7$, $\beta_w = \alpha_y = 0.5$, $\alpha_0 = -0.8$ and $\alpha_u=0.5$ via a multinomial distribution which encodes the corresponding bivariate Bernouli distribution using 4 levels.
\end{enumerate}

We generate $N$ i.i.d. samples of $(Y,W,A,Z,U)$, where the sample size $N$ takes values in $\{250, 500, 1000, 1500\}$. Using the simulated data, we consider three estimators:
\begin{enumerate}
  \item The naive estimator: We run a linear regression model $Y\sim A+W$, and take the coefficient of $A$ as an estimator for $\beta_a$. This estimator does not account for the presence of $U$.
  \item The two-stage estimator: An estimator based on the two-stage regression approach presented in Section \ref{bi} of the main text.
  \item The oracle estimator: We run the oracle linear regression model $Y\sim A+W+U$ by using unobservable $U$, and take the coefficient of $A$ as an estimator for $\beta_a$.
\end{enumerate}

\begin{figure}[H]
\centering
\includegraphics[width=0.9\textwidth]{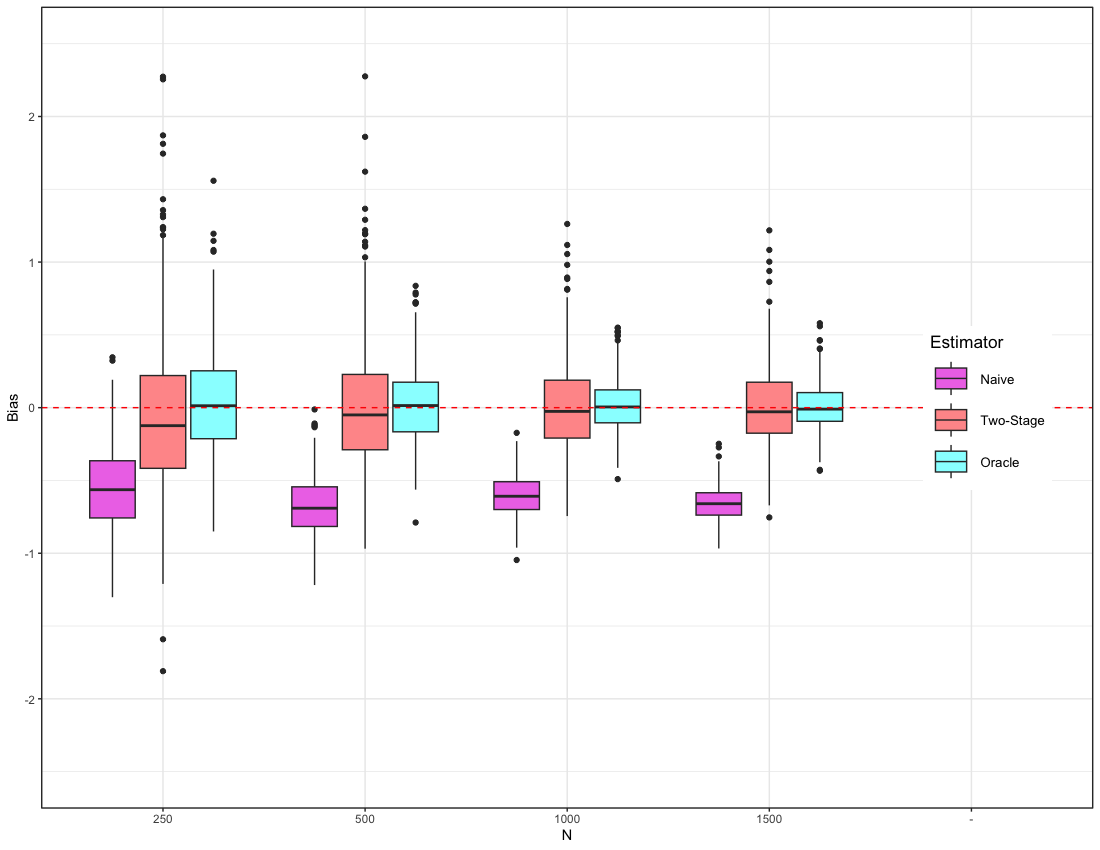}
\caption{A Graphical Summary of the Simulation Result. Each column gives boxplots of biases of the three competing estimators for $N = 250, 500,1000,1500$, respectively. The colors (purple, orange, cyan) encode the corresponding estimator (naive, two-stage, oracle). The $y$-axis represents the magnitude of bias.}
\label{fig:sim1}
\end{figure}

For inference of the two-stage estimator, we use the following percentile bootstrap approach. Specifically, (i) we randomly sample $N$ observations with replacement from the generated data of size $N$; (ii) obtain the two-stage regression estimate; and (iii) repeat steps (i) and (ii) 300 times. We then obtain a bootstrap standard error, a standard deviation of the 300 estimates, and a 95\% percent confidence interval of the form $[q_{2.5}, q_{97.5}]$ where $q_\alpha$ is the $\alpha$th percentile of the 300 bootstrap estimates. 
\\ \\
For each value of $N$, we evaluate the performance of each estimator based on 500 repetitions. The simulation results are given in Figure \ref{fig:sim1} and Table \ref{tab:sim2}. In Figure \ref{fig:sim1}, boxplots of biases of the three estimators are provided. Figure \ref{fig:sim1} shows that the naive estimator is generally biased even at small $N$, say $N=250$, while our two-stage estimator appears to yield negligible biases across all $N$. Compared to the oracle estimator, the two-stage estimator has a larger variance attributed to the uncertainty arising from the unmeasured $U$. Table \ref{tab:sim2} provides numerical summaries of our two-stage estimator. As $N$ increases, the empirical standard error and bootstrap standard error decrease, and they become similar to each other. Lastly, we find that empirical coverage from 95\% confidence intervals based on the bootstrap standard error appears to attain the nominal coverage. Based on these simulation results, the proposed two-stage estimator agrees with the asymptotic properties established in Section \ref{theori}. 
\\ \\
The codes for simulation in \texttt{R} are available at: https://github.com/NoBirdInTree/Regression-Based-PCI/tree/main.

\begin{table}[htp]
\centering
\begin{tabular}{rrrrr}
  \hline
Sample Size & Bias & Empirical S.E. & Bootstrap S.E. & Coverage \\ 
  \hline
250 & -0.07 & 0.55 & 0.88 & 0.97 \\
  500 & 0.02 & 0.49 & 0.61 & 0.95 \\
  1000 & 0.00 & 0.31 & 0.33 & 0.95 \\
  1500 & 0.02 & 0.27 & 0.28 & 0.94 \\
   \hline
\end{tabular}
\caption{Summary Statistics of the Two-Stage Estimator. ``Bias'' column gives the empirical bias of 500 estimates. ``Empirical S.E.'' column gives the standard deviation of 500 estimates. ``Bootstrap S.E.'' column gives the average of bootstrap standard errors of 500 estimates. ``Coverage'' column gives the empirical coverage rate of 95\% confidence intervals based on the bootstrap S.E.}
\label{tab:sim2}
\end{table}

\newpage
\subsection{Statistical Properties of the Two-stage Estimators}
\label{theori}
For notational brevity, let $N$ be the number of observations and $O_i=(Y_i,A_i,W_i,Z_i^T,X_i^T)$ be the observed data of unit $i$. We focus on the case where $A_i$ and $W_i$ are univariate, but $Z_i$ and $X_i$ can be multivariate, with dimensions $p_z$ and $p_x$ respectively. In cases where the first- and second-stage regression models use identity or log link functions, the models without interaction terms can be written as follows:
\begin{align*} 
    & \text{The first stage} : g_1(E[W|A,Z,X]) =\alpha_0^* + \alpha_a^*A + \alpha_z^{*T}Z + \alpha_x^{*T}X \numberthis \label{eq1} \\
    & \text{Proximal Control Variable} : S(\alpha^*) = \alpha_0^* + \alpha_a^*A +\alpha_z^{*T}Z  + \alpha_x^{*T}X \numberthis \label{eq2}\\
    & \text{The second stage} : g_2(E[Y|A,Z,X]) = \beta_0^*+\beta_a^*A+\beta_u^*S(\alpha^*) + \beta_x^{*T}X \numberthis \label{eq3}
\end{align*}
where $g_1$ and $g_2$ are link functions chosen according to the nature of $W$ and $Y$, and $S=S(\alpha^*)$ stresses $S$ is a function of parameters $\alpha^*$. The asterisk superscript ($*$) is defined in Table \ref{table:3}. The displayed regression models only involved main effects, but the derivation follows similarly with models involving interaction or transformed variables.

Table \ref{table:1} lists the regression models for cases involving logit links. Although the notation below does not apply to cases where the logit link is involved, the inference of the parameters can be derived following the same approach.

Equations \eqref{eq1}, \eqref{eq2} and \eqref{eq3} are equivalent to the following:
\begin{align*}
    & \text{The first stage} : E[W|A,Z,X] =g_1^{-1}(\alpha_0^* + \alpha_a^*A +\alpha_z^{*T}Z + \alpha_x^{*T}X) \numberthis \label{eq4} \\
    & \text{Proximal Control Variable} : S(\alpha^*) = \alpha_0^* +\alpha_a^*A +\alpha_z^{*T}Z  + \alpha_x^{*T}X\numberthis \label{eq5}\\
    & \text{The second stage} : E[Y|A, Z, X] = g_2^{-1}(\beta_0^*+\beta_a^*A+\beta_u^*S(\alpha^*)+ \beta_x^{*T}X) \numberthis \label{eq6}
\end{align*}
With \eqref{eq4}, \eqref{eq5} and \eqref{eq6}, with canonical links being used in the maximum likelihood estimation (MLE), we can construct the following estimating functions associated with the first- and second-stage regressions:
\begin{align*}
    &\text{The first stage} : \Psi_1(O ; \alpha^*) 
    = 
    \begin{bmatrix}
        1 \\ A \\ Z \\ X
    \end{bmatrix}
    \Big\{ 
        W - g_1^{-1} \big( \alpha_0^* + \alpha_a^* A + \alpha_z^{*T} Z \big)
    \Big\} 
    \in \reals^3
    \\
    &\text{The second stage} : \Psi_2(O ; \alpha^*, \beta^*) 
    = 
    \begin{bmatrix}
        1 \\ A \\ S \\X
    \end{bmatrix}
    \Big\{
        Y - g_2^{-1} \big( \beta_0^* + \beta_a^* A + \beta_u^*S (\alpha^*) + \beta_x^{*T}X \big)
    \Big\} 
    \in \reals^3
    \\
    &\text{where } S(\alpha^*) = \alpha_0^* +\alpha_a^*A +\alpha_z^{*T}Z + \alpha_x^{*T}X
\end{align*}
These two estimating functions satisfy $E\{ \Psi_1(O ; \alpha^*) | A, Z, X\} = 0$ and $E\{ \Psi_2(O ; \alpha^*, \beta^*) | A, Z, X\} = 0$.
For the three cases in the main text, $\Psi_1(O_i ; \alpha^* )$ and $\Psi_2(O_i ; \alpha^*, \beta^*)$ are represented as follows:
\begin{itemize}
    \item[(i)] Identity $Y$ Link and Identity $W$ Link Case: $g_1$ and $g_2$ are identity functions  
    \begin{align*}
        &E[W|A, Z, X] =\alpha_0^* + \alpha_a^*A +\alpha_z^{*T}Z + \alpha_x^{*T}X
	 \numberthis \label{a11} \\
	&S(\alpha^*)= \alpha_a^*A +\alpha_z^{*T}Z  + \alpha_x^{*T}X \numberthis \label{a12} \\
	&E[Y | A, Z, X] = \beta_0^*+\beta_a^*A+\beta_u^*S(\alpha^*) + \beta_x^{*T}X \numberthis \label{a13} \\
    & \Psi_1(O ; \alpha^*) 
    = 
    \begin{bmatrix}
        1 \\ A \\ Z \\ X 
    \end{bmatrix}
    \Big\{
        W -  \big( \alpha_0^* + \alpha_a^* A + \alpha_z^{*T} Z + \alpha_x^{*T}X \big)
    \Big\}  
    \\
    & \Psi_2(O ; \alpha^*, \beta^*) 
    = 
    \begin{bmatrix}
        1 \\ A \\ S(\alpha^*) \\ X
    \end{bmatrix}
    \Big\{
        Y - \big( \beta_0 + \beta_a^* A + \beta_u^*S(\alpha^*) + \beta_x^{*T}X \big)
    \Big\} 
    \end{align*}

    \item[(ii)] Log $Y$ Link and Log $W$ Link Case: $g_1$ and $g_2$ are log functions  
    \begin{align*}    
    &\log(E[W|A,, Z, X]) =\alpha_0^* + \alpha_a^*A +\alpha_z^{*T}Z  + \alpha_x^{*T}X\numberthis \label{a21} \\
	&S(\alpha^*)= \alpha_0^* +\alpha_a^*A +\alpha_z^{*T}Z  + \alpha_x^{*T}X \numberthis \label{a22} \\
	&\log(E[Y|A, Z, X]) = \beta_0^*+\beta_a^*A+\beta_u^*S(\alpha^*) + \beta_x^{*T}X \numberthis \label{a23} \\
    & \Psi_1(O ; \alpha^*) 
    = 
    \begin{bmatrix}
        1 \\ A \\ Z \\X
    \end{bmatrix}
    \Big\{
        W -   {}exp \big( \alpha_0^* + \alpha_a^* A + \alpha_z^{*T} Z + \alpha_x^{*T} X \big)
    \Big\}
    \\
    & \Psi_2(O ; \alpha^*, \beta^*) 
    = 
    \begin{bmatrix}
        1 \\ A \\ S(\alpha^*) \\ X
    \end{bmatrix}
    \Big\{
        Y - {}exp\big( \beta_0^* + \beta_a^* A + \beta_u^*S(\alpha^*) + \beta_x^{*T}X \big)
    \Big\} 
    \end{align*}

 \item[(iii)] Logit $Y$ Link and Logit $W$ Link Case: $g_1$ and $g_2$ are logit functions. Let $\alpha_{-y}^*=(\alpha_0^*, \alpha_a^*, \alpha_z^{*T}, \alpha_x^{*T})^T$, we have  
 \begin{align*}
	&\text{logit}(f(W=1|A,Z, X, Y)) =\alpha_0^* + \alpha_a^*A +\alpha_z^{*T}Z  + \alpha_x^{*T}X + {\alpha}_y^*Y
	 \numberthis \label{a31} \\
	&S(\alpha_{-y}^*) = \alpha_0^* + \alpha_a^*A +\alpha_z^{*T}Z + \alpha_x^{*T}X \numberthis \label{a32} \\
	&\text{logit}(f(Y=1|A,Z,W)) = \beta_0^*+\beta_a^*A+\beta_u^*S(\alpha^*) + \beta_x^{*T}X + {\beta}_w^*W \numberthis \label{a33}
 \\
 & \Psi_1(O ; \alpha^*) 
    = 
    \begin{bmatrix}
        1 \\ A \\ Z  \\ X \\ Y
    \end{bmatrix}
    \Big\{
        W -   \text{expit}\big( \alpha_0^* + \alpha_a^* A + \alpha_z^{*T} Z + \alpha_x^{*T}X + {\alpha}_y^* Y \big)
    \Big\} 
    \\
    & \Psi_2(O ; \alpha^*, \beta^*) 
    = 
    \begin{bmatrix}
        1 \\ A \\ S(\alpha_{-y}^*) \\ X \\ W
    \end{bmatrix}
    \Big\{
        Y -  \text{expit}\big( \beta_0^* + \beta_a^* A + \beta_u^*S(\alpha_{-y}^*) + \beta_x^{*T}X  + {\beta}_w^* W\big)
    \Big\} 
\end{align*}

\end{itemize}

We define $\Psi(O ; \alpha^*, \beta^*) = [ \Psi_1^T(O ; \alpha^*), \Psi_2^T(O ; \alpha^*, \beta^*) ]^T$, comprising both the first-stage and second-stage estimating functions. We obtain estimators $(\hat{\alpha}^*, \hat{\beta}^*)$ by solving the estimating equation
\begin{align*}
    0 =\frac{1}{N} \sum_{i=1}^{N} \Psi(O_i ; \hat{\alpha}^*, \hat{\beta}^*) \numberthis \label{sol1}
\end{align*}
In the case of identity links, from \eqref{a11} to \eqref{a13}, it is reduced to the standard two-stage least square equations. A closed-form representation of $(\hat{\alpha}^*, \hat{\beta}^*)$ is available, which is given by 
\begin{align*}
    &
    \hat{\alpha}^*= (\mathcal{B}^T \mathcal{B})^{-1} \mathcal{B}^T\mathcal{W}
    \\
    &
    \hat{\beta}^*= (\mathcal{C}^T\mathcal{C})^{-1}\mathcal{C}^T \mathcal{Y}
\end{align*}
where
\begin{align*} 
	& \mathcal{B} =\begin{bmatrix}
1 & A_1 & Z_1 &X_1\\
\vdots & \vdots & \vdots & \vdots\\
1 & A_N & Z_N & X_N\\
\end{bmatrix} \ , 
&&\mathcal{W} =\begin{bmatrix}
W_1 \\
\vdots \\
W_N \\
\end{bmatrix} 
 \\
	& \mathcal{C} =\begin{bmatrix}
1 & A_1 & S(\hat{\alpha}^*)_1 &X_1\\
\vdots & \vdots & \vdots & \vdots\\
1 & A_N & S(\hat{\alpha}^*)_N &X_N\\
\end{bmatrix} \ , && \mathcal{Y} =\begin{bmatrix}
Y_1 \\
\vdots \\
Y_N \\
\end{bmatrix} 
\ , 
&&
S(\hat{\alpha}^*) = \begin{bmatrix}
    S(\hat{\alpha}^*)_1 \\ \vdots \\ S(\hat{\alpha}^*)_N
\end{bmatrix} = \mathcal{B}\hat{\alpha}^* = \mathcal{B}(\mathcal{B}^T\mathcal{B})^{-1}\mathcal{B}^T \mathcal{W} 
\end{align*}
In other cases, a closed-form solution of $(\hat{\alpha}^*, \hat{\beta}^*)$ may not be available. Nonetheless, one can find the solution using off-the-shelf software such as \texttt{geex} R-package \cite{geex}. \\ \\
Then, referring to standard estimating equation theory, one can establish the asymptotic normality of $(\widehat{\alpha}, \widehat{\beta})$ under regularity conditions\cite{van2000asymptotic,stefanski2002calculus}. The regularity condition from Chapter 5 of van der Vaart (2000)\cite{van2000asymptotic} is stated below for completeness:
\begin{customthm}{5.21}\label{ASN}
For each $\theta$ in an open subset of Euclidean space, let $x \mapsto \psi_{\theta}(x)$ be a measurable vector-valued function such that, for every $\theta_1$ and $\theta_2$ in a neighborhood of $\theta_0$ and  a measurable function $\dot \psi$ with $E [ \dot \psi^2 ]<\infty$,
\begin{align*}
    || \psi_{\theta_1}(x) - \psi_{\theta_2}(x) || \leq \dot \psi(x) || \theta_1 - \theta_2||.
\end{align*}
Assume that $E [ ||\psi_{\theta_0}||^2 ] <\infty$ and that the map $\theta \mapsto E[  \psi_\theta]$ is differentiable at a zero $\theta_0$, with nonsingular matrix derivative $V_{\theta_0}$. If $\mathbb{P}_n [ \psi_{\hat{\theta}_n} ] = o_p(n^{-1/2})$, and $\hat{\theta}_n \stackrel{p}{\rightarrow} \theta_0$, then
\begin{align*}
    \sqrt{n}(\hat{\theta}_n - \theta_0) = -V_{\theta_0}^{-1}  \frac{1}{\sqrt{n}} \sum_{i=1}^{n} \psi_{\theta_0}(X_i) + o_p(1). 
\end{align*}
In particular, the sequence $\sqrt{n}(\hat{\theta}_n - \theta_0)$ is asymptotically normal with mean zero and covariance matrix \\ $V_{\theta_0}^{-1} E[ \psi_{\theta_0} \psi_{\theta_0}^T ] (V_{\theta_0}^{-1})^T$.
\end{customthm}

Under regularity conditions, we have the following result:
\begin{align*}
    \sqrt{N}
    \bigg\{ 
    \begin{pmatrix}
        \hat{\alpha}^* \\ \hat{\beta}^*
    \end{pmatrix}
    - 
    \begin{pmatrix}
        \alpha^* \\ \beta^*
    \end{pmatrix}
    \bigg\}
    \stackrel{D}{\rightarrow}
    N ( 0,\mathbf{V}(\alpha^*, \beta^*)) , \quad 
    \mathbf{V}(\alpha^*, \beta^*) = \mathbf{A}(\alpha^*, \beta^*)^{-1}\mathbf{B}(\alpha^*, \beta^*) \{\mathbf{A}(\alpha^*, \beta^*)^{-1}\}^{T} \\
    \text{where }
    \mathbf{A}(\alpha^*, \beta^*)=E \bigg[ \frac{\partial}{\partial(s,t)^T}\Psi(O ; s, t) 
    \bigg|_{s=\alpha^*, t=\beta^*}\bigg]
    \text{ and }  \mathbf{B}(\alpha^*, \beta^*)= E[\Psi(O ; \alpha^*, \beta^*)\Psi(O ; \alpha^*, \beta^*)^T]
\end{align*}
Moreover, the variance matrix $\mathbf{V}(\alpha^*, \beta^*)$ can be consistently estimated by the following estimator:
\begin{align*}
	&\mathbf{V}_n(O_i,\hat{\alpha}^*, \hat{\beta}^*) = \mathbf{A}_n(O,\hat{\alpha}^*, \hat{\beta}^*)^{-1}\mathbf{B}_n(O,\hat{\alpha}^*, \hat{\beta}^*) \{\mathbf{A}_n(O,\hat{\alpha}^*, \hat{\beta}^*)^{-1}\}^{T} \\
	& \mathbf{A}_n(O,\hat{\alpha}^*, \hat{\beta}^*)=\frac{1}{N} \sum_{i=1}^N \frac{\partial}{\partial(s,t)^T}\Psi(O_i ; s, t) 
    \bigg|_{s=\hat{\alpha}^*, t=\hat{\beta}^*} \ ,  
    \
	\mathbf{B}_n(O,\hat{\alpha}^*, \hat{\beta}^*)=\frac{1}{N} \sum_{i=1}^N\Psi_i(O_i ; \hat{\alpha}^*, \hat{\beta}^*)\Psi_i(O_i ; \hat{\alpha}^*, \hat{\beta}^*)^T, 
\end{align*}
where $(\hat{\alpha}^*, \hat{\beta}^*)$ are the solutions to \eqref{sol1}. 

Let $j$ be the index corresponding to $\beta_a^*$ in the vector $(\alpha^*, \beta^*)$. Then, by taking the $j$th element of the asymptotic normality result, we establish that 
\begin{align*}
    \sqrt{N} (\hat{\beta}_a - \beta_a^* )
    =
    \sqrt{N} (\hat{\beta}_a - \beta_a )
    \stackrel{D}{\rightarrow} N(0, \sigma_a^2)
    \ , \ \sigma_a^2 = \mathbf{V}(\alpha^*, \beta^*) _{jj}
\end{align*}
where $\mathbf{V}(\alpha^*, \beta^*)_{jj}$ is the $j$th diagonal entry of $\mathbf{V}(\alpha^*, \beta^*)$. Note that $\beta_a^* = \beta_a$ for the proposed two-stage regression approaches. Moreover, $\hat{\sigma}_{a}^2 = \mathbf{V}_n(O_i,\hat{\alpha}^*, \hat{\beta}^*)_{jj}$, the $j$th diagonal entry of $\mathbf{V}_n(O_i,\hat{\alpha}^*, \hat{\beta}^*)$, is consistent for $\sigma_a^2$. Consequently, an asymptotic $100(1-\alpha)\%$ double-sided confidence interval for $\beta_a$ is given by
\begin{align*}
    \bigg( \hat{\beta}_a + z_{\alpha/2} \frac{\hat{\sigma}_a}{\sqrt{N}} , 
    \hat{\beta}_a + z_{1-\alpha/2} \frac{\hat{\sigma}_a}{\sqrt{N}}
    \bigg)
\end{align*}
where $z_{1-\alpha/2}$ is the value from the standard normal distribution for the selected confidence level $\alpha$.

While $\mathbf{B}(\alpha^*, \beta^*)$ is straightforward to estimate, the matrix $\mathbf{A}(\alpha^*, \beta^*)$ (Jacobian matrix of the estimating function) requires additional calculation. Numerical derivation may be time-consuming or risk loss of precision especially when there are many parameters. Fortunately, for the generalized linear models we covered in the manuscript, the matrix $\mathbf{A}(\alpha^*, \beta^*)$ all has closed-form expression. Note that the Jacobian matrix is
\begin{align}
\dfrac{\partial}{\partial(\alpha^*,\beta^*)^T}\Psi(O_i ; \alpha^*, \beta^*) = \begin{bmatrix}
    \dfrac{\partial}{\partial\alpha^{*T}}\Psi_1(O_i; \alpha^{*}) & 0\\
    \dfrac{\partial}{\partial\alpha^{*T}}\Psi_2(O_i;\alpha^*, \beta^*) & \dfrac{\partial}{\partial\beta^{*T}}\Psi_2(O_i;\alpha^*, \beta^*)
\end{bmatrix} \label{eq:jacobian}
\end{align}
and that the form of $\partial \Psi_2(O_i;\alpha^*, \beta^*) / \partial\alpha^*$ depends on $g_2$ but not $g_1$ as the contribution of $\alpha^*$ to $\Psi_2$ is only through $S(\alpha^*)$. Define
$$\dot S_i :=\dfrac{\partial}{\partial\alpha^{*T}}S_i(\alpha^*) =\begin{bmatrix}
    1 & & & \\
     & A_i & & \\
     & & Z_i^T & \\
     & & & X_i^T
\end{bmatrix} $$
where the empty entries are zeroes. For completeness, we list the closed-form expressions for the components of $\partial\Psi(O_i ; \alpha^*, \beta^*)/\partial(\alpha^*,\beta^*)^T$ below:

\begin{enumerate}
    \item If $g_1$ is identity link, then
    $$\dfrac{\partial}{\partial\alpha^{*T}}\Psi_1(O_i;\alpha^{*T})=  -\begin{bmatrix}
        1 \\ A_i \\ Z_i \\ X_i
    \end{bmatrix}
    \begin{bmatrix}
        1 & A_i & Z_i^T & X_i^T
    \end{bmatrix};
    $$
    \item If $g_1$ is log link, then
    \begin{align*}
    \dfrac{\partial}{\partial\alpha^{*T}}\Psi_1(O_i;\alpha^{*T})=  -
    \begin{bmatrix}
        1 \\ A_i \\ Z_i \\ X_i
    \end{bmatrix} 
    \begin{bmatrix}
        1 & A_i & Z_i^T & X_i^T
    \end{bmatrix} 
    {}exp \big( \alpha_0^* + \alpha_a^* A_i + \alpha_z^{*T} Z_i + \alpha_x^{*T} X_i \big);
    \end{align*}
    \item If $g_2$ is identity link, then
     $$ \dfrac{\partial}{\partial\alpha^*}\Psi_2(O_i ; \alpha^*, \beta^*) 
    = 
    \begin{bmatrix}
        0_{1\times (2 + p_z + p_x)}\\0_{1\times (2 + p_z + p_x)}\\\dot S_i \\ 0_{p_x\times (2 + p_z + p_x)}
    \end{bmatrix}\left\{
        Y - \big( \beta_0 + \beta_a^* A + \beta_u^*S(\alpha^*) + \beta_x^{*T}X \big)
    \right\}-\beta_u^*
    \begin{bmatrix}
        1 \\ A_i \\ S_i(\alpha^*) \\ X_i
    \end{bmatrix}\dot S_i
    $$
    and
     $$ \dfrac{\partial}{\partial\beta^*}\Psi_2(O_i ; \alpha^*, \beta^*) 
    = -\begin{bmatrix}
        1 \\ A_i \\ S_i(\alpha^*) \\ X_i
    \end{bmatrix} \begin{bmatrix}
        1 & A_i & S_i(\alpha^*) & X_i^T
    \end{bmatrix}$$
    \item If $g_2$ is log link, then
     \begin{align*} \dfrac{\partial}{\partial\alpha^*}\Psi_2(O_i ; \alpha^*, \beta^*) 
    &= 
    \begin{bmatrix}
        0_{1\times (2 + p_z + p_x)}\\0_{1\times (2 + p_z + p_x)}\\\dot S_i \\ 0_{p_x\times (2 + p_z + p_x)}
    \end{bmatrix}\left\{
        Y - {}exp\big( \beta_0 + \beta_a^* A + \beta_u^*S(\alpha^*) + \beta_x^{*T}X \big)
    \right\}-\\&\qquad \beta_u^*\begin{bmatrix}
        1 \\ A_i \\ S_i(\alpha^*) \\ X_i
    \end{bmatrix}\dot S_i{}exp\big( \beta_0^* + \beta_a^* A_i + \beta_u^*S_i(\alpha^*) + \beta_x^{*T}X_i \big)
    \end{align*}
    and
     $$ \dfrac{\partial}{\partial\beta^*}\Psi_2(O_i ; \alpha^*, \beta^*) 
    = -\begin{bmatrix}
        1 \\ A_i \\ S_i(\alpha^*) \\ X_i
    \end{bmatrix} \begin{bmatrix}
        1 & A_i & S_i(\alpha^*) & X_i^T
    \end{bmatrix}{}exp\big( \beta_0^* + \beta_a^* A_i + \beta_u^*S_i(\alpha^*) + \beta_x^{*T}X_i \big).$$
\end{enumerate}

Finally, if $g_1$ and $g_2$ are both logit link functions, we have the additional restriction of $\alpha_y^*=\beta_w^*$. If the parameters were to be estimated by solving the two estimating equations simultaneously with the restriction, the Jacobian matrix no longer has the simple form in Eq.~\eqref{eq:jacobian} because the two-stage models have shared parameters. One way to simplify the computation is that after solving the first-stage estimating equation, we use $\hat\alpha^* W$ as an offset for the second-stage model.

  The estimating equations are then
  
  \begin{align*}
      &\Psi_1(O ; \alpha^*) 
    = 
    \begin{bmatrix}
        1 \\ A \\ Z  \\ X \\ Y
    \end{bmatrix}
    \Big\{
        W -   \text{expit}\big( \alpha_0^* + \alpha_a^* A + \alpha_z^{*T} Z + \alpha_x^{*T}X + {\alpha}_y^* Y \big)
    \Big\} 
    \\
    & \Psi_2(O ; \alpha^*, \beta^*) 
    = 
    \begin{bmatrix}
        1 \\ A \\ S(\alpha_{-y}^*) \\ X
    \end{bmatrix}
    \Big\{
        Y -  \text{expit}\big( \beta_0^* + \beta_a^* A + \beta_u^*S(\alpha_{-y}^*) + \beta_x^{*T}X  + {\alpha}_y^* W\big)
    \Big\} 
\end{align*}

 and we have
\begin{align*}
    \dfrac{\partial}{\partial\alpha^{*T}}\Psi_1(O_i;\alpha^{*})&=  -\begin{bmatrix}
        1 \\ A_i \\ Z_i \\ X_i \\ Y_i
    \end{bmatrix}\begin{bmatrix}
        1 & A_i & Z_i^T & X_i^T & Y_i
    \end{bmatrix}\dfrac{{}exp(\alpha_0^* + \alpha_a^* A + \alpha_z^{*T} Z + \alpha_x^{*T}X + {\alpha}_y^* Y)}{\{1 + {}exp( \alpha_0^* + \alpha_a^* A + \alpha_z^{*T} Z + \alpha_x^{*T}X + {\alpha}_y^* Y)^T\}^2},\\
    \dfrac{\partial}{\partial\alpha_{-y}^{*T}}\Psi_2(O_i;\alpha^{*},\beta^*)&=\begin{bmatrix}
        0_{1\times (2 + p_z + p_x)}\\0_{1\times (2 + p_z + p_x)}\\\dot S_i^{-y}(\alpha_{-y}^{*T}) \\ 0_{p_x\times (2 + p_z + p_x)}
    \end{bmatrix}\left\{
        Y - \text{expit}\big( \beta_0 + \beta_a^* A + \beta_u^*S(\alpha_{-y}^{*T}) + \beta_x^{*T}X  + \alpha_y^* W\big)
    \right\}-\\&\qquad \beta_u^*\begin{bmatrix}
        1 \\ A_i \\ S_i(\alpha_{-y}^{*}) \\ X_i 
    \end{bmatrix}\dot S_i^{-y}(\alpha_{-y}^{*})\dfrac{{}exp(\beta_0^* + \beta_a^* A + \beta_u^*S_i(\alpha_{-y}^*) + \beta_x^{*T}X  + \alpha_y^* W)}{\{1 + {}exp(\beta_0^* + \beta_a^* A + \beta_u^*S_i(\alpha_{-y}^*) + \beta_x^{*T}X  + \alpha_y^* W)\}^2}\\
    \dfrac{\partial}{\partial\alpha_y^{*T}}\Psi_2(O_i;\alpha^{*},\beta^*)&= -\begin{bmatrix}
        1 \\ A_i \\ S_i(\alpha_{-y}^*) \\ X_i 
    \end{bmatrix}\alpha_y^* W \dfrac{{}exp(\beta_0^* + \beta_a^* A + \beta_u^*S(\alpha_{-y}^*) + \beta_x^{*T}X  + \alpha_y^* W)}{\{1 + {}exp(\beta_0^* + \beta_a^* A + \beta_u^*S(\alpha_{-y}^*) + \beta_x^{*T}X  + \alpha_y^* W)\}^2},\\
    \dfrac{\partial}{\partial\alpha^{*T}}\Psi_2(O_i;\alpha^{*},\beta^*) &= \begin{bmatrix}
        \dfrac{\partial}{\partial\alpha_{-y}^{*T}}\Psi_2(O_i;\alpha^{*},\beta^*) & \dfrac{\partial}{\partial\alpha_y^{*T}}\Psi_2(O_i;\alpha^{*},\beta^*)
    \end{bmatrix},\\
     \dfrac{\partial}{\partial\beta^{*T}}\Psi_2(O_i;\alpha^{*},\beta^*)&= -\begin{bmatrix}
        1 \\ A_i \\ S_i(\alpha_{-y}^*) \\ X_i 
    \end{bmatrix}\begin{bmatrix}
        1 & A_i  & S_i(\alpha_{-y}^*) & X_i^T 
    \end{bmatrix}\times\\
    &\qquad \dfrac{{}exp(\beta_0^* + \beta_a^* A + \beta_u^*S(\alpha_{-y}^*) + \beta_x^{*T}X  + \alpha_y^* W)}{\{1 + {}exp(\beta_0^* + \beta_a^* A + \beta_u^*S(\alpha_{-y}^*) + \beta_x^{*T}X  + \alpha_y^* W)\}^2},
\end{align*}

where 
$$S_i(\alpha_{-y}^*) = \alpha_0^* + \alpha_a^* A_i + \alpha_z^{*T} Z_i + \alpha_x^{*T}X_i$$
and
$$\dot S_i^{-y}(\alpha_{-y}^*) := \dfrac{\partial}{\partial\alpha_{-y}^{*T}}S_i(\alpha_{-y}^*)=\begin{bmatrix}
    1 & & & \\
    & A_i & & \\
    & & Z_i^T & \\
    & & & X_i^T 
\end{bmatrix}$$

The sandwich variance estimators are automatically produced in the R package \url{pci2s} (\url{https://github.com/KenLi93/pci2s}).

\subsection{\texttt{R} \& \texttt{SAS} Codes Demonstration for the  2-Stage GLM}
\label{demo}
In this section, we provide annotated \texttt{R}  and \texttt{SAS} codes for the two-stage GLM \textbf{Procedures 1, 2, and 3} established in the main text. Note that this is only for estimation purposes; for variance estimation, bootstrap or M-estimation methods should be used. For \texttt{R} implementation, please refer to the \texttt{pci2s} package, for which we attach a brief demonstration at the end. \\ \\
\texttt{R} Codes:
\begin{verbatim}
# Load the data into a dataframe
mydata <- data.frame(
  y = c(1, 0, 1, 0, 1, 1, 0, 1, 1, 1, 1, 1, 1, 1, 0),
  a = c(0, 0, 1, 0, 1, 1, 1, 0 ,1, 0, 1, 1, 1, 1, 0),
  z = c(3, 4, 5, 6, 1, 2, 4, 1, 1, 4, 3, 7, 1, 3, 3),
  w = c(1, 0, 0, 0, 1, 1, 1, 0 ,0, 1 ,0, 1, 0, 1, 1)
)

# Linear (Y, W) Case

# Perform the first-stage linear regression W ~ A + Z
m1 <- lm(w ~ a + z, data = mydata)

# Compute the estimated E[W|A,Z]
S <- m1$fitted.values

# Perform the second-stage linear regression Y ~ A + E[W|A,Z]
m2 <- lm(y ~ a + S, data = mydata)
summary(m2)

# Count (Y, W) Case

# Perform the first-stage Poisson regression W ~ A + Z
m1 <- glm(w ~ a + z, family = poisson(), data = mydata)

# Compute the estimated log(E[W|A,Z])
S <- m1$fitted.values

# Perform the second-stage Poisson regression Y ~ A + log(E[W|A,Z])
m2 <- glm(y ~ a + S, family = poisson(), data = mydata)
summary(m2)

# Binary (Y, W) Case

# Perform the first-stage logistic regression W ~ A + Z + Y
m1 <- glm(w ~ a + z + y, family = binomial(), data = mydata)

# Compute the estimated logit(P(W=1|A,Z,Y=1))
S <- m1$coefficients[1] + m1$coefficients[2] * mydata$a + 
  m1$coefficients[3] * mydata$z + m1$coefficients[4]

# Perform the second-stage logistic regression Y ~ A + logit(P(W=1|A,Z,Y=1)) + W
m2 <- glm(y ~ a + S + w, family = binomial(), data = mydata)
summary(m2)
\end{verbatim}

\newpage

\texttt{SAS} Codes:
\begin{verbatim}
data mydata;
    input y a z w;
    datalines;
    1 0 3 1
    0 0 4 0
    1 1 5 0
    0 0 6 0
    1 1 1 1
    1 1 2 1
    0 1 4 1
    1 0 1 0
    1 1 1 0
    1 0 4 1
    1 1 3 0
    1 1 7 1
    1 1 1 0
    1 1 3 1
    0 0 3 1
    ;
run;

* Linear (Y, W) Case;
* Perform the first-stage linear regression W ~ A + Z;
proc reg data=mydata;
    model w = a z;
    output out=first_stage p=S;
run;

* Perform the second-stage linear regression Y ~ A + E[W|A,Z];
proc reg data=first_stage;
    model y = a S;
run;

* Count (Y, W) Case;
* Perform the first-stage Poisson regression W ~ A + Z;
proc genmod data=mydata;
    model w = a z / dist=poisson link=log;
    output out=first_stage p=S;
run;

* Perform the second-stage Poisson regression Y ~ A + log(E[W|A,Z]);
proc genmod data=first_stage;
    model y = a S / dist=poisson link=log;
run;

* Perform the first-stage logistic regression W ~ A + Z + Y;
proc logistic data=mydata;
    model w(event='1') = a z y;
    ods output ParameterEstimates=logistic_model;
run;

* Extract the coefficients from the logistic model;
proc sql noprint;
    select estimate into :gamma_it from logistic_model where variable='Intercept';
    select estimate into :gamma_a from logistic_model where variable='a';
    select estimate into :gamma_z from logistic_model where variable='z';
    select estimate into :gamma_y from logistic_model where variable='y';
quit;

* Compute the estimated logit(P(W=1|A,Z,Y=1));
data new_data;
    set mydata;
    S = &gamma_it + a * &gamma_a + z * &gamma_z + &gamma_y;
run;

* Perform the second-stage logistic regression Y ~ A + logit(P(W=1|A,Z,Y=1)) + W;
proc logistic data=new_data;
    model y(event='1') = a S w;
run;
\end{verbatim}

\texttt{R} demonstration of for using \texttt{pci2s} package:
\begin{verbatim}
# ?devtools::install_github
devtools::install_github("KenLi93/pci2s")
library(pci2s)


# Load the data into a dataframe
Y <- c(1, 0, 1, 0, 1, 1, 0, 1, 1, 1, 1, 1, 1, 1, 0)
A <- c(0, 0, 1, 0, 1, 1, 1, 0 ,1, 0, 1, 1, 1, 1, 0)
Z <- c(3, 4, 5, 6, 1, 2, 4, 1, 1, 4, 3, 7, 1, 3, 3)
W <- c(1, 0, 0, 0, 1, 1, 1, 0 ,0, 1 ,0, 1, 0, 1, 1)

# Linear (Y, W) Case
p2sls_lm_results <- p2sls.lm(Y = Y, A = A, Z = Z, W = W, Xw = cbind(1, A, Z))
p2sls_lm_results$summary_first_stage
p2sls_lm_results$summary_second_stage

# Count (Y, W) Case
p2sls_loglin_results <- p2sls.loglin(Y = Y, A = A, Z = Z, W = W, Xw = cbind(1, A, Z))
p2sls_loglin_results$summary_first_stage
p2sls_loglin_results$summary_second_stage

# Binary (Y, W) Case
p2sls_logit_reg_results <- p2sls.logitreg(Y = Y, A = A, Z = Z, W = W)
p2sls_logit_reg_results$summary_first_stage
p2sls_logit_reg_results$summary_second_stage


\end{verbatim}
\end{appendices}
\end{document}